\documentclass[prx,a4paper,aps,reprint,superscriptaddress,longbibliography]{revtex4-2}
\usepackage{hyperref}
\usepackage{color}  
\usepackage{tikz}
\usepackage{tikz-cd}
\usepackage{pgf}
\usepackage{lineno}

\usepackage[normalem]{ulem}
\usepackage{qcircuit}
\usepackage{amsmath}  
\usepackage{amsthm}
\usepackage{amssymb}  
\usepackage{graphicx}
\usepackage{physics}

\usepackage[linesnumbered, ruled]{algorithm2e}
\newtheorem{theorem}{Theorem}
\newtheorem{corollary}{Corollary}

\newtheorem{proposition}{Proposition}
\newtheorem{definition}{Definition}
\newcommand{\Z}{\mathbb{Z}}
\newcommand{\mi}{\mathrm{i}}
\newcommand{\QFT}{\mathrm{QFT}}
\newcommand{\SWAP}{\mathrm{SWAP}}

\newcommand{\WQFT}{\mathrm{WQFT}_{\vec{\theta}}}

\makeatletter

\begin{document}

\title{Information compression via hidden subgroup quantum autoencoders}
\author{Feiyang Liu}
\affiliation{Department of Physics, City University of Hong Kong, Tat Chee Avenue, Kowloon, Hong Kong SAR, China}
\affiliation{Shenzhen Institute for Quantum Science and Engineering and\\ Department of Physics, SUSTech, Nanshan District, Shenzhen, China}
\author{Kaiming Bian}
\affiliation{Shenzhen Institute for Quantum Science and Engineering and\\ Department of Physics, SUSTech, Nanshan District, Shenzhen, China}

\author{Fei Meng } 
\email{feimeng@cityu.edu.hk}
\affiliation{Department of Physics, City University of Hong Kong, Tat Chee Avenue, Kowloon, Hong Kong SAR, China}
\affiliation{Shenzhen Institute for Quantum Science and Engineering and\\ Department of Physics, SUSTech, Nanshan District, Shenzhen, China}

\author{Wen Zhang}
\email{zhangwen20@huawei.com}
\affiliation{HiSilicon Research, Huawei Technologies Co., Ltd., Shenzhen, China}

\author{Oscar Dahlsten}
\email{oscar.dahlsten@cityu.edu.hk}
\affiliation{Department of Physics, City University of Hong Kong, Tat Chee Avenue, Kowloon, Hong Kong SAR, China}
\affiliation{Shenzhen Institute for Quantum Science and Engineering and\\ Department of Physics, SUSTech, Nanshan District, Shenzhen, China}
\affiliation{Institute of Nanoscience and Applications, Southern University of Science and Technology, Shenzhen 518055, China}

\date{\today}

\begin{abstract}
 {We design a quantum method for classical information compression that exploits the hidden subgroup quantum algorithm. We consider sequence data in a database with {\em a priori} unknown symmetries of the hidden subgroup type. We prove that data with a given group structure can be compressed with the same query complexity as the hidden subgroup problem, which is exponentially faster than the best known classical algorithms. We moreover design a quantum algorithm that variationally finds the group structure and uses it to compress the data. There is an encoder and a decoder, along the paradigm of quantum autoencoders. After the training, the encoder outputs a compressed data string and a description of the hidden subgroup symmetry, from which the input data can be recovered by the decoder. In illustrative examples, our algorithm outperforms the classical autoencoder on the mean squared value of test data. This classical-quantum separation in information compression capability has thermodynamical significance: the free energy assigned by a quantum agent to a system can be much higher than that of a classical agent. Taken together, our results show that a possible application of quantum computers is to efficiently compress certain types of data that cannot be efficiently compressed by current methods using classical computers.}
\end{abstract}

\maketitle

\section{Introduction}
Information compression is the ubiquitous task of reducing data size by appropriate encoding, increasing information storage and transmission efficiency~\cite{goldberg1991compression}. A prominent method of compression is to employ autoencoders~\cite{ackley1985learning,kingma2019introduction},  artificial neural networks that automatically learn to compress unlabeled data without prior knowledge of the underlying patterns.  Exploiting a bottleneck structure, an autoencoder can automatically extract essential features as compressed data in such a manner that the original data can be reconstructed~\cite{liou2008modeling,liou2014autoencoder,biamonte2017quantum}. 

Recently, the quantum version of autoencoders, to our knowledge first proposed independently in Refs.~\cite{wan2017quantum, romero2017quantum}, has similarly been applied to encode~\cite{bravo2021quantum}, compress~\cite{wan2017quantum, romero2017quantum,cao2021noise}, classify and denoise~\cite{bondarenko2020quantum} quantum data and been implemented experimentally~\cite{zhou2022preserving,   huang2020realization, zhang2022resource, ding2019experimental,  pepper2019experimental}. Quantum autoencoders can, via unitary circuits, compress data hidden in quantum superpositions, but it is natural to wonder if they can also be valuable in compressing {\em classical} data.

\begin{figure}

\includegraphics[width=\linewidth]{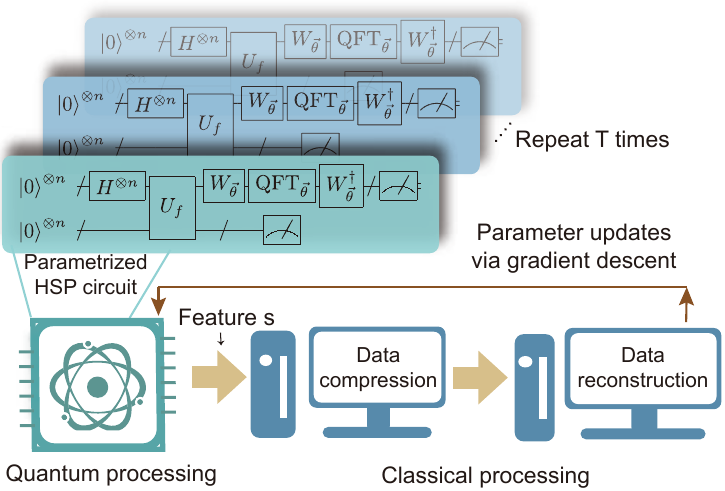}

\caption{{\bf  {The information compression algorithm.}} Our quantum auto-encoder first applies a parametrized hidden sub-group problem (HSP) circuit using an oracle $U_f\ket{i}\ket{0} = \ket{i} \ket{f(i)}$ and a parametrized quantum Fourier transform QFT$_{\vec{\theta}}$ for $T$ times to extract the hidden subgroup structure as features $s$.  {$W_{\vec{\theta}}$ is a parameterized gate that permutes qubits to search over broader range of group structures. The parameters $\vec{\theta}$ are tuned using gradient descent until the circuit identifies features $s$ representing reversible compressions of the data from the given source.} }
\label{fig:overallcircuit}
\end{figure}

Quantum algorithms are indeed able to extract certain features in data that are not efficiently accessible to classical computers. For instance, the quantum period finding algorithm~\cite{kaplan2016breaking,wang2021quantum,ben2020symmetries} is exponentially faster than the best current classical algorithms in identifying the period of a function.
More generally, the Abelian hidden subgroup problem (HSP) represents a broad class of problems (including period finding) that do not have known efficient classical algorithms, whereas efficient quantum algorithms often exist~\cite{kitaev1995quantum,ettinger2004quantum,lomont2004hidden,jozsa2001quantum, nielsen2002quantum}. 

Can that quantum speedup on HSPs be turned into an advantage for information compression? To tackle this question, we combine quantum autoencoders and the quantum HSP algorithm to create a concrete algorithm to compress data with symmetries of the HSP type.

We prove an exponential speedup in query complexity of quantum algorithms in data compression with symmetries of the hidden subgroup type, extending the quantum computational advantage in HSP to data compression. Then, we extend this algorithm to a variational quantum auto-encoder (Fig.~\ref{fig:overallcircuit}) by designing a parameterized quantum circuit for HSP, making the variational algorithm capable of finding the hidden subgroup automatically. This is achieved by establishing a parameterized quantum circuit ansatz for quantum Fourier transforms that covers a wide range of the Abelian HSP case. We demonstrate the algorithm explicitly on simple compression examples where classical computers are used to simulate small quantum computers. 

\section{Results}

We first examine the scenario with a fixed group structure in sequential data. Our analysis demonstrates that quantum algorithms can achieve exponential acceleration in compressing this specific type of sequential data.
Next, we extend our examination to include situations with indeterminate group structures. In these cases, we introduce a variational quantum algorithm capable of autonomously searching for the appropriate group structures. These structures reveal hidden subgroup symmetries within the sequential data.

\subsection{\fontsize{17}{12} Compression with respect to fixed group structure.}
\textbf{Time series data in databases and compression via symmetry. \label{sec:preliminaries_timeseries_generation_compression}} 
We consider a database $\mathcal{Q}$ storing discrete time series (or sequential data) generated from a source of length $N$. For simplicity, (detailed reason in Supplementary Material~\ref{supplementary:assigning_group_structure_via_binary_strings}), we assume $N=2^n$ for an integer $n$.  The data is represented as an ordered sequence $\{x_0, x_1,...,x_i,...,x_{2^n-1}\}$, where each data point $x_i \in \{0,1\}^m$ is a bit string of length $m$. Using binary representation, we identify any index $i$ and a binary string $\mathbf{i}=i_1i_2...i_n = \tau(i)$. Querying $\mathcal{Q}$ on $i$ yields the value $x_i$, denoted as $\mathcal{Q}(i) = x_i$. If each sequence is generated by a function $f$, such that $x_i = f(i)$ for all $i$, then the data can be compressed as a description of $f$, which we refer to as the {\em generating function} of the time series.

Without fully recognizing the exact generating function, the data in the database can still be compressed, e.g. by identifying duplicated entries (see Methods for an example). Different levels of knowledge of the function result in varying compression ratios. For instance, a data sequence generated by function $f$ with periodicity can be compressed by retaining only the first period and marking the remaining indices free to use. If an index can be overwritten, we call it free. Even without knowing the exact function, identifying periodic symmetry allows for identifying free indices and leads to data compression.

This paper explores data compression by automatically identifying symmetries in the unknown generating function, treating the problem as a hidden subgroup problem (HSP).

\medskip
\textbf{The HSP in database compression}. A concise introduction of quantum HSP algorithms can be found in Supplementary Material~\ref{supplementary:introduction_QFT_and_Application} and ~\ref{supplementary:hidden_subgroup_problems_formal_discussion}.
Suppose that the generating function $f$ has a hidden subgroup $H$, i.e. $f(i)= f(j)$ if and only if $i-j \in H$, with respect to a given group structure $G$ on the set of indices $\{i\}_{0}^{2^n -1}$.   The value $f$ may take on a given coset is not deterministic. A coset $c_0 H$ for $c_0\in G$ is defined as the set $\{c_0 + h | h \in H\}$. Then the hidden subgroup $H$ imposes a hidden pattern in the time series data generated by the function $f$, which is a redundancy that can be eliminated to achieve data compression. We call this kind of compression  {\em hidden subgroup compression}, defined as follows,

\begin{definition}[HSP in database compression.]
For a given group structure $G$ defined on a set of indices $\{\mathbf{i}\}$ and a function $f$ which hides a subgroup $H<G$, where $<$ denotes a subgroup, the objective is to compress the database $\mathcal{Q}_{f}$ storing the function values, i.e., $\mathcal{Q}_{f}(\mathbf{i}) = f(\mathbf{i})$, removing all duplicated entries by constructing a characteristic function $c(\mathbf{i}) = 1$ if the data at entry $\mathbf{i}$ can be overwritten, and $c(\mathbf{i}) = 0$ otherwise. Furthermore, a query function $q$ should be constructed that accurately retrieves $f(\mathbf{i})$ upon query: $\mathcal{Q}'_f(q(\mathbf{i})) = f(\mathbf{i})$, for any updated database $\mathcal{Q}'$ with data modified on free indices.
\end{definition}
 {For the moment, we assume that the sequence has a hidden subgroup structure with respect to a given group. This is to establish basic concepts and tools later used in the variational version. In the next section (Sec.~\ref{sec:Compression_with_respect_to_indeterminae_group_structure}), we discuss how this assumption can be relaxed to include more types of data symmetry.}

{  For simplicity, we focus on the case where there is a single generator of the hidden sub-group, which encompasses many studied HSP instances~\cite{nielsen2002quantum}. The methods we use can naturally be extended to the case of several generators, since the HSP algorithm also works for the case of several generators~\cite{ettinger2000quantum,lomont2004hidden}.}

The HSP database compression involves identifying the hidden subgroup's generator using an HSP algorithm. Efficient construction and evaluation of both the characteristic function $c$ and query function $q$ are achievable through this generator. Consider Simon's problem~\cite{simon1997power} as an illustration: the function $f$ exhibits symmetry, $f(\mathbf{i}) = f (\mathbf{j})$ if and only if $\mathbf{i} = \mathbf{j} \oplus \mathbf{s}$, where $\mathbf{s}$ is an unknown secret bit string. This $\mathbf{s}$ acts as the generator for the hidden subgroup ${\mathbf{0},\mathbf{s}}$, employing bit-wise addition $\oplus$ as the group operation. Once $\mathbf{s}$ is acquired through the HSP algorithm, constructing the characteristic function $c$ proceeds by
\begin{equation}
    c(\mathbf{i}) = \begin{cases}
        1 & \text{ if $\mathbf{i\oplus s < i}$}\, , \\
        0 & \text{otherwise,}
    \end{cases}
\end{equation}
where we say $\mathbf{i}<\mathbf{j}$ if and only if $\tau(\mathbf{i})<\tau(\mathbf{j})$.
In other words, for any pair of two indices $\mathbf{i,j}$ such that  $f(\mathbf{i}) = f(\mathbf{j})$, we keep the smaller index and set the larger index free. Therefore, we can construct the query function $q$ by querying the smaller index,
\begin{equation}
    q(\mathbf{i}) = \min \{\mathbf{i}, \mathbf{i \oplus s}\} \, .
\end{equation}
This process frees half the database capacity for new data storage. The best classical HSP algorithm, as proven in Ref.~\cite{simon1997power}, demands an exponential query count to uncover the secret key $\mathbf{s}$ from the database $\mathcal{Q}$. Conversely, the quantum HSP algorithm requires only $O(n)$ queries, presenting an exponential acceleration. Consequently, the quantum HSP's exponential speedup directly enhances database compression.

More generally, the HSP quantum algorithm can be used to solve the HSP database compression. Identifying the generator of the hidden subgroup $H$ and the characteristic function can be constructed as
$c(\mathbf{i}) = 1 $ if $ \mathbf{i} \mod H < \mathbf{i}$ and $c(\mathbf{i}) = 0$ otherwise. The query function $q$ can be constructed as $q(\mathbf{i}) = \mathbf{i} \mod H$. Conversely, if all the duplicated entries in the database are identified, the characteristic function can be used to find $H$. Using the bisection method, $O(\log |G|)$ queries to $c(\cdot)$ can identify the largest non-free index $i_f$, and $i_f + 1$ will be the generator of $H$, with $+$ being the group operation. This established the equivalence (up to polynomial overhead) between the HSP and the hidden subgroup compression, summarized in the following theorem.
\begin{theorem}[Exponential speedup in database compression.]
The query complexity of the database compression is the same as that of the corresponding HSP. 
\end{theorem}

According to the theorem, the exponential speedup of quantum algorithms over their classical counterparts in the Hidden Subgroup Problem (HSP) can be naturally extrapolated to the database compression problem. This is exemplified by period finding~\cite{shor1994algorithms,shor1999polynomial} and Simon's problem~\cite{simon1997power}. Note that the query complexity is defined as the number of times a black-box function is invoked, irrespective of the time consumed per call~\cite{nielsen2002quantum}. For the Abelian HSP, the most efficient known classical algorithm~\cite{nayak2021deterministic,ye2022deterministic} exhibits a query complexity of $O(\sqrt{\frac{|G|}{|H|}})$. In some instances, the query complexity is $\omega(\sqrt{\frac{|G|}{|H|}})$\cite{nayak2021deterministic,ye2022deterministic}, and specifically, $\Theta(\sqrt{|G|})$ for Simon's problem\cite{simon1997power}). Here, $O$ represents the asymptotic upper bound, $\omega$ denotes the asymptotic lower bound, and $\Theta$ signifies both the asymptotic upper and lower bounds. However, the quantum Abelian HSP algorithm displays a query complexity of $O(\log|G|)$, which is always polynomial~\cite{ettinger2004quantum}.

\subsection{Compression with respect to indeterminate group structure.}\label{sec:Compression_with_respect_to_indeterminae_group_structure}

\textbf{Searching over group structures.} In practice, the time series data generally does not come with a pre-defined group structure that is {\em a priori} known. Randomly guessing a group structure can lead to failures when attempting to compress the data, as patterns may not align with hidden subgroups of the group (see Supplementary Material~\ref{supplementary:searching_over_isomorphisms} for an example). The standard quantum algorithm for HSP can only be directly applied when the generating function has an unknown subgroup with respect to a known group.  {For example, all periodic data sequences with length $N$ have a hidden subgroup structure with respect to $\Z_N$. Without knowing that the group structure is $\Z_N$, the HSP algorithm cannot be applied.}

This limitation can be overcome by modifying the hidden subgroup algorithm using variational quantum circuits that adjust themselves based on parameters, allowing for the search for suitable group structures. By exploring different group types and isomorphisms of assigning them, more classes of time series data can be compressed. 
We formulate the variational hidden subgroup compression as follows,
\begin{definition}[Variational hidden subgroup compression]
Let $\mathcal{Q}_f$ be a database storing time series data $\{f(0), f(1), ..., f(2^n-1)\}$, generated by the function $f$. We say $\mathcal{Q}_f$ has a hidden subgroup structure if there exists a group structure $G=(\{0,1\}^n,*)$ defined on the indices $\{\mathbf{i}\}$ and a subgroup $H< G$, satisfying $f(\mathbf{i} * \mathbf{h}) = f(\mathbf{i})$ if and only if $\  \mathbf{h} \in H$. The variational hidden subgroup compression is to compress the time series data by varying the group structures to search for the correct operation $*$ and the hidden subgroup $H$, using queries on $\mathcal{Q}_f$.
\label{def:variational_HSP_compression}
\end{definition}

The variational hidden subgroup compression may not have a unique solution. As our goal is data compression, identifying just one solution suffices. For example, consider the time series $\{0, 1, 2, 3, 0, 1, 2, 3\}$, which exhibits multiple hidden subgroup structures, like $(\{0,4 \}, *) < \mathbb{Z}_8$ and $(\{(0,0,0),(1, 0, 0)\},*) < \mathbb{Z}_2^3$.  {Here $\Z_8$ is the group formed by integers with addition modulo $8$, and $\Z_2^3$ means $\Z_2\times \Z_2 \times \Z_2$.}

Next, we present our main results on solving this variational hidden subgroup compression problem using variational quantum circuits,  without prior knowledge of the group $G$. This is achieved by replacing the quantum Fourier transform (QFT, see Supplementary Material~\ref{supplementary:introduction_QFT_and_Application}) in a standard HSP circuit with a parameterized circuit, which recovers the QFT over any finite Abelian group $G$   {of the form $\Z_{2^{m_1}}\times \Z_{2^{m_2}} \times...\times \Z_{2^{m_q}}$},  
based on the chosen parameters. We then incorporate this parametrized hidden subgroup algorithm into an autoencoder structure. The autoencoder compresses the data through its bottleneck using the encoder and restores it using the decoder, ensuring reversible compression. During training, an automatic search finds suitable parameters to make the decoder's output as similar as possible to the input. Once training is complete, the decoder can be removed and the encoder acts as a compressor.

\medskip
\textbf{Searching over different types of Abelian groups.}
We present here a parametrized quantum circuit that is capable of implementing the quantum Fourier transform over any  Abelian group  {of the form $\Z_{2^{m_1}}\times \Z_{2^{m_2}} \times...\times \Z_{2^{m_q}}$}, with suitable choices of parameters. We will begin with examples and then present the general architecture.  {The quantum Fourier transform over a group $G$ is defined as a unitary,
\begin{equation}
\mathrm{QFT}_{\Z_N} = \frac{1}{\sqrt{N}} \sum_{i,j \in G} \exp(\frac{2 \pi \mi\, j i}{N}) \ket{j} \bra{i} \, .
\end{equation}
The quantum Fourier transform over a Cartesian product of two groups $\Z_{N_1}\times \Z_{N_2}$ is the tensor product of the respective quantum Fourier transform~\cite{nielsen2002quantum,moore2006generic}, i.e., $\QFT_{\Z_{N_1}\times \Z_{N_2}} = \QFT_{\Z_{N_1}} \otimes \QFT_{\Z_{N_2}}$.}

 {Let us now find the patterns of the quantum circuits for quantum Fourier transform in order to formulate a parametrised circuit that generalises these transforms. } 
Consider the groups $G\cong \Z_8$, $\Z_2\times \Z_4$, $\Z_4 \times \Z_2$, or $\Z_2^3$ (detailed definition of $G$ is in Supplementary Material~\ref{supplementary:assigning_group_structure_via_binary_strings}).  As proved in Supplementary Material~\ref{supplementary:efficient_circuit_QFT}, the quantum circuit for the Fourier transform over $\Z_8$ is
\begin{equation}
\begin{array}{c}
\Qcircuit @C=.3em @R=0em @!R {
\lstick{i_1}& \multigate{2}{\QFT_{\Z_8}} & \qw &                                          & & \gate{H} & \gate{R_2}   & \gate{R_3}    & \qw        &\qw            &\qw      &\qw &\qswap   &\qw&\qw\\
\lstick{i_2}& \ghost{{\QFT_{\Z_8}}}      & \qw & \push{\rule{.3em}{0em}=\rule{.3em}{0em}} & & \qw      & \ctrl{-1}    &\qw            &\gate{H}    &\gate{R_2}     &\qw      &\qw &\qw\qwx   &\qw&\qw\\
\lstick{i_3}& \ghost{{\QFT_{\Z_8}}}      & \qw &                                          & & \qw      & \qw          & \ctrl{-2}     &\qw         &\ctrl{-1}      &\gate{H} &\qw &\qswap \qwx  &\qw&\qw\\
}\, 
\end{array},
\end{equation}
where $H$ is the Hadamard gate, and $R_k = \begin{bmatrix}
    1&0\\
    0& e^{\frac{2 \pi \mi }{2^k}}
\end{bmatrix}$ is a phase rotation gate.
The  circuit for quantum Fourier transform over $\Z_4 \times \Z_2$  is
\begin{equation}
\begin{array}{c}
\Qcircuit @C=.4em @R=0em @!R {
\lstick{i_1} & \multigate{2}{\QFT_{\Z_4 \times \Z_2}} & \qw &                                          & & \gate{H} & \gate{R_2}   &\qw     & \qw        &\qw     &\qw       &\qw &\qswap        &\qw &\qw \\
\lstick{i_2} & \ghost{{\QFT_{\Z_4 \times \Z_2}}}      & \qw & \push{\rule{.3em}{0em}=\rule{.3em}{0em}} & & \qw      & \ctrl{-1}    &\qw     &\gate{H}    &\qw     &\qw       &\qw & \qswap \qwx  &\qw &\qw \\
\lstick{i_3} & \ghost{{\QFT_{\Z_4 \times \Z_2}}}      & \qw &                                          & & \qw      & \qw          &\qw     &\qw         &\qw     &\gate{H}  &\qw &\qw           &\qw &\qw \\
} \, 
\end{array}.
\end{equation}
Note that we can also have the quantum Fourier transform over $\Z_2 \times \Z_4$, which is
\begin{equation}
\begin{array}{c}
    \Qcircuit @C=.4em @R=0em @!R {
\lstick{i_1}& \multigate{2}{\QFT_{\Z_2 \times \Z_4}} & \qw &                                          & & \gate{H} & \qw          &\qw          & \qw        &\qw     &\qw       &\qw &\qw           &\qw &\qw \\
\lstick{i_2}& \ghost{{\QFT_{\Z_2 \times \Z_4}}}      & \qw & \push{\rule{.3em}{0em}=\rule{.3em}{0em}} & & \qw      & \gate{H}     &\qw          &\gate{R_2}  &\qw     &\qw       &\qw &\qswap        &\qw &\qw \\
\lstick{i_3}& \ghost{{\QFT_{\Z_2 \times \Z_4}}}      & \qw &                                          & & \qw      & \qw          &\qw          &\ctrl{-1}   &\qw     &\gate{H}  &\qw & \qswap \qwx  &\qw &\qw \\
} \, 
\end{array}.
\end{equation}
Finally, the quantum circuit for the quantum Fourier transform over $\Z_2^3$ is
\begin{equation}
\begin{array}{c}
\Qcircuit @C=.5em @R=0em @!R {
\lstick{i_1} & \multigate{2}{\QFT_{\Z_2^3}} & \qw &                                          & & \gate{H} &\qw     & \qw        &\qw     &\qw       &\qw \\
\lstick{i_2} & \ghost{{\QFT_{\Z_2^3}}}      & \qw & \push{\rule{.3em}{0em}=\rule{.3em}{0em}} & & \qw      &\qw     &\gate{H}    &\qw     &\qw       &\qw \\
\lstick{i_3} & \ghost{{\QFT_{\Z_2^3}}}      & \qw &                                          & & \qw      &\qw     &\qw         &\qw     &\gate{H}  &\qw \\
} \, 
\end{array}.
\end{equation}
The quantum circuit of the Fourier transform over $\Z_4 \times \Z_2$, $\Z_2 \times \Z_4$  and $\Z_2^3$ can be obtained by shutting down certain controlled phase rotation gates $R_k$ in the QFT circuit of $\Z_8$, followed by some swap operations among the qubits. In fact, three parameters will be sufficient to control the circuit. Denote the parameterized swap operation over all the three qubits as
\begin{equation}
    \SWAP_{{\vec{\theta}}} = \SWAP_{12}^{\theta_1} \, \,  \SWAP_{23}^{\theta_2} \SWAP_{12}^{\theta_1 \theta _2}\, ,
\end{equation}
where $\SWAP_{ij}^{\theta}$ is the unitary that swaps qubit $i$ and qubit $j$ when $\theta =1 $, and  does nothing when $\theta = 0 $. Therefore, the following circuit 
\begin{equation}\label{eq:parameterized_QFT_3qubits}
\Qcircuit @C=.25em @R=0em @!R {
& \multigate{2}{\QFT_{\vec{\theta}}}  & \qw &                                          & & \gate{H} & \gate{R_2^{\theta_1}}   & \gate{R_3^{\theta_3}}    & \qw        &\qw                       &\qw      &\qw & \multigate{2}{\SWAP_{{\vec{\theta}}}}             & \qw \\
& \ghost{\QFT_{\vec{\theta}}}         & \qw & \push{\rule{.1em}{0em}=\rule{.1em}{0em}} & & \qw      & \ctrl{-1}               &\qw                       &\gate{H}    &\gate{R_2^{\theta_2}}     &\qw      &\qw & \ghost{\SWAP_{{\vec{\theta}}}}       &\qw\\
& \ghost{\QFT_{\vec{\theta}}}         & \qw &                                          & & \qw      & \qw                     & \ctrl{-2}                &\qw         &\ctrl{-1}                 &\gate{H} &\qw & \ghost{\SWAP_{{\vec{\theta}}}}       & \qw \\
}
\end{equation}
can implement the quantum Fourier transform over all considered types of groups by suitable parameters. With $\theta_1 = \theta_2 = \theta_3 = 1$, it implements the quantum Fourier transform over $\Z_8$; with $\theta_1=1, \theta_2 =\theta_3 = 0$, it implements QFT over $\Z_4\times\Z_2$; with $\theta_1 = \theta_3 =0, \theta_2 = 1$, it implements QFT over $\Z_2 \times \Z_4$; and finally, with $\theta_1=\theta_2=\theta_3=0$, it implements the QFT over $\Z_2^3$.  For simplicity, in the main result part we use parameter $\theta_i$ to represent the controlling parameter, but in the numerical simulation, $\theta_i$ will be replaced with $\sin^2(\theta_i)$ to limit the value of the switch to be in the range of $[0,1]$.

Generalizing the above example, we have the following parametrized circuits for QFT.
\begin{definition}[Parametrized circuits for QFT]\label{def:parametrized_circuits_QFT}
     The circuit is constructed by first applying the Hadamard gate $H$ to every qubit and then consecutively applying
\begin{equation}
 V_{k}^{ \vec{\theta}}:=   (R_{n-k+1}^{n \rightarrow k})^{\theta_{k,n}}(R_{n-k}^{n-1 \rightarrow k})^{\theta_{k,n-1}}...(R_{2}^{k+1 \rightarrow k})^{\theta_{k,k+1}} H_{(k)}
\end{equation}
for $k=1,...,n-1$, where the $\theta_{i,j}$'s are the tuning parameters 
and where $R_m^{i\rightarrow j}$ is the phase rotation operator on the $j$-th qubit controlled by the $i$-th qubit and $H_{(k)}$ is the Hadamard gate applied to the $k$-th qubit. Finally, we apply the parametrized SWAP gates
\begin{equation}
    \SWAP_{{\vec{\theta}}} =  \SWAP_{n-1,n}^{\theta_{n-1,n}}\, \SWAP_{n-2,n-1}^{\theta_{n-2,n-1}}\, ... \,  \SWAP_{1,2}^{\theta_{1,2}}.
\end{equation}
We will denote these parametrized circuits for quantum Fourier transform over different types of groups as $\QFT_{\vec{\theta}}$,
\begin{equation}
    \QFT_{\vec{\theta}} = \SWAP_{\vec{\theta}}\, V_{n-1}^{ \vec{\theta}} \, V_{n-2}^{ \vec{\theta}}\, ... \, V_{1}^{ \vec{\theta}}  \, , 
\end{equation}
which has $\frac{n(n-1)}{2}$ parameters. It corresponds to the following quantum circuit
\begin{equation}
\begin{array}{c}
    \Qcircuit @C=.3em @R=0.2em @!R {
& \multigate{4}{\QFT_{\vec{\theta}}}  & \qw &                                                                        & & \qw & \multigate{4}{ V_{1}^{ \vec{\theta}}} &\qw &\qw&\qw&\qw & \qw & \multigate{4}{\SWAP_{{\vec{\theta}}}}  & \qw \\
& \ghost{\QFT_{\vec{\theta}}}         & \qw &                                                                        & & \qw &\ghost{ V_{1}^{ \vec{\theta}}}         &\multigate{3}{ V_{2}^{ \vec{\theta}}}  &\qw&\qw   &\qw        & \qw &\ghost{\SWAP_{{\vec{\theta}}}}                    & \qw \\
\raisebox{1em}\vdots & \nghost{\QFT_{\vec{\theta}}}& \raisebox{1em}\vdots & \push{\rule{.1em}{0em}=\rule{.1em}{0em}} & & \raisebox{1em}\vdots   &\nghost{ V_{1}^{ \vec{\theta}}} &\ghost{ V_{2}^{ \vec{\theta}}}   &\qw&\qw& \,\,\,\,\,\,\,\,\,\, \cdots &  &\ghost{\SWAP_{{\vec{\theta}}}}                    & \raisebox{1em}\vdots \\
& \ghost{\QFT_{\vec{\theta}}}         & \qw &                                                                        & & \qw &\ghost{ V_{1}^{ \vec{\theta}}}      &\ghost{ V_{2}^{ \vec{\theta}}}             &\qw&\qw & \qw &\multigate{1}{ V_{n-1}^{ \vec{\theta}}}&\ghost{\SWAP_{{\vec{\theta}}}}                    & \qw \\
& \ghost{\QFT_{\vec{\theta}}}         & \qw &                                                                        & & \qw &\ghost{ V_{1}^{ \vec{\theta}}}       &\ghost{ V_{2}^{ \vec{\theta}}}            &\qw&\qw  & \qw  &\ghost{ V_{n-1}^{ \vec{\theta}}} &\ghost{\SWAP_{{\vec{\theta}}}}                    & \qw \\
} \, 
\end{array}.
\end{equation}
\end{definition}

This parameterized quantum circuit $\mathrm{QFT}_{\vec{\theta}}$ can implement the quantum Fourier transform over Abelian groups  {of type $\Z_{2^{m_1}}\times \Z_{2^{m_2}} \times...\times \Z_{2^{m_q}}$}.

\begin{theorem} 
[ {Expressivity}]
\label{theorem:tunable_circuits_for_QFT}
The parametrized quantum circuit of $n$-qubits  constructed  {in Definition ~\ref{def:parametrized_circuits_QFT}}, can implement the quantum Fourier transform over 
 {groups of type $\Z_{2^{m_1}}\times \Z_{2^{m_2}} \times...\times \Z_{2^{m_q}}$}
with suitable choices of parameters, where $m_1, m_2,...,m_q$ form an integer partition of $n$.
\end{theorem}

To prove the theorem, it is direct to verify that the implementation of the QFT over $\Z_{2^{m_1}}\times \Z_{2^{m_2}} \times...\times \Z_{2^{m_q}}$, for $t=0,1,..., q-1$ can be achieved with parameters
\begin{equation}
    \theta_{k,k+1} = \theta_{k,k+2}= ...= \theta_{k, s(t+1)} = 1 \, ,
\end{equation}
for $k= s(t)+1, s(t)+2,...,s(t+1)-1$, where $s(t) := \sum_{j=1}^t m_j$ and $s(0):=0$, and we set all other parameters zero.  In particular, when the parameters $\theta_{i,j}$'s are all zeros, $\mathrm{QFT}_{\vec{\theta}}$ performs the QFT over $\Z_2^n$ and when the parameters $\theta_{i,j}$'s are all ones, it performs the QFT over $\Z_{2^n}$. 

\medskip
\textbf{Searching over bit-permutations.}
Searching across all possible group types is not enough to solve variational hidden subgroup compression, given the existence of multiple ways to assign the same group type (illustrated in Supplementary Material~\ref{supplementary:searching_over_isomorphisms}). It is necessary to explore diverse isomorphisms for the same group type. An isomorphism identifies each element in $\Z_{2^{m_1}}\times \Z_{2^{m_2}} \times...\times \Z_{2^{m_q}}$ with an element in $\{\mathbf{i}=\tau(i)| i = 0,1,...,2^n - 1 \}$—bit strings of length $n$ on which the group structure is defined. There are $2^n !$ isomorphisms, making exhaustive search computationally infeasible. Thus, we confine our algorithm to isomorphisms induced by bit-permutations, which rearrange bit-string digits.

For a bit-permutation $\pi\in S_n$, where $S_n$ is the symmetric group of $n$ elements, acting this on the group $G=(\{0,1\}^n, *)$ yields a distinct isomorphic group $G_\pi =(\{0,1\}^n, *')$. This isomorphism $\pi$ induces a unitary transformation:
\begin{equation}
U_{\pi} \ket{i_1i_2...i_n} = \ket{i_{\pi(1)}i_{\pi(2)}...i_{\pi(n)}} , .
\end{equation}
$U_\pi$ can be implemented using only SWAP gates. Supplementary Material~\ref{sec:parameterized_quantum_circuits_QFT_differnt_isomorphism} demonstrates that $(n-1)n/2$ parameters controlling SWAP gate activation can parametrize all bit-permutations at the circuit level.

\medskip

\textbf{Parameterized quantum circuits for variational HSP}.  Substituting the standard QFT module in an HSP circuit with the provided parametrized QFT circuits yields the circuit ansatz for the variational HSP, denoted as $\mathrm{HSP}_{\vec{\theta}}$, as follows:
\begin{corollary}[Parameterized quantum circuits for hidden subgroup problems]\label{theorem:Parameterized_Quantum_Circuits_for_HSP}
Upon suitable configuration of the parameters $\Vec{\theta}$, the following parametrized quantum circuits can solve the HSP over any type of finite Abelian group  {of the form $\Z_{2^{m_1}}\times \Z_{2^{m_2}} \times...\times \Z_{2^{m_q}}$}, that is assigned by bit-permutations,
\begin{equation}
\label{eq:ansatz}
\Qcircuit @C=.2em @R=0.2em @!R {
\lstick{\ket{i_1}}&\qw&\gate{H}&\qw     &\multigate{6}{U_f} &\qw &\multigate{3}{W_{\vec{\theta}}}&\qw &\multigate{3}{\QFT_{\vec{\theta}}} &\qw &\multigate{3}{W^\dagger_{\Vec{\theta}}} &\qw &\meter &\cw&\cw & \rstick{j_1}\\
\lstick{\ket{i_2}}&\qw&\gate{H}&\qw            &\ghost{U_f} &\qw&\ghost{W_{\vec{\theta}}}&\qw &\ghost{{\QFT_{\vec{\theta}}}} &\qw &\ghost{W^\dagger_{\Vec{\theta}}} &\qw&\meter &\cw&\cw & \rstick{j_2}\\
\lstick{\cdots}&\qw&\gate{H}&\qw               &\ghost{U_f} &\qw&\ghost{W_{\vec{\theta}}}&\qw &\ghost{{\QFT_{\vec{\theta}}}} &\qw &\ghost{W^\dagger_{\Vec{\theta}}} &\qw&\meter &\cw&\cw & \,\,\,\,\,\,\,\cdots\\
\lstick{\ket{i_n}}&\qw&\gate{H}&\qw            &\ghost{U_f} &\qw&\ghost{W_{\vec{\theta}}}&\qw &\ghost{{\QFT_{\vec{\theta}}}} &\qw &\ghost{W^\dagger_{\Vec{\theta}}} &\qw&\meter &\cw&\cw & \rstick{j_n}\\
\lstick{\ket{0}}&&\qw&\qw                        &\ghost{U_f} &\qw&\qw&\qw&\meter&&&&&&&\\
\lstick{\cdots}&&\qw &\qw                        &\ghost{U_f} &\qw&\qw&\qw&\meter&&&&&&&\\
\lstick{\ket{0}}&&\qw&\qw                        &\ghost{U_f} &\qw&\qw&\qw&\meter&&&&&&&\\
}
\end{equation}
where $W_{\vec{\theta}}$ is the parameterized unitary corresponding to bit-permutations and $U_f\ket{i} \ket{0} = \ket{i}\ket{f(i)}$.
\end{corollary}

\medskip

\textbf{Obtaining group structures from parameters.}  Given valid parameters $\vec{\theta}$, if the parametrized quantum circuits execute the quantum Fourier transform on $G_\sigma$, the group $G_\sigma$ can be determined from $\vec{\theta}$. This involves two steps: firstly, extracting the group type using parameters in $\QFT$, and secondly, deducing permutations from $W_{\vec{\theta}}$ parameters. With the established group structure, group operations between elements can be computed, revealing the hidden subgroup. Remarkably, these computations are achievable using classical algorithms on classical devices.

\subsection{Hidden subgroup auto-encoder.}
 {We now construct an auto-encoder that automatically finds the correct parameters of $\mathrm{HSP}_{\vec{\theta}}$,  compressing the input data into a description of $s$ and the group structure. We describe the autoencoder algorithm, the compression ratio, numerical experiments and discuss potential applications.}

\begin{figure}[t]
\centering
\includegraphics[width =\linewidth]{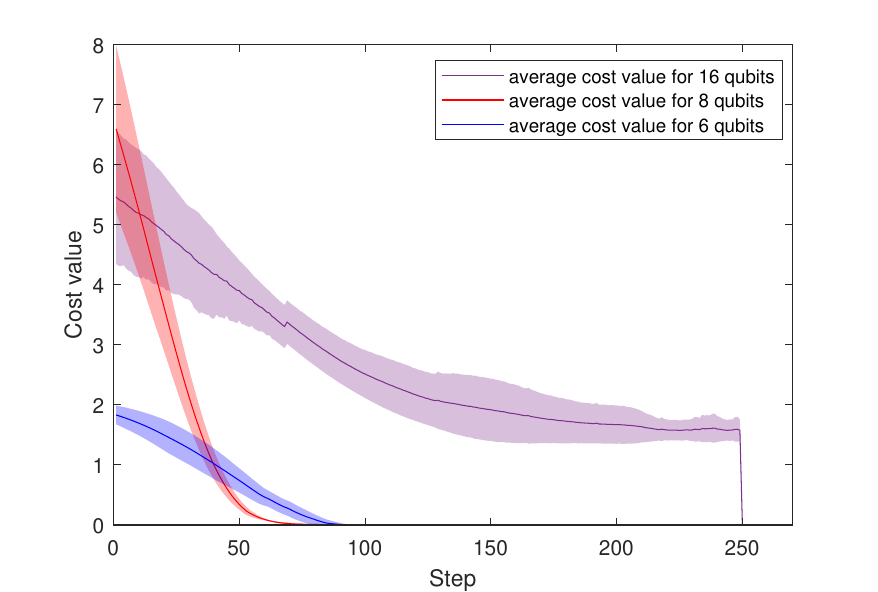}
\caption{{\bf Convergence of our algorithm under training.} The solid lines are averages and the coloured areas indicate values within one standard deviation of the mean, as obtained from 10 trainings in each case. The Ansatz of Eq.~\eqref{eq:ansatz} (with $W_{\vec{\theta}}$ being trivial) was used with the parameters initially random for each training. In one case, $2\, n = 6$ qubits are used and the data has Simon's symmetry (Eq.~\eqref{eq:simonsdata}). In a second example, $2\, n = 8$ qubits are used and the data has periodic symmetry.  {In the third example, $2\, n = 16$ qubits are used and the data has Simon's symmetry. The sudden drop at the last step appears when we choose the nearest integer for $s$.} We see that in all cases the cost function of the training data converges to 0.  {The cost function on test data from the same source is then guaranteed to also be 0 and is therefore not shown.}}
\label{fig:bigsizeperformance}
\end{figure}

\begin{figure}[t]
\centering
\includegraphics[width =\linewidth]{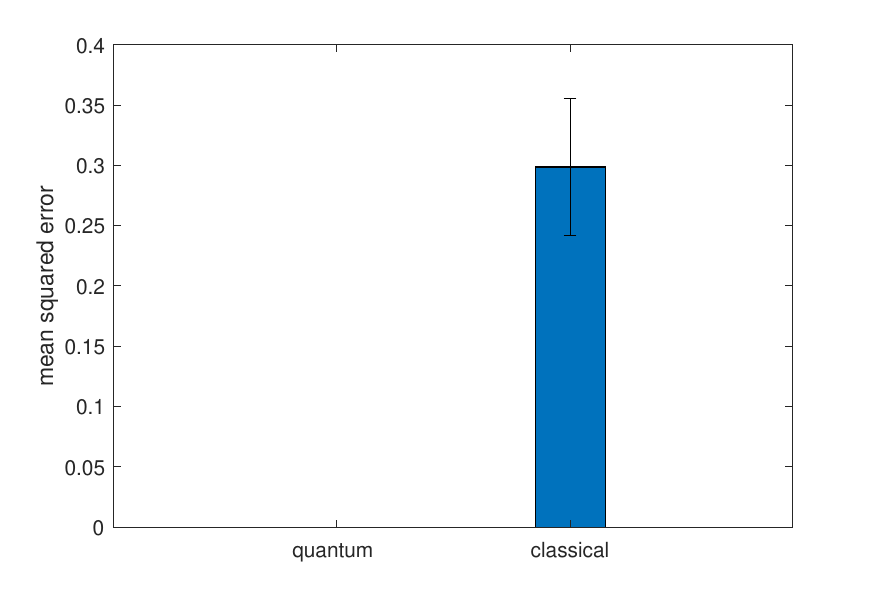}
\caption{ {{\bf Comparison of performance of a classical autoencoder and our algorithm.} The performance on test data after training is compared for the quantum case associated with $2\, n = 16$ qubits,  and the classical MATLAB autoencoder\cite{moller1993scaled}. The autoencoder has a 256-8-256 structure with 256 neurons for input and output and an 8-bit hidden layer. All the test data has the same hidden subgroup. The mean squared error is defined as $\frac{1}{256}\sum_{i\in \text{test data}} \left[ f_{in}(i) - f_{out}(i)\right]^2 \,$, where $f_{in}$ is the input data (test data) and $f_{out}$ is the output of the autoencoder. We try different initial values to train the classical autoencoder and apply the trained classical autoencoder to the test data to get the mean squared error. The mean squared error standard deviation over different trainings is also shown as a line segment on the top of the bar. The quantum autoencoder achieves negligible error on the test data, unlike the classical autoencoder.}}
\label{fig:CAE}
\end{figure}

\textbf{The  {autoencoder} algorithm.} Our primary algorithm is outlined below, with comprehensive details available in the Methods section.

We employ the quantum autoencoder design, consisting of an encoder $\mathcal{E}$ and a decoder $\mathcal{D}$. The quantum autoencoder bottleneck, designed to select specific features, generally transmits quantum states to the decoder. For our problem, we impose a classical bottleneck to ensure the classical output as the compressed data. Remarkably, a classical decoder is sufficient to reconstruct the original data, consistent with the compression being the difficult direction of a one-way function and the decoding the easy direction.  

Unlike typical autoencoders characterized by arbitrary node connections, we enforce the distinct structure denoted $\mathrm{HSP}_{\vec{\theta}}$ as defined in Eq.\eqref{eq:ansatz}, to serve as our circuit ansatz. We run $\mathrm{HSP}_{\vec{\theta}}$ for multiple times to solve a tentative hidden subgroup $H$.  Then we use classical post-processing to generate a classical message $\sigma_{classical}$ which describes the hidden subgroup $H$ and the data values $f(i)$ on different cosets $c_i H$. This classical message serves as the compressed data, which is passed to the decoder to reconstruct the data. The classical decoder determines the group structure using the message, and then for each index $i$, reconstructs $\hat{f}(i) = f(i \mod H)$, where $f(i \mod H)$ is the data value on some cosets $c_i H$ that can be decoded from the classical message $\sigma_{classical}$. To evaluate the variational autoencoder's performance, a cost function is defined using Euclidean distance between the original data and the reconstructed data,
\begin{equation}\label{eq:cost_function}
C(\Vec{\theta}) = \sum_{i\in \text{training set}} \left[ f(i) - \hat{f}(i)\right]^2 \, ,
\end{equation}
where $\hat{f}$ implicitly depends on $\vec{\theta}$. For computational efficiency, we estimate $C(\vec{\theta})$ using polynomial samples. In the Methods, we extend the binary-valued parameter $\vec{\theta}$ to the real-valued parameter $\vec{\gamma}$, enabling gradient descent.  Minimizing this cost function yields the quantum circuit for hidden subgroup compression.

\medskip
\textbf{Numerical demonstrations.}
{  We implemented the algorithm via classical simulation. In all examples examined, there is indeed convergence to error free compression, as shown in Fig.~\ref{fig:bigsizeperformance}. Nevertheless, as with gradient descent optimisation in general, we do not analytically prove that there will be convergence to 0 cost function in all cases.  

In Fig.~\ref{fig:CAE}, we demonstrate that our quantum algorithm achieves lower mean squared error on test data than a MATLAB classical autoencoder~\cite{moller1993scaled}. We do not rule out that other classical compression approaches may succeed for small input data sizes, but if a classical approach could work for very large data it would imply a classical method for efficiently solving the hidden subgroup problem, which is less likely. 

Details of the numerical experiments and links to the codes can be found in the Methods.}

\medskip
\textbf{Compression ratio of the encoder.}
 Now let us upper bound the size of the compressed data $\sigma_{classical}$ and the compression ratio. According to our circuit architecture shown in Theorem~\ref{theorem:tunable_circuits_for_QFT} and Theorem~\ref{theorem:Parameterized_Quantum_Circuits_for_HSP}, it takes $n(n-1)$ bits to specify the parameter configuration $\vec{\theta}$. Moreover, $n$ bits are needed to specify the generator of $H$, and $|G/H|m$ bits are used to store the data values $f(c)$ for each coset $c\in G/H$, where $m$ is the number of bits used to represent $f(c)$. In total, it takes no more than $n^2+|G/H| m$ bits to store the compressed data, while it takes $|G|\, m$ bits to store the full sequential data. Since $|G|=2^n$, we get a compression ratio
\begin{equation}
    \kappa \leq \frac{n^2}{2^n \, m} + \frac{1}{|H|}\, ,
\end{equation}
which is smaller than 1 for sufficiently large $n$ or $m$.
For example, let us consider the subset of all balanced binary sequences of length $2^n$, where the sequence has the same number of 0's and 1's, e.g., $\{0,0,1,1\}$. Because 0 and 1 are repeated for $2^{n-1}$ times, the subgroup $H$, if it exists, has cardinality $2^{n-1}$, which gives us a compression ratio $\kappa = \frac{n ^2 + 2}{2^n}$. For a more detailed discussion about the compression ratio, please refer to Supplementary Material~\ref{supplementary_compression_ratio}.

{ 
\medskip 
\textbf{Potential applications to real-world data.} 
Our algorithm is designed to work for input data sequences satisfying the following conditions: (i) The length of the data sequence is $2^n$ for some integer $n$. (ii) There exists an Abelian group of the type $\Z_{2^{m_1}}\times \Z_{2^{m_2}}\times ...\times \Z_{2^{m_q}}$, where $m_1+m_2+..+m_q=n$ and an isomorphism to assign this group structure to the set of indices of the data sequence up to bit-permutations, such that two data point values are the same if and only if their indices are in the same coset. Data sequences satisfying the above three assumptions can in principle be exactly solved by the variational HSP compression.

These strict assumptions can be relaxed to cover data sequences generated in reality that may violate assumptions (ii), if we allow {\em approximate} compression. For example, a data sequence of length 8, say, 01010001 that has five 0's and three 1's can never have hidden subgroup structure, since the cardinality of the subgroup must be an integer factor of the cardinality of the group. Our variational algorithm in fact allows for getting around this limitation by searching over approximate subgroup structures. When our algorithm converges to a reconstruction error that is not exactly zero but relatively small, then we get an approximate compression of the data sequence. In this way, even if the data does not have an exact group structure, we can approximate it by a sequence that is close to it. For example, 01010111 does not have a hidden subgroup, but 01010101 has and is close to the original sequence.

To this regard, our method can potentially be applied to the compression of pictures (regarded as sequential data), audio, and stock price data, where approximate reconstruction is allowed, as long as the approximation meets the respective requirements. We are presently investigating such applications.}

{ 
\subsection{Application to quantum state compression.} Our algorithm can be adapted to compress quantum states satisfying translational symmetry. Specifically, consider a set of quantum states $\mathcal{S}=\{\ket{\varphi}\,| T_{h_0} \ket{\varphi} = \ket{\varphi}\}$ that is translationally invariant under the action of the operator $T_{h_0}$, defined as
\begin{equation}
    T_{h_0} \ket{x} = \ket{x * h_0} \, , 
\end{equation}
for all computational basis $\ket{x}$, where $*$ represents the group operation of $G$ assigned to the bit strings $x$'s. Assume that there is a source that randomly generates $\ket{\varphi}^{\otimes n}$ according to a probability density $\mu_{\ket{\varphi}}$, i.e., $n$ copies of a state $\ket{\varphi}\in \mathcal{S}$ satisfying the translational symmetry. Then our variational HSP algorithm can be applied to compress the quantum states generated by the source, with two modifications. First, we do not need Hadamard gates and $U_f$, and the variational QFT is applied directly to the quantum state $\ket{\varphi}$, as shown by the following circuit,
\begin{equation}
\label{eq:ansatz}
\Qcircuit @C=.2em @R=0.2em @!R {
&\lstick{}&\qw&\qw      &\qw &\multigate{3}{W_{\vec{\theta}}}&\qw &\multigate{3}{\QFT_{\vec{\theta}}} &\qw &\multigate{3}{W^\dagger_{\Vec{\theta}}} &\qw &\meter &\cw&\cw & \rstick{j_1}\\
&\lstick{}&\qw&\qw            &\qw&\ghost{W_{\vec{\theta}}}&\qw &\ghost{{\QFT_{\vec{\theta}}}} &\qw &\ghost{W^\dagger_{\Vec{\theta}}} &\qw&\meter &\cw&\cw & \rstick{j_2}\\
&\lstick{} &\qw \inputgroupv{1}{4}{.75em}{2em}{\ket{\varphi}} &\qw                &\qw&\ghost{W_{\vec{\theta}}}&\qw &\ghost{{\QFT_{\vec{\theta}}}} &\qw &\ghost{W^\dagger_{\Vec{\theta}}} &\qw&\meter &\cw&\cw & \,\,\,\,\,\,\,\cdots\\
&\lstick{}&\qw&\qw             &\qw&\ghost{W_{\vec{\theta}}}&\qw &\ghost{{\QFT_{\vec{\theta}}}} &\qw &\ghost{W^\dagger_{\Vec{\theta}}} &\qw&\meter &\cw&\cw & \rstick{j_n}
}.
\end{equation}
Secondly, the cost function Eq.\eqref{eq:cost_function} is replaced by the infidelity between the input state and the reconstructed state. See Methods E for a more detailed description of this compression algorithm for quantum states.}

\subsection{Quantum thermodynamical advantage implied by the algorithm.}
Now we discuss the thermodynamic implications of our results. Quantum advantages for thermodynamical tasks are being sought in coherent accurate information processing~\cite{chiribella2022nonequilibrium}, in charging time of batteries~\cite{campaioli2017enhancing}, and in the efficiencies of thermal machines~\cite{hammam2022exploiting}.   Here we argue that the quantum computational advantage of HSP algorithms translates into a thermodynamic advantage in energy harvesting, enabling more work extracted from a source within a fixed amount of time.

Information can fuel energy harvesting together with heat energy; having the knowledge of the system will enable work extraction that is otherwise impossible, e.g. using Szilard engine~\cite{szilard1964decrease,bennett1982thermodynamics,dahlsten2011inadequacy}. 
The free energy of $n$ degenerate 2-level systems is $F= U - k_B T S \ln 2 =  k_B T \sum_i p_i \ln p_i$, where $T$ is the temperature of the heat bath, $k_B$ is the Boltzmann's constant,  $U=0$ is the internal energy, $S = - \sum_i p_i \log_2 p_i$ is the Shannon entropy and $p_i$ is the probability for the system being in energy level $i$. Transforming the system into the thermal state gives a free energy difference $(n-S)k_B T \ln 2$, which is the amount of work that can be extracted~\cite{bennett1982thermodynamics}.

\begin{figure}
\includegraphics[width=\linewidth]{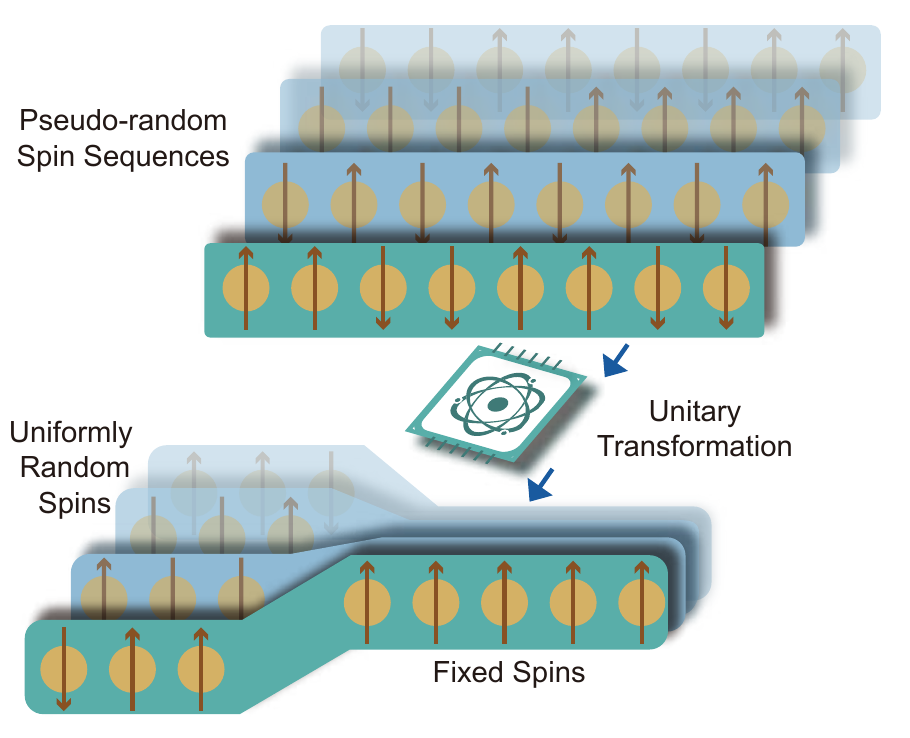}
    \caption{{\bf Compressing pseudo-random sequences encoded in spins.} After training, we can get a unitary operation that maps the pseudo-random spin sequences generated from a source to sequences whose first part is highly random spins and the second part is fixed spins. The fixed spins can then be used in work extraction.}
    \label{fig:unitary_compressor}
\end{figure}

As depicted in Fig.~\ref{fig:unitary_compressor}, we consider a source that generates pseudo-random spin sequences, where spin up encodes bit 0 and spin down encodes bit 1. Correlations among spins can be exploited to extract work, even if the marginal distributions of each spin are uniformly random~\cite{bennett1982thermodynamics}. For example, if two spins are correlated in such a way that they are pointing to either both up or both down, e.g.\ $\frac{1}{2}(\ket{00}\bra{00}+ \ket{11}\bra{11})$, then we can apply a CNOT operation and transform it to $\frac{1}{2}(\ket{0}\bra{0}+\ket{1}\bra{1})\otimes\ket{0}\bra{0}$, and can therefore deterministically extract $k_B T \ln 2$ work from the second bit. More generally, we can use a unitary transformation to convert non-uniformly distributed long spin sequences into shorter, uniformly-random spin sequences, with a fixed number of spins appended at the end, as shown in Fig.~\ref{fig:unitary_compressor}.

To demonstrate quantum energy harvesting advantage, consider a competition between classical and quantum intelligent extractors working on a pseudo-random spin source. The classical extractor uses classical neural nets or algorithms, while the quantum counterpart leverages our algorithm on a quantum computer.  Specifically, let $\mathcal{P}\subset\{0,1\}^n$ be the set of all the binary strings with length $n$ that have the same hidden subgroup structure $H<G$ for a known $G$, as specified in the Methods~\ref{sec:hidden_subgroup_as_data_compression}. Then for each run of the experiment, the source will, e.g.\ uniformly randomly, output a sequence in $\mathcal{P}$. The goal of intelligent extractors is to maximize the work extraction from this sequence within a reasonable amount of time. The competition is divided into two stages. At the first stage, the intelligent extractors are allowed to access the source for a polynomial amount of time and query and learn the patterns in the sequence generated. In the second stage, within a polynomial amount of time, the extractor will extract work from the source, and the extractor that has the largest average work extraction will win the competition.

We assert that a quantum intelligent extractor outperforms the classical version, effectively turning the quantum computational advantage into an increase in free energy. No polynomial-time classical algorithm is known for compressing this sequence, as that would imply solving the hidden subgroup problem efficiently. However, by exploiting the exponential speedup of quantum HSP algorithms, there is an efficient way to extract the largest amount of work from this pseudo-random source. Through training, our variational algorithm uncovers hidden subgroup structures. After training, we construct a unitary mapping pseudo-random $n$-spin sequences into those with the first $k_q$ spins uniformly random and the last $n-k_q$ spins all up, from which Szilard engines can extract $W_q=(n-k_q) k_B T \ln 2$ work. The classical extractor, which is unable to reveal the hidden symmetry efficiently, can only transform the sequence to $k_c \gg k_q$ spins that are uniformly random, resulting a much smaller work extraction. Thus quantum extraction strategy can extract more work $W_c \ll W_q$, demonstrating an advantage in energy harvesting using quantum algorithms.

\section{Discussion.}
We presented a novel quantum algorithm for non-linear information compression based on the hidden subgroup problem (HSP), which is subsequently extended to a variational version that can automatically adapt itself to the correct group structure. Our approach employs a versatile quantum circuit ansatz that covers diverse Abelian HSP scenarios.
This ansatz allows flexible feature extraction from input data, including period identification, which can be selectively passed through an autoencoder's bottleneck or discarded. 

Leveraging HSP's exponential advantage over classical counterparts, our method achieves an exponential advantage in query complexity for HSP-symmetric data compression. Our HSP-based compression algorithm demonstrates scalable performance ($O(n^2)$ free parameters) and convergence in numerical experiments simulating error-corrected quantum computers. Furthermore, our algorithm enhances energy extraction from pseudo-random sources, effectively allowing the assignment of higher free energy to specific sources.

Our work links the quantum HSP exponential speed-up to data compression and presents a promising new direction for research into quantum autoencoders and the HSP, paving the way for future investigations in the non-Abelian HSP context: the search for further quantum advantages in classical information compression.  Further investigations include scrutinizing the presence of HSP symmetries, both exact and approximate, in target compressible data. Experimental implementations and the impact of noise should be considered. 

It should be investigated whether recent methods on the analysis of gradient descent with variational circuits can be employed to analytically give guarantees or other insights on the convergence of the gradient descent~\cite{cerezo_cost_2021, ortiz_marrero_entanglement-induced_2021, mcclean_barren_2018}. Finally, we advocate for delving into the non-Abelian scenario, and the algorithm's potential to distinguish pseudo-random data, generated via modular exponentiation, from genuinely random data.

\section{Methods.}
\subsection{Compressing data in a database.}
\textbf{Database and logical deletion.} Compression optimizes data storage and analysis by removing duplicate data and saving resources in a database. Databases have both physical size (on hardware) and logical size (data stored). To save costs, data is logically deleted and overwritten when new data arrives~\cite{carlton2005critical,kozierok2005pc}. Our goal is to create a framework for data compression by logically removing duplicate values in a database.

More formally, let $\mathcal{Q}$ be a database that returns a value $\mathcal{Q}(i)$ upon query on index $i$. The data $\mathcal{Q}(i)$ is said to be a {\em duplicated value} if there exists an index $j$ such that $\mathcal{Q}(i) = \mathcal{Q}(j)$. We want to logically delete the duplicated value $\mathcal{Q}(i)$ by labeling the index $i$ as deleted and ready for storing new data. This logical deletion can be accomplished by a characteristic function $c$ that specifies if a position of the database is free for storing new data,
    \begin{align}
        c(i) = \begin{cases}
        1 & \mathrm{if }  \, i \, \mathrm{is \,\, free \,\, to  \,\, use,}\\
        0 & \mathrm{if }\,  i \, \mathrm{is\,\,  occupied.}
    \end{cases}
    \end{align}
With this characteristic function $c$, the database
$\mathcal{Q}$ can be freely modified on the index set $F=\{i|c(i)=1\}$, without affecting its data integrity, which is referred to as the set of free indices.  To retrieve the original data from the database after deletion, we also need a query function $q$ that helps us to retrieve the correct data in the modified database $\mathcal{Q}'$,
    \begin{equation}
        \mathcal{Q} (i) = \mathcal{Q'}(q(i)) \, ,
    \end{equation}
    for any new database state $\mathcal{Q}'$ with modifications made on the free indices $F$.

Under some conditions, a classical database can also be queried in a quantum superposition~\cite{giovannetti2008architectures}. For example, an optical device such as a compact disk storing classical data can be queried by light in a superposition of locations such that data stored at different indices can in principle be queried simultaneously (see Fig.4 in Ref.~\cite{giovannetti2008architectures}). In this paper, we regard the database $\mathcal{Q}$ as an oracle that can be queried by a quantum computer via a unitary that maps $\ket{i}\ket{0}$ to $\ket{i}\ket{\mathcal{Q}(i)}$: $\mathcal{Q}$ serves as $U_f$ in our circuit Ansatz in Eq.\eqref{eq:ansatz}. The quantum computer can prepare a superposition state $\frac{1}{\sqrt{N}} \sum_i \ket{i} \ket{0}$  to query the database and obtain a superimposed data set $\frac{1}{\sqrt{N}} \sum_i \ket{i} \ket{\mathcal{Q}(i)}$ as output, where $N$ is the number of indices in the database.

\medskip
\textbf{Compression via symmetry.} As long as there are symmetries, the data stored in the database can be compressed even if the exact expression of the generating function is not fully recognized.  Different levels of knowledge of the generating function will produce different levels of compression ratio. For example, consider the data sequence $\{1,4,9,16,25,16,9,4,1,4,9,16,25,16,...\}$ generated by a function $f$. On the first level, it is obviously periodic, with $f(i) = f(i +  8 k )$, for all integer $k$. Identifying this periodic pattern in the database helps us to reduce its redundancy by only keeping the record of its first period $\{ 1,4,9,16,25,16,9,4 \}$, and marking all other indices free to use by setting $c(i)=1$ for all $i>8$. On the second level, the numbers have reflection symmetry within one period, i.e. $f(i) = f(j)$ whenever the two indices satisfy $i+j = 10$. This reflection symmetry helps us to further compress the data to $\{1,4,9,16,25\}$. On the third level, we identify the exact expression of the generating function and the data can be further compressed to the description of the function $f(i)= i^2$ for $i = 1,2,3,4,5$, and all the indices in the database can be marked free to use. But even if we do not get the exact expression of the generating function, the periodic structure and the reflection symmetry in the data can be used to compress the data.  This type of compression can be achieved by identifying the symmetry in the data.

\medskip
\textbf{Introduction to the HSP.} The HSP is to identify hidden symmetries in functions. The well-known quantum algorithm to solve this problem has an advantage in query complexity compared with its classical counterparts~\cite{ettinger2004quantum}.  
We will show that solving the HSP can be leveraged for data compression, efficiently identifying duplicated data values.

\begin{definition}[Abelian hidden subgroup problem]
Given an Abelian group $ G $ and a subgroup $H$, suppose that a function $f: G \rightarrow \{0,1\}^m$ satisfies
\begin{equation} \label{eq:hidden_subgroup_definition}
    f(i_{1}) = f(i_{2}) \iff i_{1}-i_{2} \in H \, .
\end{equation}
We then say that $H$ is a hidden subgroup of $f$, and $f$ has the symmetry of the hidden subgroup $H$. The HSP is to find $H$, given an oracle of $f$, where the oracle is a black box that returns $f(i)$ upon each query of $i$.
\end{definition} 
An example of an HSP is to find the period of a sequence. For example, $\{1,4,9,16,25,1,4,9,16,25,1,4,9,16,25\}$ is a sequence of length $15$ which has a period of length $5$, because $f(i) = f(j )$ if and only if $i - j = 0, 5,$ or  $10$. Identifying the correct period is equivalent to identifying the correct hidden subgroup of $\Z_{15}$, and in this example, the hidden subgroup is $\{0, 5, 10\}<\Z_{15}$.

\subsection{Group structures in databases. \label{sec:hidden_subgroup_as_data_compression}}

\medskip

\textbf{Efficiently assigning group structures to the set of indices.}
To formulate database compression as an HSP, we need to establish a group structure for the indices in the database. Let us begin with examples and then present the general theory.

For instance, consider indices $i=0,1,2,3,...,7$ represented as bit strings of length 3, e.g., $\mathbf{i}=000,001,010,...,111$. Different types of group structures can be assigned: $\Z_2 \times \Z_2 \times \Z_2$, $\Z_4\times \Z_2$, or $\Z_8$. 

One approach defines the group operation as bit-wise addition modulo 2, making the group isomorphic to $\Z_2 \times \Z_2 \times \Z_2$. Alternatively, the group operation can be binary addition modulo $8$, resulting in a group isomorphic to $\Z_8$. Another method splits the bit string into two parts and defines the group operation $*$ between two bit strings $abc$ and $xyz$ as
\begin{equation}
    abc * xyz = \mathrm{Append}[ab + xy \mod 4 , c\oplus z].
\end{equation}
where ``Append" is the string operation that appends the second string to the end of the first string. This approach yields  {a group isomorphic}  to $\Z_4 \times \Z_2$, combining addition modulo $4$ for the first part and bit-wise addition $\oplus$ for the second part.

Further flexibility in assigning group structures comes from isomorphisms induced by permuting bit string digits. For instance, the group structure $\Z_4$ can be assigned to indices (bit strings of length 2) through the one-to-one correspondences (isomorphisms):
\begin{equation}
    \begin{matrix}
        \Z_4 & = & \{& 0, & 1, & 2, & 3  & \}  \\
        G_0 & = & \{& 00, & 01, & 10, & 11 &\} \\
        G_1 & = & \{& 00, & 10, & 01, & 11 &\} \\
    \end{matrix} \, ,
\end{equation}
where $G_0$ represents the canonical assignment and $G_1$ is obtained by permuting the first and second bits.

More generally, the group structure is established in three steps. Consider assigning a group $G$ isomorphic to $\Z_{2^{m_1}} \times \Z_{2^{m_2}} \times ... \times \Z_{2^{m_q}}$, with $m_1,m_2,...,m_q$ forming an integer partition of $n$ such that $m_1 +m_2+...+m_q=n$.
\begin{enumerate}
    \item Index Mapping: Identify index $i$, as an integer, with its binary representation $\mathbf{i}=\tau^{-1}(i)$.
    \item Bit-Permutation and Partition: Apply a bit-permutation $\sigma$ on binary digits of $\mathbf{i}$, $\sigma \in S_{n}$, and divide the permuted bit string $\sigma(\mathbf{i})$ into $q$ parts, each of length $m_1, m_2, ..., m_q$. Denote parts as $\mathbf{i}_1\mathbf{i}_2\dots\mathbf{i}_q$, with $\mathbf{i}_k$ as the $k$-th part of $\sigma(\mathbf{i})$.
    \item Group Operation: Define $G$ by specifying its operation $*$. For indices $i$ and $j$, first compute $\mathbf{i}_1\mathbf{i}_2...\mathbf{i}_q$ and $\mathbf{j}_1\mathbf{j}_2...\mathbf{j}_q$ and then calculate $\mathbf{c}_1\mathbf{c}_2...\mathbf{c}_q$ via
\begin{equation}
     \mathbf{c}_k = \tau^{-1}[\tau(\mathbf{i}_k) + \tau (\mathbf{j}_k) \mod 2^{m_k}].
\end{equation}
Define the group operation $*$ between $i$ and $j$:
\begin{equation}
     i * j = \tau(\mathbf{c}_1\mathbf{c}_2...\mathbf{c}_q).
\end{equation}
\end{enumerate}

Note that $\sigma$ can be selected from $S_{2^n}$, but a \textit{bit-permutation} is chosen here due to its simplicity. As permutations of digits in a bit string, bit-permutation is easily implemented with SWAP operations to shuffle digits.  While arbitrary permutations require exponential parameters, bit-permutations need only polynomial parameters. For a more detailed discussion on assigning group structures to the index set $\{i\}$ and the parametrization of bit-permutations, refer to Supplementary Material~\ref{supplementary:assigning_group_structure_via_binary_strings}.

\subsection{Quantum autoencoder for hidden subgroup compression. \label{sec:quantum_autoencoder_HSC}}
\textbf{Parameterized quantum autoencoder for finding suitable parameters}. We adopt a variational quantum algorithm approach, defining a cost function and employing gradient descent to determine the parameters 
 in $\mathrm{HSP}_{\vec{\theta}}$. Specifically, we employ the quantum autoencoder architecture. This comprises an encoder $\mathcal{E}$ and a decoder $\mathcal{D}$. The quantum autoencoder bottleneck, which selectively retains features, passes quantum states to the decoder. To achieve classical data compression, we ensure a classical bottleneck that guarantees classical compressed data. This assumes a classical connection between the encoder and decoder, and measurement of the bottleneck yields the compressed data. See the illustration below:
\begin{equation}
\begin{array}{c}
\Qcircuit @C=.3em @R=0.2em @!R {
 & & \qw                    & \multigate{3}{\mathcal{E} }  &\qw& \qw&\meter        & \qw         &\qw&\qw&\qw & \qw   & \multigate{3}{\mathcal{D}}            & \qw \\
 & & \qw                    &\ghost{\mathcal{E}}           &\qw&&              &            &&  &\qw   & \qw          &\ghost{\mathcal{D}}                    & \qw \\
 & & \raisebox{1em}\vdots   &\nghost{\mathcal{E}}          &  \raisebox{1em}\vdots&&   &            && &  \raisebox{1em}\vdots &     &\nghost{\mathcal{D}}                   & \raisebox{1em}\vdots \\
 & & \qw                    &\ghost{\mathcal{E}}           &\qw&&              &            && & \qw &\qw              &\ghost{\mathcal{D}}                    & \qw \\
} \, \end{array}.
\end{equation}
Note that the bottleneck may consist of multiple bits. The encoder captures a classical feature from the time series, treated as compressed data. This compressed data is then sent through the bottleneck to the decoder, which can reconstruct the original time series. Considering the  exponential speedup of quantum circuits over classical algorithms in detecting hidden subgroups within input bits, it is reasonable to anticipate a similar exponential advantage over classical neural networks for variational hidden subgroup data compression.

\medskip

\textbf{Real valued parameters and cost function.} 
An important note is that both the cost function shown in Eq.\eqref{eq:cost_function} and parameters are integers, rendering the problem combinatorial and unsuited for direct gradient descent. To address this, we generalize by making both parameters and the cost function real numbers, discussed as follows.

To avoid ambiguity, we use $\vec{\gamma}$ to denote the generalized, real-valued parameter configuration, and keep using $\vec{\theta}$ to denote the binary valued parameters configuration.  When parameters become real, the circuit generally does not implement any quantum Fourier transform. To obtain the group structure, we discretize $\vec{\gamma}$ back to $\vec{\theta}$.

We randomly generate the configuration $\vec{\theta}$ of binary valued parameters according to the parameters $\gamma_i$; specifically, we randomly generate $\theta_i \in \{0,1\}$ according to the probability distribution
\begin{equation}
    \Pr(\{\theta_i = 1\}) = \gamma_i\, ; \, \, \, \Pr(\{\theta_i = 0\}) = 1- \gamma_i \, .
\end{equation}
In this way, the probability of getting a specific parameter configuration $\vec{\theta}=(b_1, b_2,...., b_q)$ is 
\begin{equation}\label{eq:probability_distribution_theta}
    \Pr(\vec{\theta}) = \prod_{i = 1}^q \Pr(\{\theta_i = b_i\}) \, .
\end{equation}
For each binary configuration $\vec{\theta}$ generated in this way, we can then use the classical algorithm to obtain the compressed data. In general, this compressed data cannot exactly reconstruct the original data, at the intermediate step in the training process. This gives us a positive, integer-valued cost $C(\vec{\theta})$. Repeating this process many times, we can efficiently obtain the estimation of the average value of the cost function, 
\begin{equation}
    \mathbb{E}_{\Vec{\gamma}}[C] = \sum_{\vec{\theta}\in \{0,1\}^q} \Pr(\vec{\theta}) C(\vec{\theta}) \, .
    \label{eq:cost_function}
\end{equation}
This average cost takes real values, and it is amenable to minimization through gradient descent. Minimizing this average cost function we will get the appropriate parameters that solve the HSP.  

In the following, we will introduce the detailed structure of the encoder and the decoder, followed by a discussion of how parameters are updated to achieve successful compression.

\medskip
\textbf{Encoder structure overview.} In contrast to conventional autoencoders with arbitrary node connections and learnable weights, we impose a special structure $\mathrm{HSP}_{\vec{\theta}}$ defined in Eq.\eqref{eq:ansatz} as the circuit ansatz. This structure leverages quantum circuits' computational prowess in identifying Abelian hidden subgroups. Then we have a well-organized encoder (and decoder) as described below.

We use a quantum-classical hybrid structure for the encoder, where the quantum part is used to speed up the Fourier transform and the classical part is used to post-process the data sampled from the quantum circuit to calculate the hidden subgroup.  The procedure is summarized in Algorithm~\ref{alg:encoder} and explained as follows.

The key idea is to apply the parameterized quantum circuit $\mathrm{HSP}_{\vec{\gamma}}$ many times and then use a classical algorithm to find out the hidden subgroup. Specifically, we first apply the parameterized quantum circuits $\mathrm{HSP}_{\vec{\gamma}}$ for HSP with real valued parameters $\vec{\gamma}$, and obtain a bit string $\mathbf{j}=j_1j_2...j_n$. Repeat this process $K$ times and then we have $K$ such bit strings $\{ \mathbf{j}^{(k)}\}_{k=1}^K$. These bit strings would belong to the orthogonal group $H^\perp$ (see Supplementary Material~\ref{supplementary:introduction_QFT_and_Application} for an introduction) if the parameters $\vec{\gamma}$ correctly configure the parameterized circuit $\mathrm{HSP}_{\vec{\gamma}}$ exactly as the circuit solving the HSP of $f$, which is not the case as long as $\vec{\gamma}$ is not binary numbers. 

Then we randomly generate a configuration of binary parameters $\vec{\theta}$ according to the probability distribution defined in Eq.\eqref{eq:probability_distribution_theta}. This probability distribution does not guarantee that the parameter configuration generated gives a quantum circuit that solves a hidden subgroup. For example, $\theta_3 = 1$ while $\theta_1=0$ in Eq.\eqref{eq:parameterized_QFT_3qubits} does not correspond to any quantum Fourier transform. 
So we need to check the validity of the parameters $\vec{\theta}$ generated. 
For each instance of the parameters, we check if the parameterized circuit $\mathrm{HSP}_{\vec{\gamma}}$ corresponds to a quantum circuit that solves the HSP for some certain group $G$.  {At the end of the training we pick the valid QFT closest to the trained circuit. (This procedure is efficient in that there are poly(n) parameters and one decision per parameter).}

\begin{algorithm}[H]
  \SetAlgoLined
  \KwIn{Oracle $U_f:\ket{i}\ket{0} \mapsto \ket{i} \ket{f(i)}.$\\
        A parameter configuration $\vec{\gamma}$ with $\gamma_i\in[0,1]$.\\
        A parameter configuration $\vec{S}=\{S_1,S_2,...,S_n\}$ with $S_i\in  {[0,1]}$ \\
        An accuracy threshold $\epsilon$ and a learning rate $\beta$.}
  \KwOut{Classical message $\sigma_{classical}$ as a description of the hidden subgroup $H$ and the data values $f(i)$ on different cosets. }
  \While{The pretraining cost function $C_E$ in Eq.\eqref{eq:cost_fuction_pretraining} satisfies $\norm{\nabla C_E(\vec{\theta}, \vec{S})}\geq \epsilon$}{
    Apply the parameterized circuit for hidden subgroup $\mathrm{HSP}_{\vec{\gamma}}$ using the oracle $U_f$, and then measure the first register to get a bit string $\mathbf{j}=j_1j_2...j_n$. Repeat it $K = poly(n)$ times and obtain a set of bit strings $\{ \mathbf{j}^{(k)}\}_{k=1}^K$.\; 
    For each component $\gamma_i$ of $\vec{\gamma}$, randomly generate  $\theta_i = 0, \, \text{or} \, 1$ according to the probability distribution  $\Pr(\theta_i=1) = \gamma_i$. This gives us a binary-value parameter configuration $\vec{\theta}$. Repeat this step until we obtain legitimate $\vec{\theta}$ such that $\mathrm{HSP}_{\vec{\theta}}$ corresponds to a hidden subgroup algorithm for a certain group $G$.\;
    Use this group $G$ to calculate the linear congruence equation systems (e.g.\ Eq.\eqref{eq:linear_congruence_main_text}).\;
    Use $\vec{S}$ to generate  {the corresponding continuous bit string $\mathbf{s} = s_{m_1}s_{m_2}...s_{m_q}$ according to the equation Eq.\eqref{eq:probability_distribution_S_i} }and then evaluate the pretraining cost function $C_E$ defined in Eq.\eqref{eq:cost_fuction_pretraining}.\;
    Change each $\gamma_i$ and each $S_i$ slightly, repeat the line 7-10, and calculate the gradient $\nabla_{\vec{\gamma},\vec{S}}\, C_E$, where $\gets$ means assigning new values.\;
    }
    Now $\Vec{S}$ gives us a tentative generator for the hidden subgroup $H$ and $\vec{\gamma}$ gives us a tentative group structure $G$. Calculate the cosets $G/H$.\;
    For each coset $c\, H$ with representative $c$, query the oracle $U_f$ via state $\ket{c}\ket{0}$, and measure the second register to get $f(c)$.\;
    Encode the parameter $\vec{\theta}$, the generator $\mathbf{s}$ of the hidden subgroup $H$, and ordered pairs $(c,f(c))$ into the classical message $\sigma_{classical}$. Output $\sigma_{classical}$.
  \caption{{\bf Encoder.}} 
  \label{alg:encoder} 
\end{algorithm}

If a legitimate parameter configuration $\vec{\theta}$ is sampled, we then use the classical algorithm (see Supplementary Material~\ref{supplementary:classical_algorithm})  to solve the corresponding group structure $G$. With the group structure $G$ specified, we can find out which congruence equations (see Supplementary Material~\ref{supplementary:Simons_Algorithm}, Eq.\eqref{eq:linear_congruence_appendix}) are used to find out the generator $\mathbf{s} = s_1s_2 \cdots s_n$ of the subgroup $H$. Since the linear congruence equation systems always have solutions, we can always get such a subgroup $H<G$.  However,  $H$ is not guaranteed to be the actual hidden subgroup of $f$ at this intermediate step. We have to optimize the parameters to increase the probability of finding the correct hidden subgroup of $f$. This optimization is carried out by passing the results to the decoder, calculating the average cost function, and then using gradient descent. In fact, the parameter $\vec{\theta}$ can be used to describe the group structure. We then encode $\vec{\theta}$, the generator of $H$, and the data points $(c,f(c))$ for $c\in G/H$ into a classical message $\sigma_{classical}$; this classical message is regarded as the compressed data.

\medskip
\textbf{Pre-training of the encoder.}
In practical training, we discovered that finding the generator of the hidden subgroup by directly applying Euclid's algorithm to solve linear congruences (e.g.\ Eq.\eqref{eq:linear_congruence_main_text}) is ineffective. Instead, we employ a pre-training method to make the autoencoder functional. 
During intermediate training steps, the correct Fourier transformation remains undiscovered, and thus many incorrect bit strings are sampled (all bit strings are possible in general).
This results in the congruence equations having only the trivial solution, i.e., all bits being zero. Consequently, the classical post-processing continuously outputs the wrong generator (e.g.\ the secret key in Simon's algorithm) of the hidden subgroup as $s=0$. With this incorrect generator, the cost function remains large, and gradient descent often fails to improve the situation, leading to a barren plateau phenomenon with the output stuck at $s=0$.

To circumvent such barren plateaus, we introduce a pre-training technique to assist the encoder in identifying a non-trivial generator $s$, although this $s$ is not guaranteed to be the correct hidden subgroup's generator. The autoencoder training is thus split into two stages. First, the encoder is pre-trained to output a non-trivial $s$ (i.e., $s\neq0$). In the second step, the hidden subgroup's generator $s$ is sent to the decoder to reconstruct the data sequence and be evaluated by the overall cost function shown in Eq.~\eqref{eq:cost_function}. The autoencoder is then trained as a whole to obtain the correct generator $s$. During each iteration in the overall training, we apply the pre-training technique described below to ensure the encoder outputs a non-trivial generator $s\neq0$.

To help the encoder output a non-trivial generator $s$, we incorporate a new parameter set $\vec{S}$, where each $S_i$ takes values in the interval  {$[0,1]$. The set will be used to calculate the continuous bit string $\mathbf{s}=s_{m_1}s_{m_2}...s_{m_q}$, where $m_1+m_2+\dots m_q =n$, as follows:
\begin{equation}\label{eq:probability_distribution_S_i}
\begin{split}
s_{m_i}&= S_{m_i+m_{i-1}+\dots+m_{q}}\times 2^{m_{i}-1}+\\&S_{m_i-1+m_{i-1}+\dot+m_{q}}\times 2^{m_{i}-2}+\dots \\
&+S_{1+m_{i-1}+\dots +m_{q}}\times 2^{0}
\end{split}
\end{equation}
By Eq.\eqref{eq:probability_distribution_S_i} we will get each continuous $s_{m_i}$, which depends on the corresponding $\Z_{2^{m_i}}$ groups. To make sure $\mathbf{s}$ is valid, we choose the nearest integer for each $s_{m_i}$ in $\mathbf{s}$ before feeding it to the  decoder.}

Now we train both $\vec{\theta}$ and $\vec{S}$ to find a non-trivial generator $s$.  Take Simons' problem as an example.  {Here $m_1 = m_2 = \dots = m_q = 1$, and $q = n$, so  we will write $s_{m_i}$ as $s_i$ in Simon's problem for simplicity} (see Supplementary Material~\ref{supplementary:Simons_Algorithm} for a detailed introduction to Simon's algorithm). Suppose that the $k$-th sampled bit string after applying the intermediate HSP algorithm $\mathrm{HSP}_{\vec{\theta}}$ is $j^{(k)}$, then we have the following linear congruence relations to solve for the generator $s$ (proof in Supplementary Material~\ref{supplementary:Simons_Algorithm}),
\begin{eqnarray}
  j^{(1)}_1 s_1 + j_2^{(1)} s_2 + \cdots + j_n^{(1)} s_n & = & 0 \mod
  2 \nonumber\\
  j^{(2)}_1 s_1 + j_2^{(2)} s_2 + \cdots + j_n^{(2)} s_n & = & 0 \mod
  2 \label{eq:linear_congruence_main_text} \\
  & \vdots & \nonumber \\
  j^{(k)}_1 s_1 + j_2^{(k)} s_2 + \cdots + j_n^{(k)} s_n & = & 0 \mod 2 \nonumber
\, .
\end{eqnarray}
When $\mathbf{s}$ is not a solution to the linear congruence system, we can use $C_L(\mathbf{s})$ to quantify the discrepancy and therefore the cost,
\begin{equation}
    C_L(\mathbf{s}) = \sum_{i=1}^n (j^{(k)}_is_i \mod 2) \, .
\end{equation}
Denoting the probability distribution of output bit string $\mathbf{j}=j_1j_2...j_n$ as $P(\mathbf{j})$, we define a cost function $C_E$ for the pretraining of the encoder,
 {
\begin{align}\label{eq:cost_fuction_pretraining}
C_E =    \sum_{k =1}P(\mathbf{j}^{(k)})C_L(\mathbf{s})+C(\vec{S}),
\end{align}}
where $C(\vec{S}) = \sum_i \sin^2 (\pi S_i) $ is a penalty term introduced to push each $S_i$ takes value $0$ or $1$.
Minimizing this pretraining cost function $C_E$ we can get a non-trivial generator $s$. When the cost value $C_E$ reaches $0$, it means we find a plausible generator $s$, and then it is sent to the decoder for further processing and training.

\begin{algorithm}[H]
  \SetAlgoLined
  \KwIn{Classical message $\sigma_{classical}$ as a description of the hidden subgroup $H$ and the data values $f(i)$ on different cosets.}
  \KwOut{A reconstructed data sequence $\{\hat{f}(0),\hat{f}(1),\dots,\hat{f}(2^n-1)\}$.}
  Define an empty list $l$.\;
  Decode from $\sigma_{classical}$ the binary valued parameter $\vec{\theta}$, the generator of the hidden subgroup $H$, and the ordered data pair $(\mathbf{c},f(\mathbf{c}))$ for $\mathbf{c}\in G/H$. \;
  Calculate the group structure from the parameter $\vec{\theta}$, obtaining the group operation $*$.\;
  For each hidden subgroup element $h\in H$ and each representative $\mathbf{c}\in G/H$ of the cosets, store the value $f(\mathbf{c})$ in the position of $\tau(\mathbf{c}*\mathbf{h})$ of the list $l$.\;
  Output $l=\{\hat{f}(0),\hat{f}(1),\dots,\hat{f}(2^n-1)\}$.
  \caption{{\bf Decoder.}} 
  \label{alg:decoder} 
\end{algorithm}

\medskip
\textbf{Detailed structure of the decoder.} 
The encoder outputs a description of the subgroup $H$, the group structure $G$ and the values  $(\mathbf{c},f(\mathbf{c}))$ for the quotient group $\mathbf{c}\in G/H$. Using this information, the decoder can reconstruct the original time series by, i) creating a list of length $N$, and ii) for each $\mathbf{h}\in H$ and each $\mathbf{c} \in G/H$, store the value $f(\mathbf{c})$ into the position $\tau({\mathbf{c}*\mathbf{h}})$ of the list, where $*$ is the group operation of $G$ that is obtained from the compressed data. A more efficient decoder does not generate the full-time series but it outputs $f(i)$ upon a query on $i$. 

\medskip
\textbf{Full structure of the autoencoder and the description of the training process.} Now, we can combine the encoder and the decoder to make an autoencoder that automatically finds the hidden subgroup. Given a parameter configuration $\vec{\gamma}$, we run the encoder once and then pass its output to the decoder to get the reconstructed data. Repeat $T=O(n)$ times and calculate the average cost $\mathbb{E}_{\vec{\gamma}}C$. Then we slightly change the $\gamma_i$ to $\gamma_i'$  and calculate the updated average cost $\mathbb{E}_{\vec{\gamma}'}C$. Then the partial derivative of the average cost function can be estimated by $
    \frac{\partial \, \mathbb{E}_{\vec{\gamma}}C }{\partial \, \gamma_i} \approx \frac{\mathbb{E}_{\vec{\gamma}'}C - \mathbb{E}_{\vec{\gamma}}C}{ \gamma_i' - \gamma_i} \, .
$ 
Consecutively change the value of $\gamma_i$ and we can calculate the gradient $\nabla \mathbb{E}_{\vec{\gamma}}C = (\partial_{\gamma_1} \, \mathbb{E}_{\vec{\gamma}}C,\dots, \partial_{\gamma_q} \, \mathbb{E}_{\vec{\gamma}}C)$. If the gradient is smaller than a certain predefined threshold, then the parameter $\vec{\gamma}$ is good enough for the hidden subgroup compression. We just output the classical message from the encoder $\sigma_{classical}$ and the problem is solved. Otherwise, we need to update the parameters, $
    \vec{\gamma} \gets \vec{\gamma} - \beta \, \nabla \, \mathbb{E}_{\vec{\gamma}}C \, ,
$
where $\beta$ is the learning rate and $a \gets b$ means assigning the value $b$ to the variable $a$. 

In the more sophisticated scaled conjugate gradient (SCG) algorithm~\cite{moller1993scaled}, which is adopted e.g.\ by MATLAB, the step-size is a function of quadratic approximation of the cost function. Taking the classical autoencoder cost function $C_{ca}(\vec{\theta})$ as an example, at the $k$th iteration $C_{ca}(\vec{\theta}_k)$, the next point $\vec{\theta}_{k+1}$ is determined as:
\begin{equation}
\vec{\theta}_{k+1} = \vec{\theta}_{k}-C'_{ca}(\vec{\theta}_k){C''_{ca}}(\vec{\theta}_k)
\end{equation}
To calculate this, SCG defines a middle point $ \vec{\theta}_{t,k}$ between $\vec{\theta}_{k}$ and $\vec{\theta}_{k+1}$:
$\vec{\theta}_{t,k} = \vec{\theta}_{k}+\sigma_{k}\vec{d}_k,$  where $0<\sigma_{k}\leq 1$ is the step-size and $\vec{d}_k$ is the conjugate direction of the $k$-th parameter. Thus, the second-order information is computed by using the following first-order information and the conjugate directions:
\begin{equation}
s_k = C_{ac}''(\vec{\theta}_{k})\vec{d}_k \approx \frac{C_{ac}'(\vec{\theta}_{t,k})-C_{ac}'(\vec{\theta}_{k})}{\sigma_{k}} \\
\end{equation}
with $
\alpha_{k} = -{\vec{d}_{k}}^{T}C_{ac}'(\vec{\theta}_{k})/{{\vec{d}_{k}}^{T}s_k}$, 
where $s_k$ denotes the second-order information. Then the next point can be calculated as $\vec{\theta}_{k+1} = \vec{\theta}_{k}+\alpha_k\vec{d}_k$

The above procedure as a whole is summarized in Algorithm~\ref{alg:auto_encoder} and is numerically tested in the Methods~\ref{sec:numerics}. An important observation is that this algorithm not only compresses the data; as a by-product, it also solves the hidden subgroup of the generating function $f$ without {\em a priori} knowledge of the group structure $G$.  In this sense, Algorithm~\ref{alg:auto_encoder} is also a variational quantum algorithm solving the variational HSP.

\begin{algorithm}[H]
  \SetAlgoLined
  \KwIn{Oracle $U_f:\ket{i}\ket{0} \mapsto \ket{i} \ket{f(i)}.$\\
An initial parameter configuration $\vec{\gamma}$ with $\gamma_i\in[0,1]$.\\
Precision $\epsilon >0$ and learning rate $\beta$.}
  \KwOut{Classical message $\sigma_{classical}$ as a description of the hidden subgroup $H$ and the data values $f(i)$ on different cosets.}
  \While{$\norm{\nabla \mathbb{E}_{\vec{\gamma}}C}\geq \epsilon$}
{Apply the Encoder algorithm with $\vec{\gamma}$ to get $\sigma_{classical}$.\;
 Apply the Decoder algorithm with $\sigma_{classical}$ to get the reconstructed data $\{\hat{f}(0),\hat{f}(1),\dots,\hat{f}(2^n-1)\}$.\;
 Calculate the cost function $C=\sum_i \norm{\hat{f}(i) - f(i)}$.\;
 Repeat Line 6-8 for $T=O(n)$ times, calculate the average cost $\mathbb{E}_{\vec{\gamma}} C$.\;
 Change the parameter values $\vec{\gamma}$ slightly, repeat Step 1-4, and calculate the gradient $\nabla \mathbb{E}_{\vec{\gamma}}C$.\;
Update the parameters $\vec{\gamma} \gets \vec{\gamma} - \beta \, \nabla \, \mathbb{E}_{\vec{\gamma}}C$, where $\gets$ means assigning new values.}
Output $\sigma_{classical}$ as the compressed data.
  \caption{{\bf Training of autoencoder.}} 
  \label{alg:auto_encoder} 
\end{algorithm}

\subsection{Numerical experiments}\label{sec:numerics}
In this section, we will numerically check how our quantum algorithm performs by using a classical computer to simulate it. First, we will describe how we generate the training data: data strings with a certain symmetry and corresponding group structures. In line with Eq.\eqref{eq:hidden_subgroup_definition}, for the numerical implementation, we can generate the simplest data series based on $f(i) = f(j)$ if and only if $i = j+s$, where $s$ is the secret bit string which is a member of a hidden subgroup and ``$+$" is the group addition specified by the group structure. In particular, we shall feed our algorithm data that either: (i) is periodic, with a hidden subgroup of $\Z_8$, (ii) satisfies Simon's symmetry, with a hidden subgroup of $\Z_2 \times \Z_2 \times \Z_2$, or (iii) has the symmetry of a hidden subgroup of $\Z_2\times \Z_4$.

 As an example, we now describe the generating process in the 6 qubits case, 
 where 3 qubits are used as data registers and 3 qubits are used as ancillas. Then the length of the data sequence is $2^3 = 8$. Taking Simon's problem as an  example, the data series set could be  
\begin{equation}\label{eq:simonsdata}
\left\{{
\begin{array}{cccccccc}
0 & 0 & 1 & 1 & 2 & 2 & 3 & 3,\\
0 & 1 & 0 & 1 & 2 & 3 & 2 & 3,\\
0 & 1 & 1 & 0 & 2 & 3 & 3 & 2,\\
0 & 1 & 2 & 3 & 0 & 1 & 2 & 3,\\
0 & 1 & 2 & 3 & 1 & 0 & 3 & 2,\\
0 & 1 & 2 & 3 & 2 & 3 & 0 & 1,\\
0 & 1 & 2 & 3 & 3 & 2 & 1 & 0
\end{array}
}
\right\}
\end{equation}
where each row is a time series that we could choose to compress. The $(i,j)$-th entry of the table is $f_i(j)$. The rows are sorted by the value of $s$. For example, in the first row, the symmetry is the repeating of data in the sense that $f_1(1) = f_1(2)$, $f_1(3)=f_1(4)$, $f_1(5)=f_1(6)$, $f_1(7)=f_1(8)$. In terms of bits, $f_1(000) = f_1(000+001), f_1(010)=f_1(010+001), f_1(100)=f_1(100+001), f_1(110)=f_1(110+001)$, where ``$+$" denotes bit-wise addition modulo 2 and is usually denoted as ``$\oplus$". $s$ is the generator of a hidden subgroup of $\Z_2 \times \Z_2 \times \Z_2$. Other rows have the same type of symmetry generated by bit-wise addition, except $s$ is different.  
There are multiple possible data sequences that have the same hidden subgroup symmetry; in this example, $4!=24$ distinct sequences exist, as there are $4!$ distinct sequences within one ``period", e.g. $\{0,1,2,3\}$, $\{0,1,3,2\}$, $\{0,3,2,1\}\dots$. We can randomly choose 19 types of data series to train the circuit and 5 types of data series to test the circuit. In the simulations undertaken here testing the performance on the test data was perfect, since the circuit converged to the correct HSP circuit.

Following the above approach, we can generate other matrices of training data according to different group structures and secret bit strings.

We show in Fig.~\ref{fig:CAE} and Fig.~\ref{fig:bigsizeperformance} how our Algorithm~\ref{alg:encoder} combined with  Algorithm~\ref{alg:decoder} perform in terms of compressing the data thus generated, for $2n=6$, $2n=8$  {and $2n=16$} qubit quantum circuits. Fig.~\ref{fig:bigsizeperformance} demonstrates the convergence of our algorithm when compressing data sequences that either possess Simon's symmetry (with respect to group $\Z_2 \times \Z_2 \times \Z_2$) or periodic symmetry (with respect to group $\Z_8$). In either case, the training successfully converges to the corresponding group structure, sending the expected number through the bottleneck, namely, the secret key $s$ for Simons' problem or the period $r$ for the periodic symmetry case.  {Before sending the data to  the decoder, we round $\mathbf{s}$ to be composed of integers, which for the $2n=16$ case leads to a small sudden drop in the cost function at the end of the training.} In the case of Simon's symmetry, Fig.~\ref{fig:CAE} illustrates an accuracy advantage compared to the classical autoencoder. To compress classical data series, we build a fully connected classical autoencoder with a  {256-8-256} structure, with  {256} neurons for input and output, and an  {8}-bit hidden layer due to $s$ having  {8} bits. The training was executed on the MATLAB platform with the built-in scaled conjugate gradient training algorithm~\cite{moller1993scaled}. To enable a comparison with the quantum approach, the training data and  {test data} for the classical autoencoder are also chosen from the  {128!} distinct sequences but in the form of classical bits. Thus, in this numerical experiment, the quantum approach effectively compresses the time series, while the classical counterpart  {results in a higher mean squared error on test data}. 

{ 
\subsection{Compressing quantum states via finding symmetry}

We present here a preliminary method to apply our variational QFT ansatz to compress quantum states satisfying a certain type of symmetry. We present first the non-variational version of quantum state compression and then present the variational version. See Supplementary Material E for a more pedagogical discussion.

\medskip
\textbf{Non-variational quantum states compressions}. 
Consider $n$-qubit quantum states $\ket{\varphi}$, which lives in an $N=2^n$ dimensional Hilbert space, with basis $\{\ket{x}\}_{x=0}^{N-1}$. We can define a translational operator $T_{h_0}$, such that 
\begin{equation}
    T_{h_0} \ket{x} = \ket{x+h_0}\, ,
\end{equation}
where $+$ is the group operation of $\mathbb{Z}_N$, i.e.,  addition modulo $N$, and $h_0=2^m$ for some integer $m$ such that $h_0$ is a factor of $N$. Thus the group generated by $h_0$, denoted by $H=\{0, h_0, 2h_0, ..., (2^{n-m}-1) h_0\}$ is not the whole group $\mathbb{Z}_{N}$. Define the {\em coset state} as
\begin{equation}
    \ket{c+H} := \frac{1}{\sqrt{|H|}} \sum_{h \in H} \ket{c+h} \, ,
\end{equation}
for $c\in C=\{0,1,2,...,2^m-1\}$ being the representative of each coset. It is direct to verify
\begin{equation}
    T_{h_0} \ket{c+ H} = \ket{c+H}\, ,
\end{equation}
for all $c\in C$. 
These coset states form an orthonormal basis for the invariant subspace of this translation operator $T_{h_0}$, whose dimension is $\frac{|G|}{|H|} = 2^{n-m}$.
This suggest that an $n$-qubit quantum state invariant under $T_{h_0}$ can be compressed to an $(n-m)$-qubit quantum state.

The quantum state compression problem can be formulated as follows. A source will randomly generate an $n$-qubit quantum state $\ket{\varphi}\in \mathcal{S}=\{\ket{\phi}| T_{h_0} \ket{\phi} = \ket{\phi}\}$ with probability density $\mu_{\ket{\varphi}}$, when we access it. We know that $T_{h_0}$ will act on the state according to the group addition of $\Z_N$ but we do not know the value of $h_0$. The goal is to compress the $n$-qubit states by identifying their symmetry, given accesses to the quantum sates sampled from the source. 

For this non-variational quantum state compression, the group structure is given as an input to the algorithm and therefore we can directly apply the HSP algorithm over the group $\mathbb{Z}_N$ to solve for $h_0$. When $h_0$ is identified, the compression operation can be constructed.

Specifically, suppose that the source generates a state $\sum_c \alpha_c \ket{c+H}$ each time; but the state generated may vary from time to time. Then we can use four steps to solve for $h_0$ and therefore construct the compression protocol.
(i) Apply $\mathrm{QFT}_{\mathbb{Z}_N}$ to the state and then measure in computational basis to get a bit string $j$. (ii) Repeat the first step for many times to get many bit strings $j$ that satisfies $j = k \, h_0$ for some integer $k$. In Supplementary Material E we prove that $j$'s are sampled from the orthogonal group of $H$. (iii) Solve the smallest common divisor among the sampled $j$'s and you get $h_0$. (iv) Construct  the following CPTP map that will compress the state generated by the source,
\begin{equation}
    \mathcal{E} (\rho)= \sum_{i, j =0}^{|C|-1} \ket{c_i}\bra{c_j} \bra{\hat{c_i}}\rho \ket{\hat{c_j}}\, , \label{eq:quantum_compression_E}
\end{equation}
where $\ket{c_i}$'s are basis vectors of a Hilbert space of dimension $|C|=2^{n-m}$ and $\ket{\hat{c}_i}:=\ket{c_i + H}$'s are vectors of the Hilbert space of dimension $2^n$. The following CPTP map $\mathcal{D}$ will reconstruct the state,
\begin{equation}
        \mathcal{D} (\Tilde{\rho})= \sum_{i, j =0}^{|C|-1} \ket{\hat{c}_i}\bra{\hat{c}_j}\bra{c_i}\Tilde{\rho} \ket{{c_j}}\, .\label{eq:quantum_compression_D}
\end{equation}

\textbf{Variational quantum state compression.} Mathematically, we can formulate the following quantum state compression problem where our variational algorithm can be used to compress it. Define the generalized translational operator as
\begin{equation}
    T_{h_0}\ket{x} = \ket{x * h_0} \, ,
\end{equation}
where $*$ is the group operation in the unknown Abelian group $G$. A source will randomly generate quantum state $\ket{\varphi}^{\otimes n} \in \mathcal{S} $ (multiple copies are used for evaluating the cost in training, see Supplementary Material E for more explanations) according to a probability density $\mu_{\ket{\psi}}$, where the state $\ket{\varphi}$ is promised to be translational invariant under the action of $T_{h_0}$.  Given access to the quantum states generated from this random source but without knowledge of the Abelian group structure $G$ and the hidden subgroup $H$, the goal is to design $\mathcal{E}$ and $\mathcal{D}$ that compress and decompress the quantum states. 

Then our variational HSP algorithm can be applied, with a slight modification in the cost function. Instead of reconstruction error of classical bit strings, our new optimization goal is to minimize the  average infidelity between the input state to the reconstructed state,
\begin{equation}
C_F = 1-\int \bra{\psi} \mathcal{D}\circ\mathcal{E}(\ket{\psi}\bra{\psi})\ket{\psi} \mathrm{d} \mu_{\ket{\psi}} \, . 
\label{eq:infedility_cost}
\end{equation}

The variational algorithm for this quantum states compression problem takes 5 steps. We list the sketch of the protocol here.
(i) First, apply the variational QFT ansatz to the one copy of the random state $\ket{\varphi}$. (ii) Measure the state after the parametrized QFT in the computational basis to get a bit string $j$. Repeat and get a set of bit strings $j$. (iii) Use steps 3,4,5 of Algorithm 1 in the main text to solve for a tentative generator $h_0$ of the hidden subgroup. (iv) Use $h_0$ to construct the hidden subgroup $H$ and then use Eq.\eqref{eq:quantum_compression_E} and Eq.\eqref{eq:quantum_compression_D} to  construct the encoding map $\mathcal{E}$ and the decoding map $\mathcal{D}$. (v) Evaluate the infidelity between the input state $\ket{\varphi}\bra{\varphi}$ and the reconstructed state $\mathcal{D}\circ\mathcal{E}(\ket{\varphi}\bra{\varphi})$
. (vi) Finally, use gradient descent to update the parameters in the QFT ansatz to minimize the the infidelity cost Eq.\eqref{eq:infedility_cost}. }

\section{Competing Interests}
The authors declare no competing interests.

\medskip

\section{Author Contributions}
WZ raised the potential of using quantum methods for the practically important problem of non-linear information compression. OD outlined an HSP autoencoder scheme. LF coded and ran numerical experiments. BK contributed several ideas and methods concerning the generalization of HSP beyond initial examples. MF, BK, and LF derived the variational circuit form for the generalized QFT (Theorem 2). MF designed the database compression set-up and proved the exponential speed up with BK (Theorem 1).  {MF and OD extended the main algorithm for the task of quantum state compression.} All authors contributed to discussions throughout and to the writing of the manuscript.  

\section{Data availability}
Source code and dataset can be found on Github \href{https://github.com/unknownfy/parameterized-Abelian-hidden-subgroup-algorithm}{https://github.com/unknownfy/parameterized-Abelian-hidden-subgroup-algorithm}. 

\section{Acknowledgements.}
We gratefully acknowledge discussions with  {Zhaohui Wei, Zizhu Wang, Yuxuan Du, Mile Gu, Jayne Thompson}, Yupan Liu, Jon Allcock, Shyam Dhamapurkar, Nana Liu,  Sirui Ning, Swati Singh, Tian Zhang,  and support from HiSilicon.  {This work was in part carried out using the computational facilities of CityU Burgundy, managed and provided by the Computing Services Centre at City University of Hong Kong (https://www.cityu.edu.hk/)}. OD acknowledges support from the National Natural Science Foundation of China (Grants No. 12050410246, No.1200509, No.12050410245) and the City University of Hong Kong (Project No. 9610623).

\bibliography{reference}

\pagebreak

\onecolumngrid
\appendix

\section{Assigning group structures to time series by Abelian groups on binary strings.}\label{supplementary:assigning_group_structure_via_binary_strings}

\textbf{Classification of finite Abelian groups.} To incorporate hidden subgroup structures into the time series data $\{f(0), f(1), f(2),... \}$, we have to adopt a group structure to a set of indices $i=0,1,2,3,...$, making it a group $G=(\{i\},*)$ with $*$ being the group operation and $i$'s being the group elements.  In fact, the fundamental theorem for finite Abelian groups states that any finite Abelian group is isomorphic to 
\begin{equation}\label{eq:group_classification}
    G \cong \Z_{p_1^{m_1}} \times \Z_{p_2^{m_2}} \times ... \times 
 \Z_{p_q^{m_q}}  \, ,
\end{equation}
where $\Z_{p}$ is the shorthand notation of the additive integer modulo group $\Z/p\Z$, $p_i$'s are prime integers and $m_i$'s are positive integers that are both not necessarily distinct.  Moreover, we must have 
\begin{equation}\label{eq:cardinality_relation}
    |G| = p_1^{m_1} \times p_2^{m_2} \times ...\times p_q^{m_q} \, ,
\end{equation}
which means $p_i$'s must be factors of the group cardinality $|G|$. 

\medskip
\textbf{Restricting our discussion to groups whose elements are binary strings}. In principle, we can impose any finite Abelian group structure $G$ of the form \eqref{eq:group_classification} to the set of indices $\{ i|i=0,1,...,|G|-1\}$, as long as Eq.\eqref{eq:cardinality_relation} holds. However, the implementation of the algorithm at the circuit level makes a difference. Quantum algorithms are implemented on quantum circuits, which contain single-qubit and two-qubit unitaries acting on a set of qubit systems. The input and output are encoded in the qubit states, which are essentially bit strings. This implies that only those groups whose cardinality are powers of $2$ are the simplest to encode and compute on a quantum circuit; otherwise, procedures of translating group elements into binary numbers and group operations into binary operations must be applied, which will increase the difficulty of the algorithm implementation on the circuit level and introduce unnecessary complexity to our discussions. 

Our work aims to demonstrate the basic ideas of using a quantum hidden subgroup algorithm to compress time series data, so we focus our discussion on groups that satisfy $|G|=2^n$.  {There are strong reasons to believe that} our results can be generalized to arbitrary finite Abelian groups,  { and we explicitly construct examples of such circuit ansatzes (see the appendix~\ref{appedix:generalAbelian}). }

For simplicity, we assume the time series data has fixed length $2^n$, which means that the time is running from $i=0$, to $i = 2^n-1$. This allows an exact identification $\tau:\{0,1\}^n \rightarrow \{0,1,2,...,2^n-1\}$ from any binary string of length $n$ to the index $i$,
\begin{equation}\label{eq:tau}
    i = \tau (i_1 i_2...i_{n-1}i_n) = i_1 2^{n-1} + i_2 2^{n-2} + \cdots +  \cdots i_{n-1} 2 + i_{n} \, ,
\end{equation}
where $i_q$ is the $q$-th digit of the binary string $i_1 i_2...i_{n}$. And given the number $i$, the bit string $i_1i_2...i_n = \tau^{-1}(i) $ can be calculated by division. For example, $0=\tau{(000)}, 2=\tau{(010)}$ and $6=\tau{(110)}$. With this map $\tau$, we identify the index $i$ in its decimal format to its binary representation $i_1i_2...i_n$. We will use both representation whenever there is no ambiguity.

Now, with $|G|=2^n$, only groups in the following form
\begin{equation}\label{eq:group_type}
    G  \cong \Z_{2^{m_1}} \times \Z_{2^{m_2}}\times...\times \Z_{2^{m_q}} 
\end{equation}
will be allowed by relation \eqref{eq:group_classification}, where $m_i$'s are positive integers that are not necessarily distinct and satisfy
\begin{equation}
    m_1 + m_2 +...+m_q = n \, .
\end{equation}

\medskip
\textbf{Group structure over binary strings.}
The next question is, how to impose a group structure $G$ to the bit strings $\{0,1\}^n$ and therefore to all the set of indices $\{i\}$?  Let us begin with a simple example of assigning the group structure $G_0$ that is isomorphic to $\Z_4$ to the set of time values (in its binary representation) ${00,01,10,11}$. Since $G_0$ is isomorphic to $\Z_4$, we can specify the group $G_0$ by establishing a one-to-one correspondence between elements in $G_0$ to elements in $\Z_4$, say
\begin{equation}
    \begin{matrix}
        \Z_4 & = & \{& 0, & 1, & 2, & 3  & \}  \\
        G_0 & = & \{& 00, & 01, & 10, & 11 &\} 
    \end{matrix} \, ,
\end{equation}
where we map $00$ to $0$, $01$ to $1$, $10$ to $2$ and $11$ to $3$. However, this is not the only way to establish one-to-one correspondence; other alternative isomorphisms are
\begin{equation}\label{eq:alternatives_Z_4}
    \begin{matrix}
        \Z_4 & = & \{& 0, & 1, & 2, & 3  & \} \\
        G_1 & = & \{& 01, & 10, & 11, & 00 &\} \\
        G_2 & = & \{& 10, & 11, & 00, & 01 &\} \\
         & \vdots &  & &  &  &  & 
    \end{matrix}\, .
\end{equation}
In total there are $4!=24$ ways of specifying the one-to-one correspondence and thus assign the group structure $\Z_4$, among which $G_0$ is the most canonical one since the correspondence is given by the binary representation $\tau$ defined in Eq.\eqref{eq:tau}. Among these correspondences, there is a special class of isomorphisms that can be induced by permutations of the bit strings. For example, let us exchange the first bit and the second bit and then we have the following correspondence
\begin{equation}\label{eq:isomorphism_permutation_Z4}
        \begin{matrix}
        \Z_4 & = & \{& 0, & 1, & 2, & 3  & \} \\
        G_0 & = & \{& 00, & 01, & 10, & 11 &\}  \\
         G_3 & = & \{& 00, & 10, & 01, & 11 &\} 
    \end{matrix} \, .
\end{equation}

More generally, a group structure on the set of indices can be assigned by establishing a one-to-one correspondence between elements of the index (and thus the bit strings of length $n$) and elements in the group of $\Z_{2^{m_1}} \times \Z_{2^{m_2}}\times...\times \Z_{2^{m_q}}$. In total, there are $2^n!$ ways of assigning the group structure, which is the number of different one-to-one correspondences. $2^n!$ is a huge number, and in the following, we will try to characterize a smaller subset of the groups that can be easily implemented on the circuit level.

\medskip
\textbf{Specifying the group via binary operations}. Another way of assigning the group structure on the set of indices is via the group operation.
A group is defined by specifying the elements in the group and specifying the operations between any two elements. Because the group elements are fixed to be the bit strings, what is needed to define the group structure is to specify the group operation $*$ between two bit strings.

Restricting the length of time series to $2^n$ and thereby the group $G$ of the form \eqref{eq:group_type} has an advantage; it enables us to implement the group operation in binary circuits (either composed by classical or quantum bits) easier. First of all, we are able to allocate $N$ qubits to encode the group elements in a one-to-one correspondence. Secondly, group actions on the group elements can be easily implemented by bit-wise XOR or AND, or their combinations, with the assistance of ancillas. In fact, bit-wise XOR is equivalent to bit-wise addition modulo 2, which will be denoted by ``$\oplus$". For example, bit-wise addition modulo 2 gives $11 \oplus 01 = 10$, while in contrast, the addition in binary numbers modulo 4 gives $11 + 01 =00$, which is the binary equivalence of $3+1=0 \mod 4$.  We will denote the standard addition in binary number systems as  ``+”. Note that the standard addition ``+" modulo $2^{m}$ reduces to bit-wise addition ``$\oplus$" when $m=1$.

Using the bit-wise addition modulo 2 ``$\oplus$" and the addition ``$+$" in binary numbers modulo $2^m$, we are able to assign the group structure of the form \eqref{eq:group_type} on the set of binary strings for arbitrary $m_1,...,m_q$. Note this does not mean that all the $2^n!$ ways of specifying the group structure can be constructed in this way. For example, if the time takes values of $i=0,1,2,3$ and we want to assign $\Z_4$ by defining the group operation using only the bit-wise addition, addition ``+" modulo $4$, or any combination of the two, then we must have $00$ isomorphic to $0\in \Z_4$, because $00$ is the only bit string that does not change the second bit string upon operations of the bit-wise addition, the addition ``+" modulo 4, or any of their combinations. This implies that those groups whose identity element is not $00$ cannot be assigned in this way, e.g.\ the group $G_1$ and $G_2$ shown in Eq.\eqref{eq:alternatives_Z_4}.

Then, we will use an example to show the canonical way of assigning the group structure by binary operations. For example, if the time takes $i=0,1,2,3,...,7$, then we can represent the time by a bit string of length 3, say, $000,001,010,...,111$. 3 types of group structures can be defined, i.e. $\Z_2 \times \Z_2 \times \Z_2$, $\Z_4\times \Z_2$ or $\Z_8$. One way is to define the group operation as bit-wise addition modulo 2 such that $011 \oplus 101 =110$, and then the group is isomorphic to $\Z_2 \times \Z_2 \times \Z_2$. A second way to define the addition as the standard addition of binary numbers modulo $8$, say, $010+011=101$, and then the group of binary strings is isomorphic to $\Z_8$. A third way is to divide a bit string into two parts, where the first part consists of 2 bits and the second part consists of only 1 bit. Given two bit strings $abc$ and $xyz$, define the group operation $*$ as 
\begin{equation}
    abc * xyz = \mathrm{Append}[ab + xy \mod 4 , c\oplus z] \, ,
\end{equation}
where ``Append" is the string operation that appends the second string to the end of the first string. In this way, the group is isomorphic to $\Z_4 \times \Z_2$, where we divide the bits into two parts and the first part is added using $+$ modulo $4$ and the second part is added using bit-wise addition $\oplus$.

More generally, a group $G=(\{0,1\}^N,*)$ of the form \eqref{eq:group_type} can be constructed by dividing the bit strings into $q$ parts, and the group operation $*$  is defined by first adding the $q$ parts using the standard addition ``+" modulo $2^{m_q}$ respectively, and then gluing the resulting $p$ pieces of bit strings together. 
This definition of group operation is the canonical way of assigning the group structure to the bit strings but it is not the only way. 
In fact, the way to specify a group $G$ in the form of \eqref{eq:group_type} to the bit strings of length $n$ is not unique, and this is discussed in Supplementary Material~\ref{supplementary:searching_over_isomorphisms}.

Having defined the group $G=(\{0,1\}^n,*)$, we can regard the generating function $f: \{0,1,...,2^n\} \mapsto \{0,1\}^m$ as a group function, mapping from the group $G$ to bit strings of length $m$. Here we identify the time $i$ to its binary representation $i_1i_2...i_n = \tau(i)$ and thus the group elements $i_1i_2...i_n$. In this way, the generating function can also be written as $f:\{0,1\}^n \rightarrow \{0,1\}^m$. 
In the following sections, we restrict our discussion to the groups in the form of \eqref{eq:group_type}, unless otherwise stated.

\section{Compression ratio of HSP database compression.\label{supplementary_compression_ratio}}
Directly storing the function values $\{f(0), f(1),....,f(|G|-1)\}$ in the database requires
\begin{equation}
\kappa_{\mathrm{direct}}(f) =  |G| \, m = N\, m
\end{equation}
bits, where $|G|$ is the cardinality of $G$, and $m$ is the number of bits needed to specify a time series entry $f(i)$. Exploiting any hidden subgroup structure, however, will reduce the memory required to a compressed amount 
\begin{equation}
   \kappa_{\mathrm{comp}} (f) = \frac{|G|}{|H|} \,  m
\end{equation}
bits. To see this, consider the coset decomposition
\begin{equation}
    G = c_0 H \cup c_1 H \cup ...\cup c_k H \, .
\end{equation}
where $c_0, c_1, ..., c_k$ are representative elements in different cosets. For an Abelian group $G$, these cosets form a quotient group $G/H$. The hidden subgroup structure implies that $f$ takes constant values within each coset. Then we only need to store the function values over the representative elements of the quotient group $G/H$, whose cardinality is $|G|/|H|$. In other words, the characteristic function $c$ is specified by 
\begin{align}
    c(i) = \begin{cases}
    0, & \text{if $i \in \{c_0,c_1,...,c_k\}$} \\
    1, & \text{otherwise.}
    \end{cases}
\end{align}
And the query function $q$ is specified by
\begin{align}
    q(i) = i \mod H \, ,
\end{align}
where the $\mod H$ operation is defined such that $i = j \mod H$ if and only if $i-j \in H$.

\section{Introduction to quantum Fourier transform and its applications in hidden subgroup problems.\label{supplementary:introduction_QFT_and_Application}}

\subsection{Period finding and compression via quantum Fourier sampling.}

It was shown that the quantum Fourier sampling can be used to solve the Abelian hidden subgroup efficiently~\cite{shor1994algorithms,kitaev1995quantum,simon1997power}. Here we retrieve several key results, and a more detailed introduction using group representation theory can be found in Supplementary \ref{supplementary:hidden_subgroup_problems_formal_discussion}. Interested readers can also refer to Ref.~\cite{lomont2004hidden} for a comprehensive review of the development of HSP.

We begin with a simple problem and then present the general results.  For example, consider the group $\Z_8$ assigned in the canonical way by identifying the group element $i\in \Z_8$ to the index $i$. Suppose the time series data of length $8$ is $\{0,1,0,1,0,1,0,1\}$, then the period of the generating function $f$ is $2$ and the hidden subgroup is $H= \{0,2,4,6\}$, the quotient group is $G/H = \{0,1\}$ and we can compress the time series data to $\{f(0)=0, f(1)=1\}$. More generally, impose the group structure $G= \Z_N$ to the  time series $i= 0, 1,2,...,N -1$ with $N=2^n$ and let the generating function $f$ has a period $r$, meaning that
\begin{equation}
    f(i)= f(i +r) \, ,
\end{equation}
for any time $i$ and $i+r \mod N$. Furthermore, we assume that the period $r$ divides $N$, such that the generating function $f$ has a subgroup 
\begin{equation}\label{eq:expression_H}
    H = \langle r\rangle = \{0,r, 2r,... (M-1) r\} \,
\end{equation}
where $\langle r\rangle$ means the group generated by $r$ and $M = N/r =|H|$. And the cosets are denoted as $ c_1 + H, c_2 + H, ..., c_r + H$, where $c_1=0, c_2=1,...,c_r=r-1$ are representative elements in different cosets that form the quotient group $G/H$,
\begin{equation}
    G/H = \{0, 1, 2,..., r-1\}\, .
\end{equation}
Exploiting the coset decomposition, we have 
\begin{equation}\label{eq:coset_decomposition_summation}
    \sum_{i\in G} \ket{i}\ket{f(i)} = \sum_{c \in G/H} \sum_{h \in H} \ket{c + h} \ket{f(i+h)} \, .
\end{equation}
Our goal is to find this period $r$ and the time series data can therefore be compressed to $\{f(0), f(1),...,f(r-1)\}$.

In general, this period $r$ can be found by exploiting the quantum Fourier transform.
\begin{definition}[Quantum Fourier Transform (QFT)]
The quantum Fourier transform for a cyclic group $\Z_N$ is a unitary transformation 
\begin{equation}
    \mathrm{QFT}_{\Z_N} = \frac{1}{\sqrt{N}} \sum_{i,j \in G} \chi_j(i) \ket{j} \bra{i} \, .
\end{equation}
where $\chi_j: \Z_N \rightarrow \mathbb{C}^*$ is a group homomorphism to the multiplicative group of non-zero complex numbers $\mathbb{C}^*$ that is labeled by $j$ and takes the form of 
\begin{equation}
    \chi_j (i) = \exp(\frac{2 \pi \mi\, j i}{N})\, .
\end{equation}
\end{definition}
In particular, the QFT transforms $\ket{0}$ to $\frac{1}{\sqrt{N}} \sum_j \ket{j}$, which is an equal superposition of all bit strings $\ket{j}$. This particular transformation can also be achieved by applying the Hadamard gate on each qubit.

Then the quantum algorithm of finding the period $r$ was first invented by Shor~\cite{shor1994algorithms} and we retrieve it as follows. First we prepare two registers $\ket{0}\ket{0}$, each of which consists $n$ qubits. Apply the QFT on the first register and we have
\begin{equation}
    \ket{0}\ket{0}\xrightarrow{\QFT_{\Z_N} \otimes \mathbb{I}} \frac{1}{\sqrt{|G|}} \sum_{i\in G} \ket{i} \ket{0} = \frac{1}{\sqrt{N}} \sum_{i=0}^{N-1} \ket{i} \ket{0} \, .
\end{equation}
Then apply the quantum oracle $U_f: \ket{i}\ket{j} \mapsto \ket{i} \ket{f(i)+j}$ and we have
\begin{equation}
    \xrightarrow{ U_f} \frac{1}{\sqrt{N}} \sum_{i=0}^{N-1} \ket{i}\ket{f(i)}  =\frac{1}{\sqrt{N}}\sum_{c\in G/H}  \sum_{h\in H} \ket{c +h} \ket{f(c)}\, ,
\end{equation}
where the equality is due to the coset decomposition of the group $G$ shown in Eq.\eqref{eq:coset_decomposition_summation}. Upon measuring the second register with output $f(c_0)$, the post-measurement state of the first register is
\begin{equation}
    \xrightarrow[\text{on the 2nd}]{\text{Measurement}} \frac{1}{\sqrt{|H|}}\sum_{h \in H} \ket{c_0 + h} = \frac{1}{\sqrt{M}}\sum_{s=0}^{M-1} \ket{c_0 + s \, r} \, ,
\end{equation}
where the equality is due to Eq.\eqref{eq:expression_H}. Apply the quantum Fourier transform again, and we have the state of the first register
\begin{equation}
    \xrightarrow{\QFT_{\Z_N}} \ket{\phi}\equiv \frac{1}{\sqrt{MN}} \sum_{j=0}^{N-1} e^{\frac{2 \pi \mi c j}{N}} \sum_{s=0}^{M-1} e^{\frac{2 \pi \mi s j}{M}} \ket{j},
\end{equation}
where we have used the relation $N = M r$. Because the geometric series $\sum_{s=0}^{M-1} \exp(\frac{2 \pi \mi s j}{M}) $ is non-zero only if $M$ divides $j$, then we have
\begin{equation}
    \ket{\phi} = \frac{1}{\sqrt{r}} \sum_{t = 0}^{r-1} \exp(\frac{2 \pi \mi c t M}{N})\ket{t M} \, .
\end{equation}
Denote
\begin{align}
    H^\perp & \equiv \{j\in G| \chi_j (h)  =1 \, \text{for all} \, h \in H \} \\
    & = \{0, M, ..., (r-1) M\} \, ,  
\end{align}
as the orthogonal subgroup of $H$, we have
\begin{equation}
    \ket{\phi} = \frac{1}{\sqrt {H^\perp}} \sum_{j \in H^\perp} \chi_j (c_0 ) \ket{j} .
\end{equation}

Now, measuring the first register in the computational basis and repeat the above process many times we obtain samples from the set $H^\perp=\{0, M, ...., (r-1) M\}$. Finally, we can use the classical algorithm to find the greatest common divisor of the sampled data, which gives the value of $M$ from with high probability. In fact, it cost only $O(n^2)$ steps for the Euclidean algorithm to find the value of $M$, where $n= \log N$ is the number of bits in $N$. 

\subsection{Simon's problem and compression via general quantum Fourier transform.~\label{supplementary:Simons_Algorithm}}
If the group structure of the time series is not $\Z_{2^n}$, we should use a more general quantum Fourier transform to find the hidden subgroup. Take Simon's problem~\cite{simon1997power} as an example, where the group structure is $G=\Z_2^n=\Z_2 \times ...\times \Z_2$. As discussed in Supplementary Material~\ref{supplementary:assigning_group_structure_via_binary_strings}, this group $G$ can be assigned to the time series using bit-wise addition modulo 2 ``$\oplus$", such that $G=(\{0,1\}^n,\oplus)$. In Simon's problem, the generating function $f:\{0,1\}^n\rightarrow \{0,1\}^m$ is promised to have the following symmetry
\begin{equation}
    f(i) = f(i \oplus s), \, \, \, \forall \, i \in \{0,1\}^n \, ,
\end{equation}
where $s$ is a fixed bit string to be found. Because $s \oplus s = 0$, the hidden subgroup becomes $ H = \{0,s\}$. And Simon's problem is equivalent to finding the hidden subgroup $H$. To solve Simon's problem, one way is to check all the bit-strings, which is computationally intractable. Another way is to use a quantum algorithm, which is exponentially faster. The quantum algorithm is based on the Fourier transform of the group $\Z_2^n$ instead of the cyclic group $\Z_{2^n}$. In fact, the quantum Fourier transform for a general finite Abelian group is as follows.

\begin{definition}[General quantum Fourier transform for finite Abelian groups.]
Let $G\cong \Z_{N_1} \times \Z_{N_2} \times ... \times \Z_{N_q}$, then elements $i \in G$ can be identified with a tuple $i = (i_{(1)}, i_{(2)},...,i_{(q)})$, with each of its entry $i_{(q)} \in \Z_{N_q}$. Then the quantum Fourier transform for group $G$ is defined to be
\begin{equation}
    \QFT = \frac{1}{\sqrt{|G|}} \sum_{i, j \in G} \chi_j (i) \ket{j} \bra{i} \, ,
\end{equation}
where $\chi_j: G\rightarrow \mathbb{C}^*$ is a group homomorphism to the multiplicative group of non-Zero complex numbers $\mathbb{C}^*$ that is labeled by $j$ and takes the form of 
\begin{equation}
    \chi_j (i) = e^{\frac{2 \pi \mi j_{(1)} i_{(1)}}{N_1}} \, e^{\frac{2 \pi \mi j_{(2)} i_{(2)}}{N_2}} \,...\, e^{\frac{2 \pi \mi j_{(q)} i_{(q)}}{N_q}} \, .
\end{equation}
\end{definition}

Here $\chi_j$ is known as the character of the group $G$. For a more detailed discussion about the character and the group representation theory, please refer to Supplementary Material~\ref{supplementary:hidden_subgroup_problems_formal_discussion}.

For Simon's problem, the group $G\cong \Z_2^n$ and for each bit-string $i\in G$, $i_{(q)}$ is the $q$-th bits $i_q$, so we have 
\begin{equation}
    \chi_j(i) = (-1)^{j_1 i_1} (-1)^{j_2   i_2} ...(-1)^{j_n   i_n} \equiv (-1)^{j \cdot i} \, ,
\end{equation}
where $j_t$ and $i_t$ are the $t$-th bits of the bit-string $j$ and $i$ respectively, and $j \cdot i = j_1 i_1 + j_2 i_2 +...+j_n i_n$. Then the quantum Fourier transform for the Simon's problem is
\begin{equation}
\QFT_{\Z_2^n} = \frac{1}{\sqrt{2^n}} \sum_{j, i\in \{0,1\}^n}   (-1)^{j \cdot i} \ket{j}\bra{i}\, .
\end{equation}
Similar to the algorithm for period finding, we prepare two registers in state $\ket{0}$, each of which consists $n$ qubits. Then we apply the quantum Fourier transform in the first register to generate a equal superposition of all the bit-strings,
\begin{equation}
    \ket{0}\ket{0}\xrightarrow{\QFT_{\Z_2^n} \otimes \mathbb{I}} \frac{1}{\sqrt{G}} \sum_{i \in G} \ket{i} \ket{0} \, .
\end{equation}
Next, we apply the quantum oracle $U_f$ to generate a equal superposition of all values of the function $f$,
\begin{equation}
    \xrightarrow{U_f} \frac{1}{\sqrt{G}} \sum_{i \in G} \ket{i} \ket{f(i)} = \frac{1}{\sqrt{|G|}} \sum_{c\in G/H} (\ket{c} + \ket{c \oplus s}) \ket{f(c)}\, .
\end{equation}
Now, we measuring the 2nd register and conditioning on the output $f(c_0)$, we have the post-measurement state of the first register as 
\begin{equation}
    \xrightarrow{\text{measurement}} \frac{1}{\sqrt{|2|}} (\ket{c_0} + \ket{c_0\otimes s}) \equiv \ket{c_0 + H}\, ,
\end{equation}
where $\ket{c_0+H} = \frac{1}{\sqrt{|H|}} \sum_{h\in H} \ket{c_0+h}$.
Then apply the QFT again, and we have
\begin{equation}
    \xrightarrow{\QFT_{\Z_2^n}} \ket{\phi}\equiv \frac{1}{\sqrt{|G|}} \sum_{j \in G}\frac{(-1)^{j \cdot c_0}}{\sqrt{2}}\left[1+(-1)^{j\cdot s}\right] \ket{j}\, .
\end{equation}
We define the orthogonal group of $H$ as 
\begin{equation}
    H^\perp\equiv \{j| (-1)^{j\cdot s} = 1\}= \{j|\chi_j(i) = 1 \,\, \forall \, i \in H \} \, ,
\end{equation}
which contains exactly the elements that have non-zero amplitudes in $\ket{\phi}$. This means,
\begin{equation}
    \ket{\phi} = \frac{1}{\sqrt{|H^\perp|}} \sum_{j \in H^\perp} (-1)^{j \cdot c_0} \ket{j} = \frac{1}{\sqrt{|H^\perp|}} \sum_{j \in H^\perp} \chi_j(c_0) \ket{j} \, ,
\end{equation}
Measure the first register in the computational basis and we get a bit-string $j \in H^\perp$. Repeat the above process for $k$ times and we get a set of random bit-strings $\{j^{(1)},j^{(2)},...,j^{(k)}\}$ drawn uniformly from $H^\perp$. Now we should solve $s$ using this sample of bit-strings in $H^\perp$.
Because $1+(-1)^{j\cdot s}$ is non-zero if and only if $j \cdot s = 1 \mod 2$, thus solving the linear congruence equations
\begin{eqnarray}
  j^{(1)}_1 s_1 + j_2^{(1)} s_2 + \cdots + j_n^{(1)} s_n & = & 0 \mod
  2 \nonumber\\
  j^{(2)}_1 s_1 + j_2^{(2)} s_2 + \cdots + j_n^{(2)} s_n & = & 0 \mod
  2 \nonumber \\
  & \vdots & \nonumber \\
  j^{(k)}_1 s_1 + j_2^{(k)} s_2 + \cdots + j_n^{(k)} s_n & = & 0 \mod 2
\label{eq:linear_congruence_appendix}
\end{eqnarray}
and we can get the string $s$ by obtaining each of its digits $s_t$, resulting the solution of the Simon's problem. More generally, Ref.~\cite{lomont2004hidden} gives a detailed introduction on how to construct the linear congruence equations for all kinds of finite Abelian groups.

\subsection{General Abelian hidden subgroup algorithm.\label{sec:quantum_algorithm_general_hidden_subgroup}}

Now we briefly introduce the quantum hidden subgroup algorithm~\cite{nielsen2002quantum, jozsa2001quantum, lomont2004hidden}. 

Given a finite Abelian group $G$ defined on the set of indices of the data sequence and a generating function $f$ that hides the subgroup $H<G$, we have the following quantum algorithm to find the hidden subgroup $H$. First, we prepare two registers in $\ket{0}\ket{0}$ and then apply the general quantum Fourier transform $\QFT_G$ with respect to the group $G$ on the first register, generating an equal superposition of all bit-strings
\begin{equation}
    \ket{0}\ket{0}\xrightarrow{\QFT_G \otimes \mathbb{I}} \frac{1}{\sqrt{|G|}} \sum_{i\in G} \ket{i} \ket{0}\, .
\end{equation}
Then apply the quantum oracle $U_f$ to evaluate the function $f$,
\begin{equation}
    \xrightarrow{U_f} \frac{1}{\sqrt{|G|}} \sum_{i\in G} \ket{i} \ket{f(i)} = \frac{1}{\sqrt{|G/H|}} \sum_{c \in G/H } \ket{c+H} \ket{f(c)} \, ,
\end{equation}
where $\ket{c+H} = \frac{1}{\sqrt{|H|}} \sum_{h\in H} \ket{c+h}$ and the equality is due to $f$ being constant in each coset. Measure the second register and condition on the outcome $f(c_0)$ such that the post-measurement state of the first register is $\ket{c_0 + H}$.
Next, we apply the quantum Fourier transform $\QFT_G$ again on the first register and we have
\begin{equation}
  \ket{c_0 + H}  \xrightarrow{\QFT_G} \frac{1}{\sqrt{|H^\perp|}} \sum_{j \in H^\perp} \chi_j (c_0) \ket{j} \, ,
\end{equation}
where $\chi_j(c_0)$ is the character of $c_0$ from a representation of the group $G$ labeled by $j$, and $H^\perp =\{j| \chi_j(h) = 1 \, \forall \, h \in H\}$ is the orthogonal group of $H$. 
Measuring the first register in the computational basis and we get a sample $j$ from the orthogonal group $H^\perp$. Repeat the process for $k$ times and we obtain a set of random elements drawn uniformly from $H^\perp$. Using this set of sampled elements we can efficiently compute the generator of the group $H$ by a classical algorithm. A more formal introduction to the quantum hidden subgroup algorithm with proofs can be found in Supplementary Material~\ref{supplementary:hidden_subgroup_problems_formal_discussion}.  The quantum Fourier transform is the fundamental basis for this protocol, and it can be efficiently implemented in quantum circuits (see next subsection, Supplementary Material~\ref{supplementary:efficient_circuit_QFT}).

\subsection{Circuit ansatz of quantum Fourier sampling  over arbitrary finite Abelian group}
\label{appedix:generalAbelian}

In this section, we show how to approximately implement the quantum Fourier sampling over groups of cardinality $p=p_1^{m_1} p_2^{m_2}...p_q^{m_q}$, where $p_i$'s are prime numbers. In other words, we will extend the circuit ansatz proposed in the main text to implement QFT over groups whose cardinality is not $2^n$ for some integer $n$. For simplicity, we will not consider bit-permutations. 

Here we implement quantum Fourier sampling instead of quantum Fourier transform, which is easier to implement and sufficient for the HSP algorithm. Quantum Fourier transform sends one quantum state to the transformed quantum state, which is a unitary operation. In contrast, quantum Fourier sampling samples from the measurement outcome of the transformed state in the computational basis to get classical bit strings, where classical post-processing can be used to get the correct sample even if the quantum state prior the measurement is not exactly the quantum Fourier transformed one. This means if the circuit we implement differs from the Fourier transform by a unitary transformation that permutes the computational basis, then we do not need to implement this unitary but can correct the permutation via classical post-processing. We will explain more when we present the concrete circuit for quantum Fourier sampling. 

Brief remarks on notation: (i) throughout this section  numbers represented as $p$ are not necessarily prime unless otherwise stated; (ii) as in the rest of the paper the notation $N=2^n$ for some integer $n$, means that $N$ represents the dimension of the Hilbert space formed by $n$ qubits. 

Let us begin with a very concrete example that constructs explicitly the corresponding circuits, and then present the general theory. 
Consider an Abelian group with $6$ elements, where the group structure can be either $\Z_6$ or $\Z_2 \times \Z_3$. We will first present the respective approximate QFT circuits given in Ref.~\cite{hales2000improved} and then present a parametrized ansatz that can produce $\QFT_{\Z_6}$ or $\QFT_{\Z_2\times\Z_3}$ under suitable parameter values.

\medskip
\textbf{QFT circuit for $\Z_6$}.
L. Hales and S. Hallgren proposed a method to construct quantum circuits of QFT over any $\Z_N$ for integer $N$ in Ref.~\cite{hales2000improved}. Here we apply their method to construct the QFT circuit over $\Z_6$. 

The first question we need to address is how to store the 6 group elements using qubits. We use a 6 dimensional subspace of the 8 dimensional Hilbert space formed by $3$ qubits to store group elements of $\Z_6$. In particular, we use the subspace spanned by $\ket{0}, \ket{1},...,\ket{5}$ and only allow input states of the form $\ket{\psi} = \sum_{j = 0}^5 \alpha_j \ket{j}$ with general coefficients $\alpha_j$.

Then we claim that the quantum circuit that sends $\ket{\psi}$ to $\QFT_{\Z_6} \ket{\psi}$ is of the following form,
\begin{equation}
\label{circuit: QFT6}
        \Qcircuit @C=1.0em @R=1.0em @!R { 
	 &\lstick{} & \qw & \multigate{5}{\mathrm{MUL}} & \multigate{5}{\mathrm{QFT}_{\Z_{2^6}}}  & \qw & \qw & \meter \\
	 	 &\lstick{} & \qw & \ghost{\mathrm{MUL}} & \ghost{\mathrm{QFT}_{\Z_{2^6}}}   & \qw & \qw & \meter\\
	 	 &\lstick{} & \qw & \ghost{\mathrm{MUL}} & \ghost{\mathrm{QFT}_{\Z_{2^6}}}   & \qw & \qw
    \inputgroupv{1}{3}{.75em}{2em}{\ket{\psi}} & \meter\\
	 	 &\lstick{} & \multigate{2}{\mathrm{QFT}_{\Z_{2^3}}} & \ghost{\mathrm{MUL}} & \ghost{\mathrm{QFT}_{\Z_{2^6}}}   & \qw & \qw & \meter\\
	 	 &\lstick{} & \ghost{\mathrm{QFT}_{\Z_{2^3}}} & \ghost{\mathrm{MUL}} & \ghost{\mathrm{QFT}_{\Z_{2^6}}}   & \qw & \qw & \meter\\
	 	 &\lstick{} & \ghost{\mathrm{QFT}_{\Z_{2^3}}} & \ghost{\mathrm{MUL}} & \ghost{\mathrm{QFT}_{\Z_{2^6}}}   & \qw & \qw
		  \inputgroupv{4}{6}{.75em}{2em}{\ket{0}} & \meter\\
\\ },
\end{equation}
where the multiplication gate MUL is defined as, 
\begin{align}\label{eq:multiplication}
    \mathrm{MUL}: \ket{j}\ket{k}& \mapsto \ket{j + 6\ k} \, .
\end{align} Later we will show that MUL can be implemented by controlled phase rotation gates together with quantum Fourier transform.
The above circuit is composed of two registers, where the first register uses $3$ qubits to store the data state $\ket{\psi}$ and the second register is composed of three ancillas initialized to $\ket{0}$. 

Now we prove that this circuit ~\eqref{circuit: QFT6} can indeed approximately implement the quantum Fourier sampling over $\Z_6$. First, the gate $\mathrm{QFT}_{\Z_{2^3}}$ creates a superposition state $\sum_{k=0}^7 \ket{\psi} \ket{k}$ in the second register. Then the gate MUL will induce the following state transition:
\begin{equation}
    \sum_{k=0}^7 \ket{\psi} \ket{k}  \longrightarrow \sum_{j=0}^{5}\sum_{k = 0}^{7} \alpha_j \ket{j + 6 \ k}.
\end{equation}
Finally, applying the quantum Fourier transform over $\QFT_{\Z_{2^6}}$ (or $\QFT_{2^6}$ for short), m state 
\begin{align}
    \mathrm{QFT_{{2^6}}} \sum_{j,k}\alpha_j \ket{j + 6k} = & \sum_{j,k,l, t} \alpha_j \exp{2\pi i \frac{j+6k}{2^6}\Big(\lfloor \frac{64}{6}l \rceil + t \Big)} \ket{\lfloor \frac{64}{6}l \rceil + t } \nonumber \\
    =& \sum_{j,k,l, t} \alpha_j \exp{ 2\pi i \Big( \frac{j}{6} \frac{3}{32}\lfloor \frac{32}{3}l \rceil+ \frac{jt}{64} + \frac{3k}{32}\lfloor \frac{32}{3}l \rceil + \frac{3}{32}kt  \Big) }\ket{\lfloor \frac{64}{6}l \rceil + t } \label{eqn: C30},
\end{align}
where $\lfloor i \rceil $ is the integer around $i$, defined as $\lfloor i \rceil = \underset{j}{\text{argmin}} \{ ||i-j|| \mid j\in \Z \}$.

To see why Eq.\eqref{eqn: C30} is an approximation to $\QFT_{\Z_6}$, we present the following analysis. Since $\frac{3}{32}\lfloor \frac{32}{3}l \rceil  -l < 0.032$, the term $\frac{3}{32}\lfloor \frac{32}{3}l \rceil$ is approximately $l$. Thus, we can replace $\lfloor \frac{32}{3}l \rceil$ in Eq.~\eqref{eqn: C30} with $l$ with a small error that will vanish as the number of ancillary qubits grows (here we use only 3 qubits in \eqref{circuit: QFT6} but can use more). The rigorous error analysis can be found in Ref.~\cite{hales2000improved}. After the approximation, it is direct to verify that
\begin{align}
    \mathrm{QFT_{\Z_{2^6}}} \sum_{j,k}\alpha_j \ket{j + 6k} \approx& \sum_{j,k,l, t} \alpha_j \exp{2\pi i \Big(
        \frac{jl}{6} + \frac{jt}{64} + kl + \frac{3}{32}kt
    \Big)}\ket{\lfloor \frac{64}{6}l \rceil + t }\\
    = & \sum_{j,l,t} \alpha_j \exp{2\pi i \Big(
        \frac{jl}{6} + \frac{jt}{64}\Big)} \sum_{k}e^{\frac{3\pi i kt}{16}}  \ket{\lfloor \frac{64}{6}l \rceil + t }. \\
    \approx & \sum_{j,l} \alpha_j \exp{2\pi i \Big(
        \frac{jl}{6} \Big)} \sum_{t \in \{0,1\}} \ket{\lfloor \frac{64}{6}l\rceil +t }. \label{eq:approx_QFT_Z6}
\end{align}
The final approximation is due to the fact that the term $\sum_{k}e^{\frac{3\pi i kt}{16}}$ is close to $0$ unless $t$ is close to $0$. Thus, there are only small $t$ that survive after the summation $\sum_{k}e^{\frac{3\pi i kt}{16}}$. When the $t$ is small, the term $\frac{jl}{6}$ is much greater than the $\frac{jt}{64}$.
Comparing the result Eq.~\eqref{eq:approx_QFT_Z6} to the theoretical result
\begin{equation}
    \mathrm{QFT_{\Z_6}} \sum_{j}\alpha_j \ket{j} = \sum_j \alpha_j e^{2 \pi i \frac{jl}{6}}\ket{l},
    \label{eq:QFT_Z6}
\end{equation}
we could find that the two formulas are very similar. If we measure the whole circuit out that produces \eqref{eq:approx_QFT_Z6}, the probability of getting $\ket{\lfloor \frac{64}{6}l\rceil +t }$ is approximate to the ideal probability of getting $\ket{l}$ in Eq.~\eqref{eq:QFT_Z6}. Thus, we could approximately sample $l$ in Eq.~\eqref{eq:QFT_Z6} by sampling $\lfloor \frac{64}{6}l\rceil +t$ in Eq.~\eqref{eq:approx_QFT_Z6} and then solving the $l$ out,
\begin{equation}
    \lfloor \frac{64}{6}l\rceil +t \to l .
\end{equation}



The theoretical promise of the approximation which supports the intuition is shown in Ref.~\cite{hales2000improved}.  
Let $\mathcal{D}_{\mathrm{QFT}_{\Z_6}}$  be the probability distribution over the measurement outcomes (measured in the computational basis) of the quantum state  $\QFT_{\Z_6} \ket{\psi}$. Let $\mathcal{D}$ be the distribution of the measurement outcome obtained in circuit ~\eqref{circuit: QFT6}, which we aim to guarantee it as an approximation to $\QFT_{\Z_6} \ket{\psi}$. In fact,
Ref.~\cite{hales2000improved} proves general results of the distance between the target distribution 
$\mathcal{D}_{\mathrm{QFT}_{\Z_6}}$ and the distribution approximated by the circuit $\mathcal{D}$. It is shown
that the two distributions are close if the number of ancilla qubits is large enough. Concretely, let $n_s$ be the number of the ancilla qubits, and $s = 2^{n_s}$. In addition, let $n_q$ be the number of the qubits in the whole circuit, and $q = 2^{n_q}$. 
If $s=\Omega\left(\frac{\ln ^2 p}{\epsilon^2}\right)$ and $q=\Omega\left(\frac{p s}{\epsilon}\right)$, then Ref.~\cite{hales2000improved} proves that
\begin{equation}
    \left\|\mathcal{D}_{F_p|\alpha\rangle}-\mathcal{D}\right\|_1<\epsilon.
\end{equation}



Now, we analyze the explicit implementation of the MUL gate to see the fundamental components that can be used to construct the variational ansatz.
In order to implement $\mathrm{MUL}$, we need an ancillary register consisting of $3$ qubits initialized as $\ket{0}$. This register is used to store the multiplication outcome, such that $\ket{x}\ket{y}\rightarrow \ket{x+6 y}$ can be stored properly. We denote $\ket{x}$ as the computational basis of the $6$-qubit quantum system formed by this ancillary register and the register storing the input quantum state $\ket{\psi}$.
Defining a controlled phase rotation gate $\mathrm{CZ}^\frac{1}{p}$ as $\mathrm{CZ}^\frac{1}{p} \ket{xy} = e^{\frac{2\pi i }{2^n}pxy}\ket{
xy}$, we observe that 
\begin{align}
    \label{eq:CZ QFT appendix}  \mathrm{CZ}^\frac{1}{6} (\mathrm{QFT}_{\Z_{2^6}} \ket{x} \otimes \ket{y}) =&  \mathrm{CZ}^\frac{1}{6} \sum_z \exp{\frac{2\pi i}{2^6}xz} \ket{z}  \ket{y}\\
    =& \sum_z \exp{\frac{2\pi i}{2^6}xz}\mathrm{CZ}^\frac{1}{6} \ket{z,y}\\
    =& \sum_z \exp{\frac{2\pi i}{2^6}(x+6y)z}\ket{z,y}\\
    =& (\mathrm{QFT}_{\Z_{2^6}} \ket{x+6y}) \otimes \ket{y}\\
    =& (\mathrm{QFT}_{\Z_{2^6}}~ \mathrm{MUL} \ket{x})\otimes \ket{y} \, ,
\end{align}
holds for any $x$ and $y$.
Equivalently, we have
\begin{equation}
    \mathrm{CZ}^{\frac{1}{6}} \mathrm{QFT}_{\Z_{2^6}} =\mathrm{QFT}_{\Z_{2^6}}  \mathrm{MUL} \, .
\end{equation}
This implies that we can update the circuit \eqref{circuit: QFT6} as,
\begin{equation}
\label{circuit: QFT6 gate level}
\Qcircuit @C=1.0em @R=1.0em @!R { 
	 	 \lstick{}& \multigate{5}{\mathrm{QFT}_{\Z_{2^6}}} & \multigate{8}{\mathrm{CZ}^{\frac{1}{6}}} & \qw & \meter\\
	 	 \lstick{}& \ghost{\mathrm{QFT}_{\Z_{2^6}}} & \ghost{\mathrm{CZ}^{\frac{1}{6}}} & \qw & \meter\\
	 	 \lstick{}& \ghost{\mathrm{QFT}_{\Z_{2^6}}} & \ghost{\mathrm{CZ}^{\frac{1}{6}}} & \qw & \meter
    \inputgroupv{1}{3}{.75em}{2em}{\ket{0}} \\
	 	 \lstick{}& \ghost{\mathrm{QFT}_{\Z_{2^6}}} & \ghost{\mathrm{CZ}^{\frac{1}{6}}} & \qw & \meter\\
	 	 \lstick{}& \ghost{\mathrm{QFT}_{\Z_{2^6}}} & \ghost{\mathrm{CZ}^{\frac{1}{6}}} & \qw & \meter\\
	 	 \lstick{}& \ghost{\mathrm{QFT}_{\Z_{2^6}}} & \ghost{\mathrm{CZ}^{\frac{1}{6}}} & \qw & \meter
    \inputgroupv{4}{6}{.75em}{2em}{\ket{\psi}}\\
	 	 \lstick{}& \multigate{2}{\mathrm{QFT}_{\Z_{2^3}}} & \ghost{\mathrm{CZ}^{\frac{1}{6}}} & \qw & \qw\\
	 	 \lstick{}& \ghost{\mathrm{QFT}_{\Z_{2^3}}} & \ghost{\mathrm{CZ}^{\frac{1}{6}}} & \qw & \qw\\
	 	 \lstick{}& \ghost{\mathrm{QFT}_{\Z_{2^3}}} & \ghost{\mathrm{CZ}^{\frac{1}{6}}} & \qw & \qw
    \inputgroupv{7}{9}{.75em}{2em}{\ket{0}}\\
\\ },
\end{equation}
where the last register is dropped off and the readout of the first and the second register now gives us the quantum Fourier sampling. 

The $\mathrm{CZ}^{\frac{1}{p}}$ gate consists of the control $R_n$ gates. Recall that the control $R_n$ gate is $\mathrm{CR}_n \ket{b_1, b_2} = \exp(\frac{2\pi i b_1b_2}{2^n}) \ket{b_1, b_2} $, where $b_1, b_2 \in \{0,1\}$. So, the $\mathrm{CZ}^{\frac{1}{p}}$ gate could be constructed by 
\begin{align}
    \mathrm{CZ}^{\frac{1}{p}} \ket{x,y} =& \exp{\frac{2\pi i p}{2^n} (\sum 2^j x_j)(\sum 2^k y_k) }\ket{x_1, \cdots, x_{n}, y_1, \cdots, y_n} ~~~~~~~~~x_j, y_k\in \{0,1\}\\
    =& \exp{\sum_{j,k}\frac{2\pi i p}{2^{n-j-k}}x_jy_k  }\ket{x_1, \cdots, x_{n}, y_1, \cdots, y_n}\\
    =& \prod_{j,k} \exp(\frac{2\pi i p}{2^{n-j-k}}x_jy_k) \ket{x,y}\\
    =& \prod_{j>k}  (R^{j\to k}_{j-k})^p\ket{x,y}.
\end{align}
Here we give an example of the CZ gates with $4$ qubits, for which both $\ket{x}$ and $\ket{y}$ are associated with $2$ qubits. 
\begin{equation}
    \Qcircuit @C=1.0em @R=0.2em @!R { \\    
    & \qw     & \qw          & \qw          & \qw           
    & \ctrl{3} \qw 
    & \ctrl{2} \qw & \ctrl{1} \qw  
    & \qw \\
    & \qw     & \qw          & \ctrl{2} \qw & \ctrl{1} \qw  
    & \qw                     
    & \qw          & \gate{R_1^p} 
    & \qw \\
    & \qw     & \ctrl{1} \qw & \qw          & \gate{R_1^p}  
    & \qw                     
    & \gate{R_2^p} & \qw          
    & \qw \\
    & \qw     & \gate{R_1^p} & \gate{R_2^p} & \qw           
    & \gate{R_3^p}            
    & \qw          & \qw          
    & \qw \\
\\ }
\end{equation}

\textbf{QFT circuit for $\Z_2\times\Z_3$.} 
By inspecting the quantum circuit for Fourier sampling over $\Z_6$ we can write the circuit for Fourier sampling over $\Z_3$:
\begin{equation}
\Qcircuit @C=1.0em @R=1.0em @!R { 
     \lstick{}& \multigate{4}{\mathrm{QFT}_{\Z_{2^5}}} & \multigate{7}{\mathrm{CZ}^{\frac{1}{3}}} & \qw & \meter\\
     \lstick{}& \ghost{\mathrm{QFT}_{\Z_{2^5}}} & \ghost{\mathrm{CZ}^{\frac{1}{3}}} & \qw & \meter\\
     \lstick{}& \ghost{\mathrm{QFT}_{\Z_{2^5}}} & \ghost{\mathrm{CZ}^{\frac{1}{3}}} & \qw & \meter
\inputgroupv{1}{3}{.75em}{2em}{\ket{0}} \\
     \lstick{}& \ghost{\mathrm{QFT}_{\Z_{2^5}}} & \ghost{\mathrm{CZ}^{\frac{1}{3}}} & \qw & \meter\\
     \lstick{}& \ghost{\mathrm{QFT}_{\Z_{2^5}}} & \ghost{\mathrm{CZ}^{\frac{1}{3}}} & \qw & \meter
\inputgroupv{4}{5}{.1em}{1.4em}{\ket{\psi}}\\
     \lstick{}& \multigate{2}{\mathrm{QFT}_{\Z_{2^3}}} & \ghost{\mathrm{CZ}^{\frac{1}{6}}} & \qw & \qw\\
     \lstick{}& \ghost{\mathrm{QFT}_{\Z_{2^3}}} & \ghost{\mathrm{CZ}^{\frac{1}{3}}} & \qw & \qw\\
     \lstick{}& \ghost{\mathrm{QFT}_{\Z_{2^3}}} & \ghost{\mathrm{CZ}^{\frac{1}{3}}} & \qw & \qw 
     \inputgroupv{6}{8}{.75em}{2em}{\ket{0}}\\
}\, ,
\end{equation}
where $\ket{\psi} = \sum_{i = 0}^{2}\ket{b_i}$ in this case. Then we can construct the circuit for Fourier sampling over $\Z_2 \times \Z_3$ by concatenating the QFT blocks over $\Z_2$ and $\Z_3$:
\begin{equation}
\Qcircuit @C=1.0em @R=1.0em @!R { 
     \lstick{}& \multigate{4}{\mathrm{QFT}_{\Z_{2^5}}} & \multigate{7}{\mathrm{CZ}^{\frac{1}{3}}} & \qw & \meter\\
     \lstick{}& \ghost{\mathrm{QFT}_{\Z_{2^5}}} & \ghost{\mathrm{CZ}^{\frac{1}{3}}} & \qw & \meter\\
     \lstick{}& \ghost{\mathrm{QFT}_{\Z_{2^5}}} & \ghost{\mathrm{CZ}^{\frac{1}{3}}} & \qw & \meter
\inputgroupv{1}{3}{.75em}{2em}{\ket{0}} \\
     \lstick{}& \ghost{\mathrm{QFT}_{\Z_{2^5}}} & \ghost{\mathrm{CZ}^{\frac{1}{3}}} & \qw & \meter\\
     \lstick{}& \ghost{\mathrm{QFT}_{\Z_{2^5}}} & \ghost{\mathrm{CZ}^{\frac{1}{3}}} & \qw & \meter
\inputgroupv{4}{5}{.1em}{1.4em}{\ket{b_2b_3}}\\
     \lstick{}& \multigate{2}{\mathrm{QFT}_{\Z_{2^3}}} & \ghost{\mathrm{CZ}^{\frac{1}{6}}} & \qw & \qw\\
     \lstick{}& \ghost{\mathrm{QFT}_{\Z_{2^3}}} & \ghost{\mathrm{CZ}^{\frac{1}{3}}} & \qw & \qw\\
     \lstick{}& \ghost{\mathrm{QFT}_{\Z_{2^3}}} & \ghost{\mathrm{CZ}^{\frac{1}{3}}} & \qw & \qw
\inputgroupv{6}{8}{.75em}{2em}{\ket{0}}\\
\lstick{\ket{b_1}}& \gate{\mathrm{QFT}_{\Z_2}} & \qw & \qw & \meter\\
\\ }.
\label{eq:QFT_Z2XZ3}
\end{equation}
Here $\ket{b_i}$ means the $i$-th bit of $\ket{\psi}$. An element in $\Z_2 \times \Z_3$ could be denoted as $(b_1, b_2b_2)$,
\begin{equation}
    \Z_2 \times \Z_3 = \{ (0, 00), (0, 01), (0, 10), (1, 00), (1, 01), (1, 10) \}.
\end{equation}
Then, the circuit ~\eqref{eq:QFT_Z2XZ3} will give the Fourier sampling over $\Z_2 \times \Z_3$ if the state is spanned by the states $\{ \ket{b_1, b_2b_3} \}$.

\begin{figure}
    \centering
    \includegraphics[width= 0.7\linewidth]{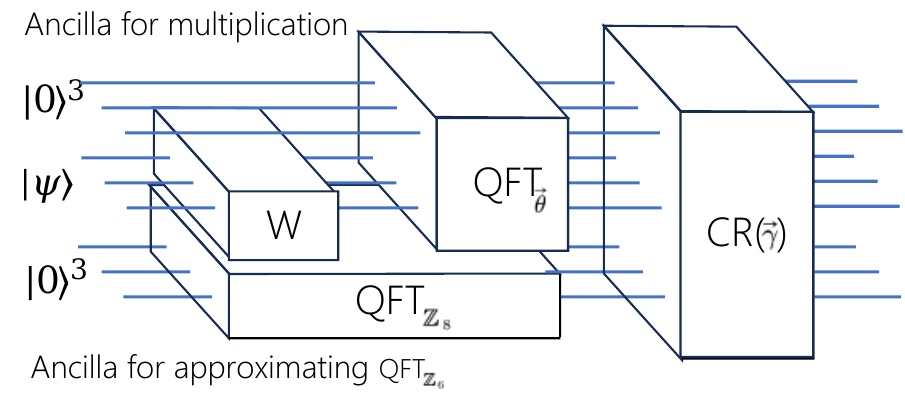}
    \caption{{\bf  The parameterized circuit for Fourier sampling over groups $\Z_{2^6}$ or $\Z_{2^2} \times \Z_{2^3} $.} The $\mathrm{CR}(\Vec{\gamma})$ is composed by $(R^{j\to k}_{j-k})^{\gamma_{j,k}}$ for $\forall j>k$. In other words, $\mathrm{CR}(\Vec{\gamma})$ is a fully  connected variational $(R^{j\to k}_{j-k})^{\gamma_{j,k}}$ layer. }
    \label{fig: paraCircuit_Z6}
\end{figure}

\medskip
\textbf{A parametrized circuit for QFT over $\Z_6$ and $\Z_2 \times \Z_3$}.
Generalising the two examples of the circuits that achieve QFT over $\Z_6$ and $\Z_2 \times \Z_3$, we can construct a parametrized circuit that can recover each case with a suitable choice of parameters.  This will shed light on how to construct the QFT over an arbitrary Abelian group. By replacing the second layer of the QFT gate with the variational QFT in the main text and parameterizing the CZ gate, we could get the variational QFT over arbitrary Abelian groups $\Z_{p_1}\times \Z_{p_2} \times \cdots \times \Z_{p_m}$. The circuit is shown in Fig.~\ref{fig: paraCircuit_Z6}.


\textbf{General Construction.}
In general, we want to construct a parameterized quantum circuit for all types of Abelian groups if the group elements are assigned properly. The elements $(g_1^{m_1}, \cdots, g_n^{m_n})$ are assigned properly if the power of generators $m_1, \cdots, m_n$ are stored in separated quantum registers. 
For example, 
\begin{equation}
    \begin{matrix}
        \Z_2 \times \Z_3 & = & \{& (0,0), & (0,1), & (0,2), & (1,0),  & (1,1), & (1,2) &\} \\
        G_1 & = & \{& 000, & 001, & 010, & 100, & 101, &110 &\} \\
        G_2 & = & \{& 000, & 001, & 010, & 011, & 100, &101 &\} \\
    \end{matrix}
\end{equation}
$G_1$ is a proper assignment but $G_2$ is not. In $G_1$, the first qubit represents the factor of $\Z_2$, while the second and the third qubits represent the factor of $Z_3$. However, in $G_2$, the second and the third qubits contain the  information of both generators $(1,0)$ and $(0,1)$. In this case, the group structure becomes mixed up when we deal with the two qubits. For simplicity we now side-step this problem by reducing the freedom of the variational circuit so that we do not vary over assignments. This reduces the expressiveness of the circuit. 

The quantum circuit of \eqref{circuit: QFT6} could be easily parameterized and promoted to a general form $\Z_p$. Then, we only need to concatenate the circuit of $\Z_{p_i}$ to construct the circuit for approximate Fourier sampling over $\prod \Z_{p_i}$. Notice that in our algorithm, the QFT always appear in Fourier sampling blocks. Thus, the approximate Fourier sampling is good enough for our algorithms.

The first thing to promote the circuit in \ref{circuit: QFT6} is to break the general gate MUL into elementary  logic gates. Making use of Ref.~\cite{ruiz2017quantum}, we can use the QFT to construct the unitary for linear combination of states 
\begin{equation}
\mathrm{QFT}_{(N)}^\dagger \left(\prod_{i=1}^{N-1} \mathrm{CZ}_{i, N}^{\frac{1}{a_i}}\right) \mathrm{QFT}_{(N)} \ket{x_1, x_2, ...,x_N} = \ket{x_1, ..., x_{N-1} }\ket{\sum_{i=1}^{N} a_ix_i},
\end{equation}
where $a_i$ can be an arbitrary integer.
This relation produces Eq.\eqref{eq:multiplication} (the MUL gate) for  $a_0=1$ and $a_1=p$ and $x_1 = i$, $x_2=j$.
The gate $\mathrm{QFT}_{(N)}$ is the QFT over the $N$-th register, $CZ^a_{b,c}$ is the control phase gate on register $b$ and $c$, 
\begin{equation}
    \mathrm{CZ}^a_{b,c}\ket{x_b, x_c} = e^{\frac{2\pi i x_b x_c}{a 2^n}}\ket{x_b, x_c},
\end{equation}
where $n$ is the number of qubits in the register. Similarly, we also have
\begin{align}
\mathrm{CZ}^\frac{1}{p} (\mathrm{QFT}_{\Z_{2^n}} \ket{x} \otimes \ket{y}) = (\mathrm{QFT}_{\Z_{2^n}}~ \mathrm{MUL} \ket{x+py})\otimes \ket{y} .
\end{align}
Substituting the $p$ as a series of variational parameters $\vec{\gamma}$, we get the parameterized MUL gate in a general form. Furthermore, using the parameterized SWAP gates and the parameterized QFT in the main text instead of the $\mathrm{QFT}_{\Z_{2^6}}$ in circuit~\ref{circuit: QFT6}, we get the approximate Fourier sampling which is shown in Fig.~\ref{fig: paraCircuit_Z6}. 


\color{black}

\subsection{Efficient circuit implementation of quantum Fourier transform.\label{supplementary:efficient_circuit_QFT}}

In this subsection, we review the results of implementing the general quantum Fourier transform in the circuit level. Note that it is direct to verify that 
\begin{equation}
    \QFT_{\Z_{2^{m_1}}\times ... \times \Z_{2^{m_q}}} = 
 \QFT_{\Z_{2^{m_1}}} \otimes ...\otimes \QFT_{\Z_{2^{m_q}}} \, .
\end{equation} 
So we only need to consider the implementation of Fourier transform over cyclic groups. In fact,  it is shown  \cite[Supplementary A.4]{mosca1999quantum} that the quantum Fourier transform $\QFT_{\Z_{AB}}$ over cyclic group $\Z_{AB}$ with co-prime $A,B$ can be implemented by $\QFT_{\Z_A}$ and $\QFT_{\Z_B}$. This means that it is sufficient to consider the circuit construction of the quantum Fourier transform over $\Z_{2^n}$ and over $\Z_N$ with $N$ being an odd number.  Since we only consider the groups whose cardinality is $2^n$, we will retrieve the circuit construction for $\QFT_{\Z_{2^n}}$ as follows, and detailed discussions for QFT over $\Z_N$ with odd number $N$ can be found in Ref.~\cite[SupplementaryA2]{lomont2004hidden}.

It can be proved~\cite[Ch.5]{nielsen2002quantum} that 
\begin{align}\label{eq:Fourier_expansion_qubits}
    &\QFT_{\Z_{2^n}}\ket{j_1 j_2 ...j_n} \\ \nonumber
    =& \frac{1}{\sqrt{2^n}} \left( \ket{0} + e^{2\pi \mi 0.j_n} \ket{1}\right) ... \left( \ket{0} + e^{2\pi \mi 0.j_1...j_n} \ket{1} \right) \, ,
\end{align}
where $0.j_1 j_2 ...j_n$ is the short hand notation of $j_1 2^{-1} + j_2 2^{-2} +...+j_n2^{-n}$.
Now, we construct the quantum circuit that generates the phase $e^{2\pi \mi 0.j_1j_2...j_n}$. Let 
\begin{equation}
    H_{(a)} =\frac{1}{\sqrt{2}}\begin{bmatrix}
    1 & 1 \\
    1 & -1 
\end{bmatrix}
\end{equation} be the Hadamard gate applied to the $a$-th qubit (corresponding to $j_a$ in the bit string $\ket{j_1j_2...j_n}$) and let 
\begin{equation}
    R_k = \begin{bmatrix}
    1 & 0 \\
    0 & \exp(\frac{2 \pi \mi}{2^k} )
\end{bmatrix}
\end{equation}
be a phase rotation operator. Now, denote $R_k^{a\rightarrow b}$ be the two-qubit operation of applying the phase rotation $R_k$ on the $b$-th qubit controlled by the $a$-th qubit, i.e. $R_k$ is applied to qubit $b$ when qubit $a$ is in state $\ket{1}$ and nothing is applied when qubit $a$ is in state $\ket{0}$. Then, it is direct to verify that
\begin{align}\label{eq:controlled_phase_rotation}
   &R_n^{n \rightarrow 1}...R_3^{3\rightarrow 1} R_2^{2\rightarrow 1}H_{(1)}\ket{j_1...j_n} \\ \nonumber
   =&\frac{1}{\sqrt{2}} \left(\ket{0} + e^{2 \pi \mi 0.j_1 j_2 ...j_n} \ket{1}\right) \otimes \ket{j_2 j_3 ...j_n} \, .
\end{align}
and similarly
\begin{align}
       &R_{n-1}^{n \rightarrow 2}...R_3^{4\rightarrow 2} R_2^{3\rightarrow 2} H_{(2)}\ket{j_1 j_2...j_n} \\ \nonumber
   =&\ket{j_1} \otimes \frac{1}{\sqrt{2}} \left(\ket{0} + e^{2 \pi \mi 0.j_2 ...j_n} \ket{1}\right) \otimes \ket{j_3 ...j_n} \, .
\end{align}
Now, consecutively apply $R_{n-k+1}^{n-k+1 \rightarrow k}...R_{2}^{k+1 \rightarrow k} H_{(k)}$ for $k=1,...,n$, and we have
\begin{equation}
    \rightarrow \frac{1}{\sqrt{2^n}} \left( \ket{0} + e^{2\pi \mi 0.j_1...j_n} \ket{1} \right)  ...  \left( \ket{0} + e^{2\pi \mi 0.j_n} \ket{1}\right) \, ,
\end{equation}
which is similar to the desired quantum Fourier transformation shown in Eq.\eqref{eq:Fourier_expansion_qubits} but with reversed qubits order. Finally, we complete the quantum Fourier transform by reversing the qubits order, using approximately $\lfloor n/2\rfloor$ SWAP operations. The total number of one-qubit and two-qubit operations used to complete the quantum Fourier transform, is $O(n^2) = O(\log^2 N)$, which is very efficient. As an example, the quantum circuit for the Fourier transform over $\Z_8$ is 
\begin{equation}
\begin{array}{c}
\Qcircuit @C=.3em @R=0em @!R {
\lstick{i_1}& \multigate{2}{\QFT_{\Z_8}} & \qw &                                          & & \gate{H} & \gate{R_2}   & \gate{R_3}    & \qw        &\qw            &\qw      &\qw &\qswap   &\qw&\qw\\
\lstick{i_2}& \ghost{{\QFT_{\Z_8}}}      & \qw & \push{\rule{.3em}{0em}=\rule{.3em}{0em}} & & \qw      & \ctrl{-1}    &\qw            &\gate{H}    &\gate{R_2}     &\qw      &\qw &\qw\qwx   &\qw&\qw\\
\lstick{i_3}& \ghost{{\QFT_{\Z_8}}}      & \qw &                                          & & \qw      & \qw          & \ctrl{-2}     &\qw         &\ctrl{-1}      &\gate{H} &\qw &\qswap \qwx  &\qw&\qw\\
} \, \end{array}.
\end{equation}

\section{Searching over possible isomorphisms.\label{supplementary:searching_over_isomorphisms}}

\subsection{Theoretical analysis.}

In constructing variational quantum circuits for HSP, not only the types of group $\Z_{2^{m_1}}\times ...\times \Z_{2^{m_q}}$, but also different isomorphisms of the same group $\{G|G\cong \Z_{2^{m_1}}\times ...\times \Z_{2^{m_q}}\}$ should be searched. The reason that we should also search over possible isomorphism is manifested in the following example where we will see that there is a hidden subgroup with respect to group $G'$ but not $G$, although both of them are isomorphic to $\Z_8$.

\medskip
\textbf{Failure of hidden subgroup compression with respect to a fixed group.} Consider the time series data $\{0,0,0,0,1,1,1,1\}$, which seems to have  no hidden subgroup at all. If we are given this sequence and we presume the group is $G$ that is isomorphic to $\Z_8$ through the canonical isomorphism $i = \tau(i_1i_2i_3) $ for $i \in \Z_8$ and $i_1i_2i_3 \in G$, 
then the algorithm fails because the time series does not have any hidden subgroup at all. This can be checked by listing the group $G=(\{0,1\}^3,*)$ and time series data as follows,
\begin{equation}
    \begin{matrix}
 G & = &\{ 000,& 001, & 010, & 011, & 100, & 101, & 110, & 111 \} \, . \\
 \Z_8 &= &\{ 0, & 1, & 2, & 3, & 4, & 5, & 6, & 7 \} \, . \\
\{f(i)\}&=&\{0, &0,&0,&0,&1,&1,&1,&1 \} \, .
\end{matrix}     \nonumber
\end{equation}
Here the first row is the bit strings as group elements in $G$ ordered in the increasing time, and the second row is the group elements $\Z_8$ that is one-to-one corresponding (isomorphic) to the elements in $G$ through the map $\tau$, and finally, the third row is the time series data.  It is then clear that the repeating elements correspond to $\{0,1,2,3\}\subset \Z_8$, 
 which is not a subgroup, because the only non-trivial subgroups of $\Z_8$ is either $\{0,4\}$ or $\{0,2,4,6\}$. This suggests that the permuted time series $\{0,0,0,0,1,1,1,1\}$ cannot be compressed by a hidden subgroup algorithm, regardless of the fact that there is an obvious repeating pattern in the data. 

 Does it mean that this time series data can never be formulated as an HSP? The answer is NO.  We will see that it has a hidden subgroup structure with respect to a slightly different group $G'$ that is isomorphic to $G$.

\medskip
\textbf{Restoring the hidden subgroup structure by an isomorphism.} Now, we will show that the compression of the permuted time series  $\{0,0,0,0,1,1,1,1\}$ can in fact be formulated as an HSP, with only a minor modification to the way of assigning group structure $\Z_8$ to the binary strings $\{0,1\}^3$. To see this, let us first review how we assigned the group $\Z_8$ to the binary strings. In our earlier example, the bit string $i_1i_2i_3$ is identified to the group elements $i\in \Z_8$ via the map $\tau$, 
\begin{align}
    i = \tau (i_1 i_2 i_3) = i_1 2^2 + i_2 2^1 + i_3 \, .
\end{align}
For two bit strings $\tau^{-1}(i)$ and $\tau^{-1}(j)$, the group operation $*$ is defined as 
\begin{equation}
  \tau^{-1}(i)*\tau^{-1}(j) =\tau^{-1}( i + j \mod 8  ) \, .
\end{equation}
This is the canonical way of assigning the group structure $\Z_8$ to the binary strings, but it is not the only way. Changing the identification map $\tau$ will give us a different specification of the group operation, resulting in a group $G'$ that is isomorphic to $\Z_8$ and therefore to $G$. For example, consider the map $\tau'$ such that
\begin{equation}
    i = \tau'(i_1 i_2 i_3) = i_3 2^2 +i_2 2 + i_1 \, ,
\end{equation}
where we swapped the role of the first bit $i_1$ and the third bit $i_3$, compared to the definition of $\tau$. Now we specify the group operation $*'$ as
\begin{equation}
    \tau'^{-1}(i) *' \tau'^{-1}(j) = \tau'^{-1}(i+j \mod 8) \, .
\end{equation}
We emphasise here that no matter how the identification map  $\tau'$ is changed, it is used to define the group operation, and the way to identify the binary strings $i_1i_2i_3$ to the set of indices  $i$ remains the same; we always have time $i=\tau(i_1i_2i_3)$ and thus the time series data remains unaltered.
Now, the group $G'=(\{0,1\}^3,*')$ and the time series data is listed as
\begin{equation}
\begin{matrix}
  G'&=&\{ 000,& 001, & 010, & 011, & 100, & 101, & 110, & 111 \} \, . \\
 \Z_8& =&\{ 0, & 4, & 2, & 6, & 1, & 5, & 3, & 7 \} \, . \\
  \{f(i)\}&=&\{0, &0,&0,&0,&1,&1,&1,&1 \} \, .
\end{matrix}     \nonumber
\end{equation}
Then the repeating elements correspond to $\{0,2,4,6\}$, which forms a subgroup of $\Z_8$. 
This means that the function $f$ now has a hidden subgroup and can be compressed using our algorithm, with the only caveat being that the Fourier transform now should be over $G'$ instead of $G$. 

This example suggests that knowing the group type is not enough; we should also search over different assignments of the same type of group to the bit strings.

\subsection{Parameterized quantum circuits.\label{sec:parameterized_quantum_circuits_QFT_differnt_isomorphism}}

\textbf{Searching over groups of the same type.}
Searching over all possible group types is not sufficient to solve the variational hidden subgroup compression, as multiple ways of assigning the same type of group exist (examples in Supplementary Material~\ref{supplementary:searching_over_isomorphisms}). We must search over diverse ways (isomorphisms) of assigning groups. This requires a parametrized quantum circuit for the quantum Fourier transforms over different isomorphisms, as detailed in this section.

To illustrate the distinction between quantum Fourier transforms over different isomorphic groups, consider groups $G$ and $G'$ both isomorphic to $\Z_8$. The isomorphisms are $i = \tau(i_1i_2i_3) = 2^2\, i_1 + 2\, i_2+ i_3$ for $i \in \Z_8$ and bit strings $i_1i_2i_3 \in G$, and $i=\tau'(i_1i_2i_3)= 2^2 \, i_3 + 2 \, i_2 +i_1$ for $\Z_8$ and $G'$. As demonstrated in Supplementary Material~\ref{supplementary:efficient_circuit_QFT}, the quantum circuit for the Fourier transform over $G$ is
\begin{equation}
\begin{array}{c}
\Qcircuit @C=.3em @R=0em @!R {
\lstick{i_1}& \multigate{2}{\QFT_{G}} & \qw &                                          & & \gate{H} & \gate{R_2}   & \gate{R_3}    & \qw        &\qw            &\qw      &\qw &\qswap   &\qw&\qw\\
\lstick{i_2}& \ghost{{\QFT_{G}}}      & \qw & \push{\rule{.3em}{0em}=\rule{.3em}{0em}} & & \qw      & \ctrl{-1}    &\qw            &\gate{H}    &\gate{R_2}     &\qw      &\qw &\qw\qwx   &\qw&\qw\\
\lstick{i_3}& \ghost{{\QFT_{G}}}      & \qw &                                          & & \qw      & \qw          & \ctrl{-2}     &\qw         &\ctrl{-1}      &\gate{H} &\qw &\qswap \qwx  &\qw&\qw\\
}\, 
\end{array}.
\end{equation}
Let the $\pi$ be a  bit-permutation acting on the digits of the bit strings such that $\pi(i_1i_2i_3)=i_3i_2i_1$. Then it gives the group isomorphism from $G$ to $G'$. At the circuit level, $\pi$ induces a unitary $U_\pi = \mathrm{SWAP}_{13}$ that transforms elements in $G$ to elements in $G'$ by swapping the first and the third qubits. Then the Fourier transform over $G'$ can be implemented by 
\begin{equation}\label{eq:QFT_Z8_Permutation}
    \QFT_{G'} = \SWAP_{13}\, \QFT_{G} \, \SWAP_{13} \, ,
\end{equation}
which implies the following quantum circuit,
\begin{equation}
\begin{array}{c}
\Qcircuit @C=.3em @R=0em @!R {
\lstick{i_1}& \multigate{2}{\QFT_{G'}} & \qw &                                          & & \qw &\qswap &\qw & \gate{H} & \gate{R_2}   & \gate{R_3}    & \qw        &\qw           &\qw      &\qw   &\qw\\
\lstick{i_2}& \ghost{{\QFT_{G'}}}      & \qw & \push{\rule{.3em}{0em}=\rule{.3em}{0em}} & & \qw &\qw \qwx   &\qw &\qw      & \ctrl{-1}    &\qw            &\gate{H}    &\gate{R_2}     &\qw      &\qw   &\qw\\
\lstick{i_3}& \ghost{{\QFT_{G'}}}      & \qw &                                          & & \qw &\qswap \qwx &\qw &\qw      & \qw          & \ctrl{-2}     &\qw         &\ctrl{-1}      &\gate{H} &\qw  &\qw\\
} \, \end{array}.
\end{equation}
Compared to the QFT over $G$, the only difference is the permutation of qubits before and after the same phase rotation gates.

Now, we give a full characterization of the quantum circuits that implement quantum Fourier transforms over  of the same type.

Let us fix $G_0$ as the canonical way of assigning the group $\Z_{2^{m_1}} \times \Z_{2^{m_2}} \times ... \times \Z_{2^{m_q}}$, without permuting the digits in the bit string. Let $\sigma$ be an isomorphism from $G_0$ to a different group $G_\sigma$,
\begin{equation}
    \begin{matrix}
        \sigma : & G_0 & \rightarrow & G_{\sigma} \\
        & i_1 i_2...i_n & \mapsto & \sigma(i_1 i_2...i_n)
    \end{matrix} \, .
\end{equation}
In fact, the isomorphism induces a unitary $U_{\sigma}$ that maps the basis $\{\ket{i_1i_2...i_n}\}$ to itself via
\begin{equation}\label{eq:unitary_isomorphism}
U_{\sigma} \ket{i_1i_2...i_n} = \ket{\sigma(i_1 i_2 ...i_n)} \, ,
\end{equation}
for all bit strings $i_1i_2...i_n$.   This unitary is well defined since $\sigma$ is a one-to-one correspondence, meaning that $ U_{\sigma^{-1}} U_{\sigma} = I $ and therefore $U_{\sigma^{-1}} = U_\sigma^\dagger$. Then it is easy to verify the following proposition,
\begin{proposition}
    The quantum Fourier transform over group $G_\sigma \cong G_0$, is unitarily conjugate to the quantum Fourier transform over $G_0$, i.e. 
    \begin{equation}
        \QFT_{G_\sigma} = U_{\sigma}^\dagger \, \QFT_{G_0} \, U_{\sigma} \, ,
    \end{equation}
    where $\sigma$ is the isomorphism from $G_0$ to $G_{\sigma}$ and $U_\sigma$ is the associated unitary defined in Eq.\eqref{eq:unitary_isomorphism}.
\end{proposition}

As also discussed In the Methods~\ref{sec:hidden_subgroup_as_data_compression}, there are $2^n !$ ways of assigning any given group type, as each isomorphism corresponds to a unique permutation of the $2^n$ bit strings of length $n$. This results in $2^n !$ permutations, denoted as $S_{2^n}$, which forms the permutation group. However, to parametrize all $2^n!$ permutation unitaries, an exponential number of parameters is necessary; it takes $O(n 2^n)$ bits to label one permutation. Thus, having a parametrized quantum circuit capable of implementing all $2^n!$ quantum Fourier transforms cannot achieve any compression at all.  We, therefore, consider searching over a subset of the $2^n!$ isomorphisms, which are given by bit-permutations.

 \medskip
\textbf{Searching over all bit-permutations.} The family of groups that we aim to cover, is generated by a subset of the permutations of  $n$ bits which is commonly referred to as bit-permutations, instead of all the permutations of all the $2^n$ bit strings. This is inspired by Eq.\eqref{eq:QFT_Z8_Permutation}, where the QFT is obtained after applying a SWAP gate on the qubits, which has a simpler structure to implement. 

Let $\pi\in S_n$ be a bit-permutation of the digits in a bit-string of length $n$, where $S_n$ is the symmetric group of $n$ elements. Acting with this permutation on the group $G=(\{0,1\}^n, *)$ and we get a different group $G_\pi =(\{0,1\}^n, *')$ that is isomorphic to $G$. On the circuit level, the isomorphism $\pi$ induces a unitary transformation,
\begin{equation}
    U_{\pi} \ket{i_1i_2...i_n} = \ket{i_{\pi(1)}i_{\pi(2)}...i_{\pi(n)}} \, ,
\end{equation}
for any bit string $i_1i_2...i_n$. A good property of $U_\pi$ is that it can be implemented by SWAP gates only.

\begin{proposition}[Parameterized circuits for bit-permutations.]
Using $n(n-1)/2$ parameters, we can implement $U_\pi$ for any $\pi \in S_n$. 
\end{proposition}
We will prove the above proposition by construction and the protocol is similar to the bubble sorting algorithm. First, we define the following  parameterized gate by SWAP gates between nearest neighbors:
\begin{equation}
\SWAP_{\vec{\lambda}_{(k)}} = \SWAP_{k,k+1}^{\lambda_{k}} \SWAP_{k-1,k}^{\lambda_{k-1}} \, ... \SWAP_{1,2}^{\lambda_1} \, ,
\end{equation}
where $\vec{\lambda}_{(k)} = (\lambda_1, \lambda_2,...,\lambda_{k})$ encodes the $k$ parameters.  After applying a given permutation $\pi \in S_n$, the $i$-th bit is permuted to the $\pi(i)$-th bit. To begin with, we map the $\pi^{-1}(n)$-th bit to the position $n$ via the parametrized $\SWAP_{\vec{\lambda}_{(n-1)}}$ with parameters $\lambda_{\pi^{-1}(n)}$, $ \lambda_{\pi^{-1}(n)+1}$, ... , $ \lambda_{n-2}$, $\lambda_{n-1}$ being 1 and all other parameters being $0$. After this operation, we successfully permute the $\pi^{-1}(n)$-th bit to the $n$-th bit. Suppose these SWAP gates induces a permutation $\sigma$ on the first $n-1$ bits, then after applying $\SWAP_{\Vec{\lambda}_{(n-1)}}$, the first $n-1$ bits become $i_{\sigma(1)}i_{\sigma(2)}...i_{\sigma(n-1)}$ which we needed to transform to $i_{\pi(1)}i_{\pi(2)}...i_{\pi(n-1)}$. Next, we apply $\SWAP_{\vec{\lambda}_{(n-2)}}$ with a similar strategy that maps the bit now in position $\sigma (\pi^{-1} (n-1))$ to the $(n-1)$-th bit. Repeating the process until we put $i_{\pi^{-1}(2)}$ at the second position, resulting in the outcome sequence of  $i_{\pi(1)}i_{\pi(2)}...i_{\pi(n-1)}$  we thus complete the implementation of the permutation $\pi$.
The overall process can be implemented with suitable parameters by the following parameterized gate $W_{\vec{\theta}}$,
\begin{equation}
    W_{\Vec{\theta}} = \SWAP_{\vec{\lambda}_{(1)}} \, \SWAP_{\vec{\lambda}_{(2)}} \, ... \SWAP_{\vec{\lambda}_{(n-1)}} \, ,
\end{equation}
where $\vec{\theta} = \{\vec{\lambda}_{(1)}, \vec{\lambda}_{(2)}, ..., \vec{\lambda}_{(n-1)}\}$ encodes the total $n(n-1)/2$ parameters. For example, the permutation $\pi\in S_4$ such that $\pi(i_1i_2i_3i_4) = i_4 i_3 i_1 i_2$ can be implemented by the following circuit
\begin{equation}
\begin{array}{c}
\Qcircuit @C=.5em @R=0em @!R {
\lstick{i_1}& \multigate{3}{W_{\vec{\theta}}} & \qw &                                          & & \qw &\qw         &\qw  &\qw         &\qw&\qw&\qw&\qw &\qswap      &\qw &\qw            &\qw&\qw&\qw&\qw  &\qswap      &\qw &\qw & \,\,\,\, i_4 \\
\lstick{i_2}& \ghost{W_{\vec{\theta}}}        & \qw &                                          & & \qw &\qswap      &\qw  &\qw         &\qw&\qw&\qw&\qw &\qswap \qwx &\qw &\qswap         &\qw&\qw&\qw&\qw  &\qswap \qwx &\qw &\qw & \,\,\,\,  i_3\\
\lstick{i_3}& \ghost{W_{\vec{\theta}}}        & \qw & \push{\rule{.3em}{0em}=\rule{.3em}{0em}} & & \qw &\qswap \qwx &\qw  &\qswap      &\qw&\qw&\qw&\qw &\qw         &\qw &\qswap \qwx    &\qw&\qw&\qw&\qw  &\qw         &\qw &\qw & \,\,\,\, i_1\\
\lstick{i_4}& \ghost{W_{\vec{\theta}}}        & \qw &                                          & & \qw &\qw         &\qw  &\qswap \qwx &\qw&\qw&\qw&\qw &\qw         &\qw &\qw            &\qw&\qw&\qw&\qw  &\qw         &\qw &\qw & \,\,\,\,  i_2\\
} \, \, \, \,\,\, \end{array},
\end{equation}
with the parameter $\vec{\theta} = \{(1),(1,1),(0,1,1)\}$.

Now we have the following parametrized quantum circuit $\WQFT$ that is able to implement quantum Fourier transform over any finite Abelian groups  {of type $\Z_{2^{m_1}}\times \Z_{2^{m_2}} \times...\times \Z_{2^{m_q}}$} and over any bit-permutations of the same group,
\begin{equation}
\begin{array}{c}
\Qcircuit @C=.3em @R=0.2em @!R {
& \multigate{4}{\mathrm{WQFT}_{\vec{\theta}}}  & \qw &  & & \qw & \multigate{4}{W_{\vec{\theta}}^\dagger }&\qw&\qw&\qw &\multigate{4}{\mathrm{QFT}_{\vec{\theta}}} &\qw&\qw&\qw & \qw & \multigate{4}{W_{\vec{\theta}}}  & \qw \\
& \ghost{\mathrm{WQFT}_{\vec{\theta}}}         & \qw &  & & \qw &\ghost{W_{\vec{\theta}}^\dagger} &\qw&\qw&\qw        &\ghost{\mathrm{QFT}_{\vec{\theta}}}  &\qw&\qw   &\qw        & \qw &\ghost{W_{\vec{\theta}}}     & \qw \\
\raisebox{1em}\vdots & \nghost{\QFT_{\vec{\theta}}}& \raisebox{1em}\vdots & \push{\rule{.1em}{0em}=\rule{.1em}{0em}} & & \raisebox{1em}\vdots   &\nghost{W_{\vec{\theta}}^\dagger} &&&  \raisebox{1em}\vdots &\nghost{\mathrm{QFT}_{\vec{\theta}}}   & & &  \raisebox{1em}\vdots &  &\nghost{W_{\vec{\theta}}}   & \raisebox{1em}\vdots \\
& \ghost{\mathrm{WQFT}_{\vec{\theta}}}         & \qw & & & \qw &\ghost{W_{\vec{\theta}}^\dagger} &\qw&\qw&\qw     &\ghost{\mathrm{QFT}_{\vec{\theta}}}           &\qw&\qw & \qw &\qw&\ghost{W_{\vec{\theta}}}                    & \qw \\
& \ghost{\mathrm{WQFT}_{\vec{\theta}}}         & \qw & & & \qw &\ghost{W_{\vec{\theta}}^\dagger} &\qw&\qw&\qw     &\ghost{\mathrm{QFT}_{\vec{\theta}}}            &\qw&\qw  & \qw  &\qw &\ghost{W_{\vec{\theta}}}                    & \qw \\
} \, \end{array}.
\end{equation}

{ 
\section{Compressing quantum states using our variational HSP algorithm.\label{appendix:compressing_quantum_states_using_variational_HSP}}

Our variational HSP algorithm, after a small adjustment, can also be applied to compress quantum states that satisfy a certain type of symmetry. This means, our work no longer gives a quantum compression algorithm for classical data but it also provides a compression algorithm for quantum data. 

In this section, we present how our algorithm can be adapted to compress quantum states and what kinds of quantum states can be compressed by our algorithm.

\medskip
\textbf{Compression of quantum states with symmetry.} Let us start with an example demonstrating how identifying symmetry can lead to quantum state compression and establish concepts and tools for quantum state compression. A general two qubit state $\ket{\psi}_{AB}$ lives in a 4-dimensional Hilbert space. If the state satisfy the swap symmetry, i.e., $\mathrm{SWAP} \ket{\psi}_{AB}=\ket{\psi}_{AB}$, then the state must live in the 3-dimensional symmetric subspace spanned by $\{\ket{00},\ket{11},\frac{1}{\sqrt{2}}(\ket{01}+\ket{10})\}$. Therefore, knowing the symmetry of the state can effectively reduce the dimensionality of the state from $4$ to $3$, and, therefore, compress the state. 

The above observation leads to the following formulation of quantum state compression. We consider lossless compression where the input state can be deterministically recovered from the compressed state without error.

\begin{definition}[Quantum sates compression via identifying symmetry]
A source will generate pure quantum states $\ket{\psi} \in \mathcal{S}$ according to a probability distribution, with a promise that these quantum states are invariant under certain symmetry operator $O$,
\begin{equation}
    O \ket{\psi} = \ket{\psi} \, , \, \, \forall \ket{\psi} \in \mathcal{S} \, .
\end{equation}
The goal is to find a CPTP map $\mathcal{E}$ that sends $\ket{\psi}$ to a quantum state $\ket{\Tilde{\psi}}$ whose Hilbert space dimension is smaller, such that there exists the other CPTP map $\mathcal{D}$ that recovers  $\ket{\Tilde{\psi}}$ to $\ket{\psi}$ perfectly. Mathematically, we have
\begin{equation}
\mathcal{D}\circ \mathcal{E} (\ket{\psi}\bra{\psi}) = \ket{\psi}\bra{\psi}, \, \, \forall \ket{\psi} \in \mathcal{S} \, .
\end{equation}
\end{definition}
Illustrating the above definition of quantum state compression in terms of quantum circuit, we have,
\begin{equation}
\begin{array}{c}
\Qcircuit @C=.3em @R=0.2em @!R {
 & & \qw                    & \multigate{3}{\mathcal{E} }  &\qw& \qw&\qw        & \qw         &\qw&\qw&\qw & \qw   & \multigate{3}{\mathcal{D}}            & \qw \\
 & & \qw                    &\ghost{\mathcal{E}}           &\qw&&              &            &&  &\qw   & \qw          &\ghost{\mathcal{D}}                    & \qw \\
 & & \raisebox{1em}\vdots   &\nghost{\mathcal{E}}          &  \raisebox{1em}\vdots&&   &            && &  \raisebox{1em}\vdots &     &\nghost{\mathcal{D}}                   & \raisebox{1em}\vdots \\
 & & \qw                    &\ghost{\mathcal{E}}           &\qw&&              &            && & \qw &\qw              &\ghost{\mathcal{D}}                    & \qw \\
} \, \end{array},
\end{equation}
where the bottleneck represents the compressed quantum state $\ket{\Tilde{\psi}}$,

This quantum state compression problem can be solved by identifying the invariant subspace of the symmetry operator $O$. Once the invariant subspace of this symmetry operator $O$ is identified, it is easy to construct the CPTP map $\mathcal{E}$ and $\mathcal{D}$. Take the SWAP symmetry of two-qubit states presented previously as an example. Denote the basis $\{\ket{00},\ket{11},\frac{1}{\sqrt{2}}(\ket{01}+\ket{10})\}$ as $\ket{\hat{0}} = \ket{00}$, $\ket{\hat{{1}}} = \frac{1}{\sqrt{2}}(\ket{01}+\ket{10})$, and $\ket{\hat{2}} = \ket{11}$ and the compression map can be constructed as
\begin{equation}
    \mathcal{E}(\rho):= \sum_{i,j=0}^2 \ket{i}\bra{j} \bra{\hat{i}}\rho\ket{\hat{j}} \, ,
\end{equation}
which project the state in to the invariant subspace of the SWAP operation, where $\ket{\hat{i}}$ is a basis vector in the 4-dimensional Hilbert space and $\ket{i}$ is a basis vector in the 3-dimensional Hilbert space.
And the recovery map can be constructed as
\begin{equation}
    \mathcal{D} (\Tilde{\rho}) := \sum_{i,j=0}^2 \ket{\hat{i}}\bra{\hat{j}} \bra{i} \Tilde{\rho}\ket{j} \, ,
\end{equation}
which embeds the invariant subspace into the two-qubit subspace.

\medskip
\textbf{Example of quantum states compression with translational symmetry defined with respect to $\mathbb{Z}_N$}. Now, let us see a special case of symmetry that will lead to quantum state compression by the HSP algorithm. Later we will discuss more general cases.
Consider an $n$-qubit quantum state $\ket{\varphi}$, which lives in an $N=2^n$ dimensional Hilbert space, with basis $\{\ket{x}\}_{x=0}^{N-1}$. We can define a translational operator $T_{h_0}$, such that 
\begin{equation}
    T_{h_0} \ket{x} = \ket{x+h_0}\, ,
\end{equation}
where $+$ is the group operation of $\mathbb{Z}_N$, i.e.,  addition modulo $N$, and $h_0$ is $2^m$ for some integer $m$ such that $h_0$ is a factor of $N$. Thus the group generated by $h_0$, denoted by $H=\langle h_0 \rangle$ is not the whole group $\mathbb{Z}_{N}$. Note here that $H=\langle h_0\rangle$ is the standard notation in group theory representing the subgroup generated by $h_0$ but not the expectation values. Define the {\em coset state} as
\begin{equation}
    \ket{c+H} := \frac{1}{\sqrt{|H|}} \sum_{h \in H} \ket{c+h} \, ,
\end{equation}
for $c\in C=\{0,1,2,...,2^m-1\}$ being the representative of each coset. For example, when $n=3$, and $m=2$, the hidden subgroup $H$ is $(\{0,4\},+)$ and the set of coset representatives is $C=\{0,1,2,3\}$. It is easy to verify that these coset states satisfy the translational symmetry,
\begin{equation}
    T_{h_0} \ket{c+ H} = \ket{c+H}\, ,
\end{equation}
for all $c\in C$. Moreover, these coset states are orthonormal to each other and it can be prove that they form a basis for the invariant subspace of this translational symmetry $T_{h_0}$, whose dimension is $\frac{|G|}{|H|} = 2^{n-m}$. If the quantum state has a translational symmetry, such that
\begin{equation}
    T_{h_0} \ket{\varphi} = \ket{\varphi} \, ,
\end{equation}
then this quantum state can be compressed to the translational invariant subspace, leading to a compression ratio of $\frac{2^{n-m}}{2^n} = \frac{1}{2^m}$.

\medskip
\textbf{Example of quantum states compression via HSP algorithm}.
How to identify the invariant subspace, given access to many copies of the quantum state that satisfying the translational symmetry with respective to the group structure $\mathbb{Z}_N$?  One method is to apply the HSP algorithm over the group $\mathbb{Z}_N$ to solve for $h_0$ and therefore obtain the subgroup $H = \langle h_0\rangle$ generated by $h_0$.

Specifically, suppose that the source generates a state $\sum_c \alpha_c \ket{c+H}$ each time, but the state generated may vary from time to time. Then, we can use four steps to solve for $h_0$ and, therefore, construct the compression protocol.
First step is to apply $\mathrm{QFT}_{\mathbb{Z}_N}$ to the state and then measure in computational basis to get a bit string $j$. Second step is to repeat the first step for many times to get many bit strings $j$ that satisfy $j = k \, h_0$ for some integer $k$; it does not matter if the states generated by the source are different for each repetition. Thirdly, we solve the smallest common divisor among the sampled $j$'s and you get $h_0$. Finally, we construct the following CPTP map that will compress the state generated by the source,
\begin{equation}
    \mathcal{E} (\rho)= \sum_{i, j =0}^{|C|-1} \ket{c_i}\bra{c_j} \bra{\hat{c_i}}\rho \ket{\hat{c_j}}\, , \label{eq:quantum_compression_E}
\end{equation}
where $\ket{c_i}$ and $\ket{c_j}$ are basis vectors of a Hilbert space of dimension $|C|=2^{n-m}$ and $\ket{\hat{c}_i}:=\ket{c_i + H}$ and $\ket{\hat{c}_j}:=\ket{c_j +H}$ are vectors of the Hilbert space of dimension $2^n$. And the following CPTP map $\mathcal{D}$ will reconstruct the state,
\begin{equation}
        \mathcal{D} (\Tilde{\rho})= \sum_{i, j =0}^{|C|-1} \ket{\hat{c}_i}\bra{\hat{c}_j}\bra{c_i}\Tilde{\rho} \ket{{c_j}}\, .\label{eq:quantum_compression_D}
\end{equation}

To verify the above compression protocol, simply check that after the quantum Fourier transform is applied to the states, only elements in the orthogonal group can be sampled:
\begin{gather}
\mathrm{QFT}_{\mathbb{Z}_N}\sum_{c\in C} \alpha_c \ket{c+H} = \sum_{c\in C} \alpha_c  \mathrm{QFT}_{\mathbb{Z}_N}\ket{c+H}     =\frac{1}{\sqrt{|H^\perp|}}  \sum_{j \in H^\perp} \left(\sum_{c\in C}\alpha_c \chi_j (c)\right) \ket{j} \, ,
\end{gather}
where we have used the following identity
\begin{equation}
\mathrm{QFT}_{\mathbb{Z}_N} \ket{c_0 + H}  = \frac{1}{\sqrt{|H^\perp|}} \sum_{j \in H^\perp} \chi_j (c_0) \ket{j} \, ,
\end{equation}
for the group character function  $\chi_j(c_0) = \mathrm{exp}(\frac{2 \pi \mathrm{i} \, c_0 \,j}{N})$, and $H^\perp =\{j| \chi_j(h) = 1, \, \forall \, h \in H\}$ is the orthogonal group of $H$ (see Supplementary Material C.3 for more details).
This implies that we can sample from the orthogonal group and use equations $\chi_j(h)=1$ for all $h$ to solve for $H$, as we did in the HSP algorithm. The above discussion suggests that HSP algorithm can be used to compress the quantum states with translational symmetry when it is promised that the symmetry is defined with repsect to the group structure $\mathbb{Z}_N$.

\medskip
\textbf{Quantum states with generalized translational symmetry and variational HSP compression.}
More generally, what if the quantum states are promised to have symmetry with respect to an Abelian group $G$ but we do not know the group structure? Then we need to learn the group structure and then figure out the hidden subgroup. For example, the quantum states could either satisfy the Simon's symmetry with repect to group $\mathbb{Z}_2 \times ...\times \mathbb{Z}_2$ discussed in the main text, or is translational invariant with respect to $\mathbb{Z}_N$, but we don not know which is the case. Then this situation is similar to what we formulated as variational HSP compression in Definition 2 of the main text. Thus our variational HSP algorithm to search over different group structures and figure out which symmetry the quantum states possess and then compress them. But the cost function should be modified slightly. For one possible modification, the optimization goal is to minimize the  average infidelity between the input state to the reconstructed state,
\begin{equation}
C_F = 1-\int \bra{\psi} \mathcal{D}\circ\mathcal{E}(\ket{\psi}\bra{\psi})\ket{\psi} \mathrm{d} \mu_{\ket{\psi}} \, . 
\label{eq:infedility_cost}
\end{equation}
However, since the infidelity is evaluated by measurements and then taking average,  in order for the variational algorithm to work, we must have access to multiple copies of the same quantum sate such that the infidelity cost can be evaluated. This lead to a slightly different assumption of the random source. Instead of generating one copy of a state $\ket{\varphi}$ satisfying the translational symmetry, it generates multiple copies $\ket{\varphi}^{\otimes n}$ each time, making it possible to measure the infidelity between the state $\ket{\varphi}$ and the reconstructed state.

Mathematically, we can formulate the following quantum state compression problem where our variational algorithm can be used to compress it. Define the generalized translational operator as
\begin{equation}
    T_{h_0}\ket{x} = \ket{x * h_0} \, ,
\end{equation}
where $*$ is the group operation in the Abelian group $G$. A source will randomly generate quantum state $\ket{\varphi}^{\otimes n} \in \mathcal{S} $ according to a probability density $\mu_{\ket{\psi}}$, where the state $\ket{\varphi}$ is promised to be translational invariant under the action of $T_{h_0}$. Given access to the quantum states generated from this random source but without knowledge of the Abelian group structure $G$ and the hidden subgroup $H=\langle h_0 \rangle$, the goal is to find a CPTP map $\mathcal{E}$ that sends $\ket{\psi}$ to a quantum state $\ket{\Tilde{\psi}}$ with smaller Hilbert space dimension, and a CPTP map $\mathcal{D}$ that recovers  $\ket{\Tilde{\psi}}$ to $\ket{\psi}$. Our variational algorithm can be used to solve this compression problem, with slight cost function change from the reconstruction error of bit strings to the average infidelity between the input and the reconstructed quantum states.

The variational algorithm for this quantum states compression problem takes 6 steps. We list the sketch of the protocol here.
(i) First, apply the variational QFT ansatz to the one copy of the random state $\ket{\varphi}$. (ii) Measure the state after the parametrized QFT in the computational basis to get a bit string $j$. Repeat and get a set of bit strings $j$. (iii) Use steps 3,4,5 of Algorithm 1 in the main text to solve for a tentative generator $h_0$ of the hidden subgroup. (iv) Use $h_0$ to construct the hidden subgroup $H$ and then use Eq.\eqref{eq:quantum_compression_E} and Eq.\eqref{eq:quantum_compression_D} to  construct the encoding map $\mathcal{E}$ and the decoding map $\mathcal{D}$. (v) evaluate the infidelity between the input state $\ket{\varphi}\bra{\varphi}$ and the reconstructed state $\mathcal{D}\circ\mathcal{E}(\ket{\varphi}\bra{\varphi})$
. (vi) Finally, use gradient descent to update the parameters in the QFT ansatz to minimize the the infidelity cost Eq.\eqref{eq:infedility_cost}. }

\section{Introduction to hidden subgroup problem and representation theory of Abelian groups.\label{supplementary:hidden_subgroup_problems_formal_discussion}}

The Fourier sampling method is used to solve the HSP over Abelian groups while it does not work for non-Abelian groups in general~\cite{lomont2004hidden}. However, there are certain special cases of non-Abelian groups for which Fourier sampling can be applied effectively to solve HSP over non-Abelian groups. For example, Fourier sampling could work for non-Abelian HSP when the subgroup is normal, or the group $G$ is Dihedral group
~\cite{grigni2001quantum, kuperberg2005subexponential}. What is more, Ref.~\cite{grigni2001quantum} discusses the use of Fourier sampling as a subroutine in solving the group that is close to an Abelian group. 
There is still a potential to extend our variational method to certain types of non-Abelian groups. In this regard, here we give a more formal and general discussion on the definition of the general HSP and the general form of Fourier sampling. While the notions we reviewed here are generally applicable to non-Abelian case, we focus on the Abelian HSP because our main result is about the Abelian HSP. 

In general, an HSP is about finding the hidden structure in a group $G$. The hidden structure depends on the function $f$ by the following  definition:
\begin{definition}[Hidden subgroup problem]
    Given a group $ G $, set $ X $, and a group function $ f: G\to X $, a subgroup $ H $ is called a hidden subgroup if it satisfies $ f(g_{1}) = f(g_{2}) $ if and only if $ g_{1}H = g_{2}H $, where the $ g_{i}H $ are cosets of $ H $. The \emph{hidden subgroup problem} is to determine $H$ given access to the function $f$ as a black box.
\end{definition} 
The definition here is a generalization of Abelian HSP. Take the period finding problem as an example, the group $ G $ has the elements of $ \mathbb{Z} $, and the group addition is the normal arithmetical $ + $. Thus, the function $ f $ is a group function which maps $ (\mathbb{Z}, +) $ to $ X= \mathbb{Z}$, and $ f: \mathbb{Z} \to \mathbb{Z} $. The hidden sub-group of interest $H=\{0,r,2r,...\}$ is generated by $r$, and $H=\{nr\}_{n=0,1,...\infty}$. 
The HSP is equivalent to finding $r$ in this example, given black-box access to $f$. 

This section is organized into three subsections. Firstly, in Supplementary Material~\ref{supplementary:math_preliminary}, we introduce group representation theory, a mathematical tool that proves to be highly useful in solving the HSP. Secondly, in Supplementary \ref{supplementary:fourier_sampling}, we present a generalized version of Fourier Sampling using the language of group representation theory. Lastly, in Supplementary \ref{supplementary:classical_algorithm}, we discuss the classical post-processing algorithm for the outcomes of Fourier sampling, also generalized in the language of group representation theory. By combining these three subsections, we generalize HSP to the non-Abelian case, from the problem description to its treatment in the Abelian case. This provides a framework to explore how our scheme can be extended to the non-Abelian case, as well as to evaluate its performance in this context.

\subsection{Group representations and characters\label{supplementary:math_preliminary}}

The quantum period finding algorithm~\cite{nielsen2002quantum}  finds this hidden subgroup, which means it solves an HSP. Other HSPs have been solved using similar quantum techniques over finite Abelian groups, while solving the HSP over non-Abelian problems still remains an open question. This paper will mainly focus on the finite Abelian case. 

The standard method for solving finite Abelian HSP is called quantum Fourier sampling. 
Before we begin describing this Fourier sampling, certain necessary representation theory preliminaries will be introduced.
\begin{definition}[group representation]
    A group homomorphism $ \rho : G \to GL(V)$ is called the group representation of group $ G $, where $GL(V)$ means the general linear group of invertible matrices acting on vector space $V$ with matrix multiplication as the group operation. 
\end{definition}

\begin{definition}[invariant subspace]
    Let $ (\rho, V) $ be the representation of $ G $, and consider a subspace $ U\subset V $. $ U $ is called a $ G $-\emph{invariant subspace} of $ V $ if for all elements $ g\in G $ and all vectors $ v\in V $, $ \rho(g)v\in U $ . 
\end{definition}
\begin{definition}[irreducible representation]
    Let $ (\rho, V) $ be a representation of group $ G $. $ \rho $ is defined as an {\em irreducible representation} (irrep) , if $ G $-invariant subspaces of $ \rho $ are only either $ V $ or $ \{0\} $.  
\end{definition}

For an Abelian group, the representation space of all the non-equivalent irreducible representations are one dimensional, i.e.the invariant vector space has only one basis vector. The irreps can then be labelled by a group element $\rho_h$, where $h\in G$. 
The following process shows how this is done.
The finite Abelian group $G$ is isomorphic to the direct sum of a series finite cyclic group $\mathbb{Z}_{p_i}$, 
\begin{equation}
    G\simeq \mathbb{Z}_{p_1}\times \mathbb{Z}_{p_2} \times\cdots \times \mathbb{Z}_{p_s}.
\end{equation}
The irreps of cyclic group $\mathbb{Z}_p$ are $\rho_t(q) = \omega_p^{qt}$, where $\omega_p = e^{\frac{2\pi i}{p}}$, and $t$ is an integer. 
Let the group element $g\in G$ be $g = a_1^{q_1} a_2^{q_2} \cdots a_s^{q_s}$, 
where $\langle a_i \rangle \simeq \mathbb{Z}_{p_i}$, and $t_i\in \mathbb{Z}$, $\langle a_i \rangle$ is the group generated by $a_i$.
Similarly, denote $h\in G$ as $h = a_1^{t_1} a_2^{t_2} \cdots a_s^{t_s}$.
Then, all non-equivalent irreps of finite Abelian groups are
\begin{equation}
    \rho_h(g) :=  \omega_p^{hg}:=
    \omega_{p_1}^{t_1q_1}\omega_{p_2}^{t_2q_2}\cdots\omega_{p_s}^{t_sq_s}~.
    \label{Eqn: use g label rho}
\end{equation}

Equation \eqref{Eqn: use g label rho} gives the precise expression of all non-equivalent irreps over finite Abelian groups. These irreps form a set of basis functions because they are orthogonal to each other\cite{serre1977linear}.
\begin{definition}[inner product over group]
    The inner product of two group functions
    $f_1, f_2 : G \to \mathbb{C}$ is defined by
\begin{equation}
    \langle f_1, f_2 \rangle_G = \frac{1}{|G|}\sum_{g\in G}f_1(g) f_2^*(g), 
\end{equation}
where the $f^*(g)$ denotes the complex conjugate of $f(g)$. 
\end{definition}
Considering irreps $\rho_h(.)$ as such functions, we have the following theorem.

\begin{theorem}[Orthogonality relation of the first kind]
The irreps of finite Abelian groups are orthogonal:
\begin{equation}
    \langle \rho_h, \rho_{h'} \rangle_G =  \delta_{h,h'}~.
\end{equation}
\end{theorem}
The details of the theorem and the proofs can be found e.g.\ in Ref.~\cite{serre1977linear}. Due to the orthogonality relation of Abelian representations, the representations $\rho_h$ have the potential to be the basis functions of all group functions $f: G\to X$. In particular, a function $f$ could be expanded in terms of  representations~\cite{tao2009fourier} as
\begin{equation*}
    f(g) = \sum_{h\in G} \langle f,\rho_h \rangle_G\rho_h(g)~.
\end{equation*}

 The Fourier transform is a basis transform that aims to get the coefficients of basis functions. For the functions over Abelian groups, the basis functions may be  $\rho_h(g)$. Thus, the Fourier Transform over an Abelian group could be defined from this expansion.
\begin{definition}[Fourier Transform]
The Fourier transform of function $f:G\to X$ over finite Abelian group is defined as
\begin{equation}
    \hat{f}(h) := \langle f,\rho_h\rangle_G = \frac{1}{\sqrt{|G|}}\sum_{g\in G} f(g) \rho_h(g)~.
\label{Eqn: FT}
\end{equation}
\end{definition}

Equation \eqref{Eqn: FT} leads to the quantum Fourier transform(QFT) over finite Abelian groups. However, it is still some distance away from the familiar FT. The next few paragraphs will show that the FT given by the Eq.~\eqref{Eqn: FT} is consistent with the QFT we are familiar with.
\begin{equation}
    QFT\ket{i} = \frac{1}{|G|} \sum_j \rho_j(i)\ket{j}~.
    \label{Eqn: QFT}
\end{equation}
Denoting the computational basis as $\ket{i} = \sum_j \delta_i(j)\vec{v_j}$, where $\delta_i(j)$ is the delta function,  
and $\{\vec{v_j}\}$ is a set of basis. The Fourier transform of $\ket{i}$ is 

 \begin{align}
     \text{FT} \ket{i}
     =& \text{FT} \sum_{j = 0}^{n-1}\delta_i(j)\vec{v_j} = \sum_{j = 0}^{n-1} \hat{\delta}_i(j) \vec{v_j}\nonumber\\
     =& \frac{1}{\sqrt{|G|}}\sum_{j}\sum_{k} \delta_i(k) \rho_j(k)\vec{v_j} \nonumber\\
     =& \frac{1}{\sqrt{|G|}}\sum_{j}\rho_j(i)\vec{v_j}\nonumber\\
     =& \frac{1}{\sqrt{|G|}}\sum_{j} \rho_j(i) \ket{j}. 
     \label{eqn: FS explaination}
 \end{align}
Eq.~\eqref{eqn: FS explaination} aims to explain why quantum FT obtains the form in Eq.~\eqref{Eqn: QFT}. By considering a state $\ket{\psi}$ as a discrete function $\phi: \Z \to \mathbb{C}$, a FT acts on state $\ket{\psi}$ could be regarded as acts on the function $\phi$, where $\phi$ gives the coefficient of the state in the computational basis. 
\subsection{Fourier sampling \label{supplementary:fourier_sampling}}

\begin{figure}
    \centering
    \includegraphics[width = .6\linewidth]{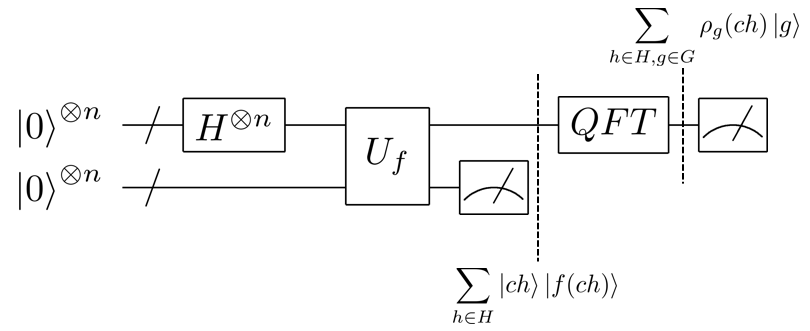}
    \caption{
        (i). The states below or above the blue line are the state after each step in Algorithm \ref{alg:FS}.  
        (ii). Because $f(h_1) = f(h_2)$ iff $h_1, h_2 \in cH$, we can use $f(cH)$ to represent the same value of $f(h)$, $h\in cH$. 
        (iii) After the measurement of the first register, a $\ket{j}$ is read out. Due to Equation \eqref{Eqn: use g label rho}, $\ket{j}$ corresponds to a representation $\rho_j$. So we will get a representation by this process.
    }
\label{fig:FS}
\end{figure}

This shows the matrix form of the FT, which  also defines the quantum (unitary) Fourier transform (QFT)~\eqref{Eqn: QFT}. With the definition of general QFT over finite Abelian group $G$, we can now describe the Fourier sampling(FS)\cite{lomont2004hidden} algorithm that solves the HSP over $G$. The process of FS is described in Algorithm \ref{alg:FS}, and shown in Fig.\ref{fig:FS}. Firstly, use Hadamard gates to get the superposition state $\sum_{x\in G} \ket{x,0}$. Secondly, apply the oracle $U_f$ which satisfies $U_f\ket{x,0} = \ket{x,f(x)}$ to get the quantum data table $\sum_x\ket{x,f(x)}$. Finally, read out the second quantum register and do QFT on the first register. Due to the property of hidden subgroup conserved function $f$, which is specifically expressed as $f(g_1) = f(g_2)$ iff $g_1H= g_2H$. Thus, $g_1,g_2\in cH$, where $c$ is a group element that label the coset $cH = g_1H = g_2H$. Therefore, the state in the first register is $\frac{1}{\sqrt{|H|}}\sum_{h\in H}\ket{ch}$. These results are the label of irreps of group G, which includes the information of hidden subgroup. The whole Fourier sampling process is described in Alg. \ref{alg:FS}.


\begin{algorithm}[H]
  \SetAlgoLined
  \KwIn{Group $G$, An oracle that calculate group function $f$.}
  \KwOut{a irrep $\rho$ such that $\rho(k) = 1$.}
  $n = \log_2 |G| $.\;
  Initialize two quantum registers. $R_1 \gets \ket{0}^{\otimes n}$, $R_2 \gets \ket{0}^{\otimes n}$.\;
  Apply hadamards, $R_1\gets H^{\otimes n}R_1 $.\;
  Apply oracle, $R_1\otimes R_2 \gets U_f ~R_1\otimes R_2$.\;
  Measure $R_2$.\;
  Do QFT to $R_1$.\;
  read out $R_1$ and get a binary number $b$.\;
  return the irrep correspond to $b$
  \caption{{\bf Fourier Sampling.}} 
  \label{alg:FS} 
\end{algorithm}

 The state $ \frac{1}{\sqrt{|H|}}\sum_{h\in H}\ket{ch}$ actually contains the information about $H$, but we can not directly extract the information by measuring it. After measuring the second register of $\sum_{x}\ket{x,f(x)}$, it will randomly choose a $c$ and get state $\frac{1}{\sqrt{|H|}}\sum_{h\in H}\ket{ch}$. Then, if we directly measure the first register, it will randomly collapse to $\ket{ch}$. Combining the 2 processes, the probability to get any state $\ket{g}$ is uniformly $Prob(\ket{g}) = \frac{1}{|G|}$. 
 
Thus, directly sampling the second quantum register is useless. However, here is a way to remove its influence. The way to remove the influence is by removing the influence of $c$, which is called the shifting in Jozsa's analysis \cite{jozsa2001quantum}. 
As it is mentioned in Equation \eqref{Eqn: QFT}, the QFT of this state is 
\begin{equation}
    \text{QFT}\frac{1}{\sqrt{|H|}}\ket{ch} = \frac{1}{\sqrt{|G||H|}}\sum_{h,g\in G}\rho_g (ch)\ket{g}.
\end{equation}
Because the representations can be reformed into an unitary matrix\cite{lomont2004hidden}, the norm of a matrix is the same as the norm of the matrix multiply a representation unitary. This fact gives us the technique to calculate the probability of state $\ket{g}$,  
\begin{equation}
    \left | \sum_h \rho_g(ch) \right |^2 = \left | \rho_g(c)\sum_h \rho_g(h) \right |^2 =  \left | \sum_h \rho_g(h) \right |^2 .
\end{equation}
After applying QFT, the information of $H$ can be extracted by measuring the first register. The measurement yields a $\ket{g}$. Then Equation \eqref{Eqn: use g label rho} is used to compute the representation $\rho_g$. 

This process samples a representation that satisfies $H\subset\ker(\rho)$, where $\ker(\rho)$ is the kernel of representation $\rho$, $\ker(\rho) := \{k~|~\rho(k) = 1\}$ if $\rho$ is a representation for an Abelian group.  

Here is the reason. The probability distribution of sampling representation is decided by the amplitude $\left|\sum_h\rho_g(ch)\right|^2$.  

\begin{equation}
    \sum_h\rho_g(ch) = \sum_{t}\exp\left(2\pi i\sum_j (q_jt_j/p_j)\right), 
    \label{Eqn: sum rho_h}
\end{equation}
where $t$ is the parameter that depicts $h = \prod a_i^{t_i}$. Because $H<G$, $H = \langle a_i^{r_i}\rangle$, Equation \eqref{Eqn: sum rho_h} is non-zero only when $q_jr_j | p_j$. In other words, the representation that could be sampled by Algorithm \ref{alg:FS} satisfy $\rho_g(h) = 1$, which means $H\subset \ker(\rho)$. Looping $m$ times Algorithm \ref{alg:FS}, $m$ representations could be sampled. 

The set of sampled representations gives the hidden subgroup $H$ via a classical algorithm.

 \subsection{Classical algorithm}\label{supplementary:classical_algorithm}

  There is also classical post-processing. 
    A set of sampled representations reflect a lot of information about the hidden subgroup. The following algorithm \cite{hallgren2003hidden} requires only $m = 4\log_2|G|$ loops to approximately get the hidden subgroup $H$ within probability $1-2\exp(-\log_2|G|/8)$. 
    The idea of the classical post-processing algorithm is to utilize the common kernel of sampled irreps. A subgroup that contains hidden subgroup $H$ is got by calculating the kernel  of the irreps sampled by Alg. \ref{alg:FS}, because they satisfy $H\subset \ker(\rho) $. By continuously intersecting those subgroups, the intersection will exponentially converge to the common part in those subgroups, which is exactly what we want, the hidden subgroup $H$. 
    The whole algorithm is shown in Alg. \ref{alg:Classical alg}.
  \begin{algorithm}[H]
  \SetAlgoLined
  \KwIn{$m > 4\log_2|G|$ irreps $\{\rho_j\}_{j = 1}^m$ which satisfy $ H\subset \ker(\rho_j) $.}
  \KwOut{hidden subgroup $H$ with in probability $1-2\exp(-\log_2|G|/8)$.}
  Initialize $N_0 = G$, $i = 1$.\;
  \While {$i\leq m+1$}{
  State ~~~~$N_i = N_{i-1} \bigcap \ker(\rho_i)$.} 
  Output $N_m$
  \caption{{\bf Classical algorithm after FS for Abelian groups.}} 
  \label{alg:Classical alg} 
\end{algorithm}

The computational complexity of FS is mainly caused by the QFT over a group, while the QFT over Abelian groups have been efficiently constructed. Kiteav proposed a model to construct QFT over $\mathbb{Z}_p$ for any $p \in \mathbb{Z}^+$\cite{kitaev1995quantum}, and Hales and Hallgren improved it in 2000\cite{hales2000improved}.
The most famous classical algorithm for computing the Fourier transform is the fast Fourier transform. The fast Fourier change is based on the decomposition of a chain of invariant subgroups for the group structure. In fact, there are several more general approaches to the classical algorithm for the Fourier transform on a finite group, which are based on subgroup tower\cite{ekert1998quantum}. The subgroup tower is a subgroup chain 
$
    G = H_t>H_{t-1}>\cdots > H_0 = \{0\}
$, 
where $H_{i+1}>H_i$ means $H_{i}$ is a subgroup of $H_{i+1}$. For example, when $G = \mathbb{Z}_{2^n}$, $H_t \simeq \mathbb{Z}_{2^t}$. From Eq.~\eqref{Eqn: FT}, we can see that the classical approach to calculate FT is based on the following recursion. 
\begin{align}
    \hat{f}(j) =& \frac{1}{\sqrt{2^n}}\sum_{k = 0}^{2^n-1} f(k)\rho_j(k)\nonumber\\
     =& \frac{1}{\sqrt{2}} \left [ \sum_{k = 0}^{2^{n-1}-1}f(2k)\frac{\omega^{2jk}_{2^n}}{\sqrt{2^{n-1}}}
     + \omega_{2^n}^j\sum_{k = 0}^{2^{n-1}-1}f(2k+1)\frac{\omega^{2jk}_{2^n}}{\sqrt{2^{n-1}}}  \right]\nonumber\\
     \label{Eqn: recursive FT} =& \frac{1}{\sqrt{2}} \left[ \hat{f}_1 (j) + \omega_{2^n}^j \hat{f}_2(j)\right] ~. 
\end{align}
    $f_1 = f|_{\mathbb{Z}_{2^k-1}}$ is the $f$ which limits its domain on subgroup $\mathbb{Z}_{2^k-1} $, so $\hat{f}_1 (j)$ is the FT over subgroup $\mathbb{Z}_{2^k-1} $. Suppose FT over $\mathbb{Z}_{2^k}$ needs $T(2^n)$ time to achieve, then Equation \eqref{Eqn: recursive FT} deduce that $T(2^n) = 2T(2^{n-1})+\mathcal{O}(2^n)$ in classical case, which gives $ T(2^n) = \mathcal{O}(n2^n) $. In quantum case, it just needs a unitary to calculate all the $\hat{f}(j)$ in Equation \eqref{Eqn: recursive FT} if we have the FT over subgroup, so $T(2^n) = 2T(2^{n-1})+\mathcal{O}(n)$. That gives the computation time of QFT over $\mathbb{Z}_{2^k} $ as $T(2^n) = \mathcal{O}(n^2)$. More explictly gave the quantum circuit of QFT base on this subgroup tower idea\cite{moore2006generic}, and also promoted it to some special kinds of non-Abelian group. With the results of efficiently QFT over Abelian group, the HSP over Abelian group has been solved.

 \end{document}